\documentclass[aps,prd,twocolumn,nofootinbib,superscriptaddress,preprintnumbers,balancelastpage,longbibliography]{revtex4-2}
\usepackage[english]{babel}

\usepackage{amsfonts}
\usepackage{amsmath,mathtools,physics,xfrac}
\usepackage[normalem]{ulem}
\usepackage{graphicx}
\usepackage{afterpage}
\usepackage{float}
\usepackage{subfigure}
\usepackage{rotating}
\usepackage{multirow}
\usepackage{tabularx}
\usepackage{booktabs}
\usepackage{fancyhdr}
\usepackage[utf8]{inputenc}
\usepackage{theorem}
\usepackage{moreverb}
\usepackage{euscript}
\usepackage{psfrag}
\usepackage{slashed}
\usepackage{mathtools}
\usepackage{makecell}
\usepackage{adjustbox}
\usepackage{dcolumn}
\usepackage{bm}
\usepackage[dvipsnames]{xcolor}
\usepackage{graphics}
\usepackage{hyperref}
\usepackage{lettrine}

\input Zallman.fd

\definecolor{darkviolet}{rgb}{0.58, 0.0, 0.83}

\hypersetup{
     colorlinks   = true,
     citecolor    = darkviolet,
     urlcolor     = darkviolet,
     linkcolor    = darkviolet
}

\usepackage{color,xcolor}

\definecolor{darkblue}{rgb}{0.0,0.0,0.75}
\definecolor{darkred}{rgb}{0.6,0.0,0}
\definecolor{darkgreen}{rgb}{0.0,0.6,0.}

\newcommand{\mchi}{m_\chi}

\newcommand\redsout{\bgroup\markoverwith{\textcolor{red}{\rule[0.5ex]{2pt}{0.4pt}}}\ULon}

\begin{document}

\preprint{SLAC-PUB-250312}

\title{Sub-GeV Dark Matter Direct Detection with Neutrino Observatories}

\author{Rebecca~K.~Leane}
\thanks{\href{mailto:rleane@slac.stanford.edu}{rleane@slac.stanford.edu}; \href{http://orcid.org/0000-0002-1287-8780}{0000-0002-1287-8780}
}
\affiliation{Particle Theory Group, SLAC National Accelerator Laboratory, Stanford, CA 94305, USA}
\affiliation{Kavli Institute for Particle Astrophysics and Cosmology, Stanford University, Stanford, CA 94305, USA}

\author{John F. Beacom}
\thanks{\href{mailto:beacom.7@osu.edu}{beacom.7@osu.edu}; \href{https://orcid.org/0000-0002-0005-2631}{0000-0002-0005-2631}
}
\affiliation{Center for Cosmology and AstroParticle Physics (CCAPP),
Ohio State University, Columbus, Ohio 43210, USA}
\affiliation{Department of Physics, Ohio State University, Columbus, Ohio 43210, USA}
\affiliation{Department of Astronomy, Ohio State University, Columbus, Ohio 43210, USA}

\date{\today}

\begin{abstract}
We present a new technique for sub-GeV dark matter (DM) searches and a new use of neutrino observatories. DM-electron scattering in an observatory can excite or ionize target molecules, which then produce light that can be detected by the photomultiplier tubes (PMTs). While individual DM scatterings are indistinguishable, the aggregate rate from many independent scatterings can be isolated from the total PMT dark rate using the expected DM annual modulation. We showcase this technique with the example of JUNO, a 20,000-ton scintillator detector, showing that its potential sensitivity in some mass ranges exceeds other techniques and reaches key particle-theory benchmarks.
\end{abstract}

\maketitle

%%%%%%%%%%%%%%%%%%%%%%%%%%%%%%%%%%%%%%%%%%%%%%%%%%%%%%%%%%%%%%%%%%%%%%%%%%%%%%%%%%%%%
%%%%%%%%%%%%%%%%%%%%%%%%%%%%%%%%%%%%%%%%%%%%%%%%%%%%%%%%%%%%%%%%%%%%%%%%%%%%%%%%%%%%%

\lettrine{I}{t is very hard} to test dark matter (DM) in the MeV--GeV mass range. A key challenge is the low recoil energies in scattering with matter. Compared with DM in the GeV mass range, the allowed cross sections are several orders of magnitude larger, making them tantalizing targets. Therefore, improving sensitivity to sub-GeV DM is a major goal of the particle physics community~\cite{Essig:2022dfa}. There has been a surge of interest in novel detection strategies, including using superconductors, superfluids, organic targets, quantum devices, and more~\cite{Hochberg:2022apz,Essig:2011nj, Graham:2012su, Essig:2015cda, Hochberg:2015pha, Hochberg:2015fth, Hochberg:2019cyy, Hochberg:2021ymx, Hochberg:2021yud, Derenzo:2016fse, Schutz:2016tid, Hochberg:2016ntt, Essig:2016crl, Hochberg:2017wce, Cavoto:2017otc, Griffin:2018bjn, Sanchez-Martinez:2019bac, Essig:2019xkx, Emken:2019tni, Hochberg:2022apz, Kurinsky:2019pgb, Blanco:2019lrf, Griffin:2020lgd, Blanco:2021hlm, Baxter:2019pnz,Kurinsky:2020dpb,Geilhufe:2019ndy,Emken:2017erx,Emken:2017qmp,Catena:2019gfa,Radick:2020qip,Gelmini:2020xir,Trickle:2020oki,Knapen:2021run,Cook:2024cgm, Simchony:2024kcn, Essig:2022dfa, Das:2022srn, Das:2024jdz, Griffin:2024cew, QROCODILE:2024zmg, Marocco:2025eqw}. 

In this \textit{Letter}, we present a new technique for sub-GeV DM searches and a new use of neutrino observatories. The basic ideas are as follows. DM-electron scattering in an observatory excites or ionizes target molecules, producing eV-range photons that can be detected with the photomultiplier tubes (PMTs). \textit{Individual} DM-electron scattering events are indistinguishable because they typically lead to less than one PMT hit. However, we show that neutrino observatories can search for the \textit{aggregate} single-hit signal rate from many independent DM scattering events. This signal rate must be compared to the observatory's dark rate, which is the total rate of PMT hits due to thermal noise and other sources. Though the DM signal rate is small compared to the dark rate, it can be isolated using the annual modulation due to Earth's orbital motion. This is a new use of neutrino observatories, which use many PMT hits per event to reconstruct signals and backgrounds. Figure~\ref{fig:schematic} illustrates the new DM signal.

We demonstrate this new technique through the example of the Jiangmen Underground Neutrino Observatory (JUNO)~\cite{JUNO:2020xtj, JUNO:2021vlw}. The first advantage is a huge target volume --- here 20,000 ton --- whereas direct-detection experiments have only reached the ton scale. The second is that the detector material --- linear alkylbenzene (LAB), a liquid scintillator ---  has an electronic excitation threshold of only a few eV, whereas noble liquids have thresholds of order 10--100~eV, due to differences in their electronic structure. The disadvantage of neutrino observatories compared to direct-detection experiments is much larger backgrounds. However, we show that the net effect is that JUNO has sensitivity to DM-electron interactions in some mass ranges that (A) exceeds all other techniques and (B) reaches some important theoretical targets. Although we focus on JUNO, our technique can be adapted to other neutrino observatories; combining several results will make it even more robust.  This new use of neutrino observatories complements familiar techniques, $e.g.$, searches for DM annihilation products~\cite{1985ApJ296679P, Srednicki:1986vj, Ritz:1987mh, Meade:2009mu, Bell:2011sn, Feng:2016ijc, Leane:2017vag, Arina:2017sng, Mazziotta:2020foa, Super-Kamiokande:2015xms, Abe:2011ts, ANTARES:2016xuh, ANTARES:2016bxz, IceCube:2016yoy, IceCube:2020wxa, Bell:2021esh, Leane:2024bvh, Nguyen:2025ygc} and the scattering of boosted~\cite{Yin:2018yjn, Ema:2018bih, Bringmann:2018cvk, Cappiello:2019qsw, Kahn:2014sra,Agashe:2014yua, Kopp:2015bfa, Lei:2020mii, An:2021qdl, Wang:2021jic, Toma:2021vlw, Aoki:2023tlb, Das:2021lcr, CDEX:2023wfz, Bell:2023sdq, Liang:2024xcx, Dutta:2024kuj}, heavy, or strongly interacting DM~\cite{Bramante:2018tos, Bai:2022nsv, Eby:2019mgs, Aggarwal:2024ngx, Bramante:2019yss, Ponton:2019hux, Church:2020env}.

%%%%%%%%%%%%%%%%%%%%%%%%%%%%%%%%%%%%%%%%
\begin{figure}[t!]
\centering
\includegraphics[width=0.9\columnwidth]{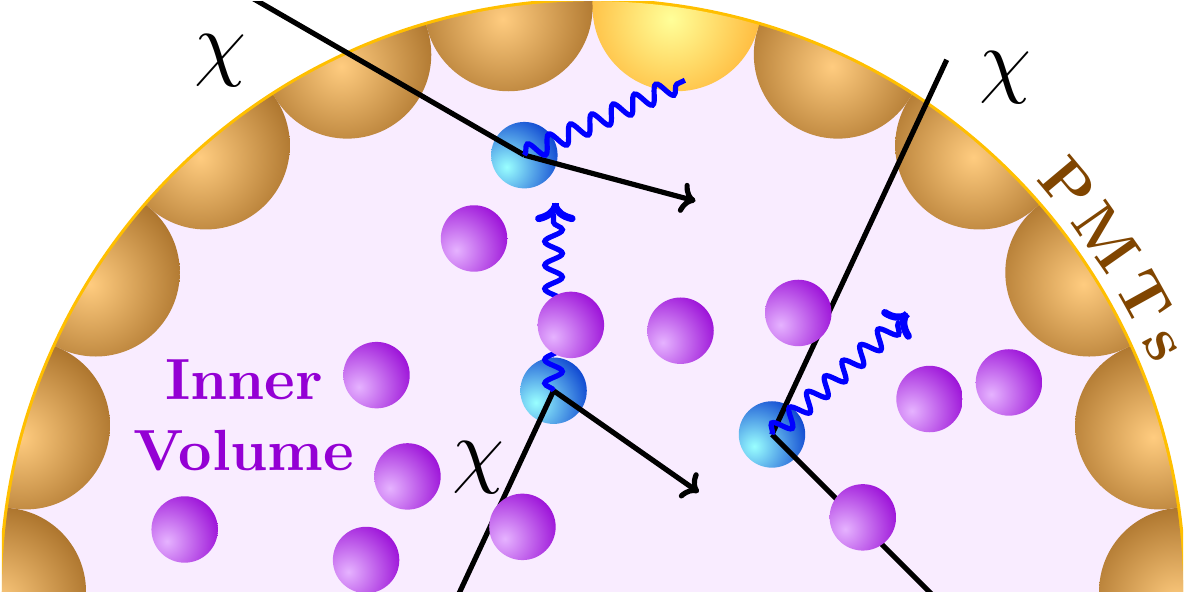}
    \caption{Schematic of the new proposal. The aggregate rate of uncorrelated DM scatterings with molecules in a neutrino observatory produces a statistical signal due to photons detected by the PMTs.}
    \vspace{-2mm}
    \label{fig:schematic}
\end{figure}
%%%%%%%%%%%%%%%%%%%%%%%%%%%%%%%%%%%%%%%%

In the following, we review the JUNO detector and DM scattering with bound electrons, then calculate the DM signal rate and sensitivity. We conclude with a  discussion of next steps. In our calculations, we make several conservative assumptions, sometimes by potentially more than an order of magnitude. In forthcoming work, we will present more details on all aspects.\\

%%%%%%%%%%%%%%%%%%%%%%%%%%%%%%%%%%%%%%%%%%%%%%%%%%%%%%%%%%%%%%%%%%%%%%%%%%%%%%%%%%%%%
%%%%%%%%%%%%%%%%%%%%%%%%%%%%%%%%%%%%%%%%%%%%%%%%%%%%%%%%%%%%%%%%%%%%%%%%%%%%%%%%%%%%%

\smallskip
\noindent\textbf{\textit{Review of the JUNO Experiment.---}}
The JUNO experiment~\cite{JUNO:2020xtj, JUNO:2021vlw}, located in southern China, is planned to begin operations in 2025. JUNO will make many important measurements in neutrino physics and astrophysics, especially probing the neutrino mass ordering by precisely measuring reactor antineutrino mixing.

To achieve its goals, JUNO needs excellent light production and detection. The central detector of JUNO is a transparent acrylic vessel containing 20,000~ton of LAB as well as other compounds to increase the production and transmission of light. These are a small admixture (2.5 g/L) of PPO (2,5-diphenyloxazole) as a fluor and a tiny admixture (3 mg/L) of bis-MSB (1,4-bis(2-methylstyryl)benzene), a wavelength shifter~\cite{JUNO:2020bcl}. The acrylic vessel is surrounded by a $\sim$3-m water buffer and then a support surface that is 78\% covered with high-quantum-efficiency PMTs~\cite{JUNO:2022hlz}. There are 17,612 20-inch PMTs and also 25,600 3-inch PMTs.

JUNO's dark rate, dominated by thermal noise from the PMTs themselves, will be high~\cite{JUNO:2022hlz, Zhang:2023dha}. Its 20-inch PMTs trigger at about 25~kHz each, while its 3-inch PMTs trigger at about 5~kHz each, leading to a total dark rate rate of about $6 \times 10^5$ kHz, $i.e.$, about $5 \times 10^{13}$ PMT hits per day or $2 \times 10^{16}$ per year. For our purposes, it will be enough for JUNO to track the total number of hits per small time interval, which we take to be a day. \\

%%%%%%%%%%%%%%%%%%%%%%%%%%%%%%%%%%%%%%%%%%%%%%%%%%%%%%%%%%%%%%%%%%%%%%%%%%%%%%%%%%%%%
%%%%%%%%%%%%%%%%%%%%%%%%%%%%%%%%%%%%%%%%%%%%%%%%%%%%%%%%%%%%%%%%%%%%%%%%%%%%%%%%%%%%%
%

\noindent\textbf{\textit{Review of DM-Electron Scattering.---}}
Upon scattering with a molecule in the scintillator, DM can excite an electron from the ground state into one of several low-lying unoccupied orbitals that can de-excite by photon emission. In the scintillator, this de-excitation photon can be absorbed and re-emitted as a longer-wavelength photon that can propagate through the volume and be detected by the PMTs. We neglect any signals from the water buffer, as its light yield is much lower.

As this process involves bound electrons, there are important dynamical effects~\cite{Essig:2015cda, Blanco:2021hlm}. First, the average speed of bound electrons is an order of magnitude greater than for DM particles. Second, bound electrons have a distribution of speeds up to high values. These effects substantially change the scattering kinematics and break the simple link between momentum transfer and energy deposition used for DM-nucleon scattering~\cite{Essig:2015cda}. Third, the molecular energy levels have a complicated structure, requiring use of a form factor $f_{ij}$, discussed below.

Still, with some simple arguments, we can understand key energy scales. The energy transferred to the electron (of mass $m_e$), $\Delta{E_e} = -\Delta E_\chi$, is related to the momentum lost by the DM (of mass $m_\chi$ and incoming velocity $\vec{v}$), $\vec{q}$, via energy conservation~\cite{Essig:2015cda},
\begin{equation}
    \Delta{E_e} = -\frac{|m_\chi\vec{v}-\vec{q}|^2}{2\mchi}+\frac{1}{2}m_\chi v^2 =\vec{q} \cdot \vec{v}-\frac{q^2}{2m_\chi}\,,
    \label{eq:deltae}
\end{equation}
where we neglect the tiny molecular recoil. The maximum energy transfer is then~\cite{Essig:2015cda}
\begin{equation}
    \Delta{E_e}\leq \frac{1}{2}m_\chi v^2 \simeq \frac{1}{2} \,\text{eV}\times\left(\frac{m_\chi}{\text{MeV}}\right)\,.
    \label{eq:maxe}
\end{equation}
The high speeds of bound electrons relative to DM enhance the likelihood of excitations sufficient to overcome this threshold. For LAB (with an excitation threshold of about 5 eV), this corresponds to a typical minimum mass of about 10 MeV, above which most DM particles in the halo have sufficient kinetic energy to excite electrons. Below roughly this mass, only particles in the higher-velocity tail of the distribution contribute, making the excitation rate increasingly suppressed.\\

%%%%%%%%%%%%%%%%%%%%%%%%%%%%%%%%%%%%%%%%%%%%%%%%%%%%%%%%%%%%%%%%%%%%%%%%%%%%%%%%%%%%%
%%%%%%%%%%%%%%%%%%%%%%%%%%%%%%%%%%%%%%%%%%%%%%%%%%%%%%%%%%%%%%%%%%%%%%%%%%%%%%%%%%%%%

\noindent\textbf{\textit{DM Signal Rate.---}}
We now calculate the aggregate DM signal rate in JUNO. The reaction rate factor, $\sigma v$, for DM particles to excite molecular electrons from level $i\rightarrow j$ (we assume $i$ is the ground state) is~\cite{Essig:2015cda}
\begin{align}
    (\sigma v)_{ij} = \int \frac{d^3 \vec{q}}{(2\pi)^3} \,(2\pi)\delta(\Delta E_{ij} - \omega ) \frac{\abs{ \mathcal M_\text{free} }^2 \abs{f_{ij}(\vec{q}) }^2 }{16 m_\chi^2 m_e^2},
\end{align}
where $\Delta E_{ij}$ is the difference in molecular energy levels, $\mathcal M_\text{free}$ is the amplitude for DM scattering on free electrons, and the energy transfer to the molecular target is $\omega=\Delta E_\chi$, per Eq.~(\ref{eq:deltae}). There are two form factors, $F_\text{DM}$ and $f_{ij}$. The first parametrizes the momentum dependence of the amplitude and depends on the particle DM model, such that
\begin{equation}
    \frac{\abs{\mathcal M_\text{free}}^2 }{16 m_\chi^2 m_e^2 } \equiv \frac{\pi \bar\sigma_e}{\mu_{\chi e}^2} F_\text{DM}^2(q),
\end{equation}
where $\mu_{\chi e}$ is the reduced mass of the DM-electron system and $\bar \sigma_e$ is the DM-electron scattering cross section, and the bar implies evaluation at the conventional fixed momentum transfer of $q = \alpha m_e$.

The second form factor, $f_{ij}$, parametrizes the molecular response to excitation. LAB is a class of organic compounds with the chemical formula C$_6$H$_5$C$_n$H$_{2n+1}$; in JUNO, n = 10--13~\cite{Lombardi:2019epz}. LAB contains a benzene ring, which features alternating double bonds, resulting in a system of conjugated $\pi$-bonds, which are covalent. In six-membered aromatic rings, six electrons participate in this conjugated $\pi$-electron system. These electrons are effectively delocalized, allowing them to move freely around the aromatic ring. Light production occurs due to electronic transitions where these $\pi$-electrons are excited into higher-energy molecular orbitals and then de-excite by photon emission. 

To the best of our knowledge, the low-excitation threshold properties of LAB have not been exploited for sub-GeV DM detection, but similar aromatic organic targets have, $e.g.$, EJ-301~\cite{Blanco:2019lrf}, which also has a benzene ring. In Ref.~\cite{Blanco:2019lrf}, the form factor of EJ-301 was simply taken to be that for benzene, as the $\pi$-electrons in the benzene ring are expected to dominate the rate. While the lowest-energy transition is suppressed for benzene alone (its symmetry imposes selection rules that suppress some electronic transitions), it will not be for LAB, because the long alkyl chain attached to the benzene ring lowers the symmetry of the molecule, changing the form factor. This symmetry breaking lifts the suppression of the lowest-energy excitation, enhancing the response at low energies. Therefore, taking the benzene form factor underestimates the rate at low momentum transfer, and is a conservative choice. We also adopt this  approach in our analysis, using the benzene form factor alone~\cite{Blanco:2019lrf}. We expect that a proper treatment of the LAB form factor will improve sensitivity. We will do this in future work.

To obtain the total DM scattering rate, $R_{\rm scatter}$, we integrate over the DM velocity distribution, $g_\chi (\vec{v})$,
\begin{align}
    R_{\rm scatter} = \sum_{i, j} N^e_{\rm LAB} \frac{\rho_\chi}{m_\chi} \int d^3 \vec{v} \, g_\chi(\vec{v}) (\sigma v)_{ij}\,,
\end{align}
where $\rho_\chi=0.4$~GeV/cm$^3$ is the local DM mass density and $N^e_{\rm LAB} \simeq  3 \times 10^{32}$ is the number of LAB molecules in JUNO's central detector multiplied by the six $\pi$-electrons in LAB's benzene ring. For the total rate, we approximate $g_\chi(\vec{v})$ to be spherically symmetric, assuming a Maxwell-Boltzmann distribution with a cutoff at the escape velocity of the Milky Way, 
\begin{equation}
    g_\chi(v) = \frac{1}{K} \exp\left(- \frac{ \abs{ \vec{v} + \vec{v_E} }^2 }{v_0^2} \right) \Theta(v_\text{esc} - \abs{\vec{v} + \vec{v_E} } )\,,
    \label{eq:vel}
\end{equation}
where $v_E = 240\text{ km/s}$ is taken as the typical speed of Earth, $v_0 = 230\text{ km/s}$ is the local mean speed, and $v_\text{esc} = 600\text{ km/s}$ is the Galactic escape speed. Last, $K$ is a normalization factor, obtained from enforcing the integral of Eq.~(\ref{eq:vel}) over $d^3 \vec{v}$ to be unity. We also define~\cite{Essig:2015cda,Blanco:2019lrf} 
\begin{align}
    \eta(v^{ij}_\text{min}) &= \int \frac{4 \pi v^2 dv }{v} g_\chi(v) \Theta(v - v^{ij}_\text{min}(q))\,,
    \\
    v^{ij}_\text{min}(q) &= \frac{\Delta E_{ij} }{q} + \frac{q}{2 m_\chi}\,,
\end{align}
where $\eta(v^{ij}_\text{min})$ is the integrated velocity distribution of DM particles above the threshold $v_{\rm min}^{ij}(q)$, which is the minimum speed DM needs for the electron to gain an energy $\Delta E_{ij}$ following a momentum transfer $q$.

The expected signal rate in JUNO is therefore
\begin{align}
    R_{\rm sig} = \xi \frac{ N^e_{\rm LAB} \rho_\chi \bar \sigma_e}{8\pi m_\chi  \mu_{\chi e}^2} \sum_{i,j} \int \frac{d^3 \vec{q} }{q} \, \eta(v^{ij}_\text{min}(q)) F_\text{DM}^2(q)  \abs{ f_{i  j}(\vec{q}) }^2,
    \label{eq:rate}
\end{align}
where we multiply the DM scattering rate, $R_{\rm scatter}$, by an efficiency factor $\xi \simeq 2.2\times10^{-2}$. We obtain this as 
\begin{equation}
    \xi=Q\times X\times Y\,,
\end{equation}
where $Q$ is the fluorescence quantum yield of LAB~\cite{1972JMoSt..13..140L}, $X$ is the wall coverage of the PMTs, and $Y$ is the quantum efficiency
of the PMT photocathode integrated over the emission spectrum of JUNO's active medium, taking into account absorption and emission of photons from the primary scintillating target by the fluor and wavelength shifter. For JUNO, we use $Q=0.1$, $X=0.78$, and $Y=0.28$~\cite{1972JMoSt..13..140L, Beretta:2025rwj, Zhang:2020mqz, Li:2015phc}. In Eq.~(\ref{eq:rate}), a key input that substantially changes the rate compared to a naive estimate (based on the product of the DM number density, velocity, number of targets, and cross section) is the effect of the molecular form factor $f_{ij}$~\cite{Blanco:2019lrf}, which increases the rate by a few orders of magnitude due to coherence of the electrons' wavefunctions.

In Eq.~(\ref{eq:rate}), we conservatively assume the rate purely arises from excitation and fluorescence of LAB. In forthcoming work, we will consider the additional signals from ionization.\\

%%%%%%%%%%%%%%%%%%%%%%%%%%%%%%%%%%%%%%%%%%%%%%%%%%%%%%%%%%%%%%%%%%%%%%%%%%%%%%%%%%%%%
%%%%%%%%%%%%%%%%%%%%%%%%%%%%%%%%%%%%%%%%%%%%%%%%%%%%%%%%%%%%%%%%%%%%%%%%%%%%%%%%%%%%%

\noindent\textbf{\textit{DM Sensitivity from Annual Modulation.---}}
In many dedicated DM detectors, the backgrounds are small after cuts, $e.g.$, requiring that events arise within a fiducial volume, occur in a target energy range, or have certain characteristics. Here, such cuts are not possible because the only observable is the total rate of PMT hits. This rate is dominated by the background (the dark rate, $B \sim 2 \times 10^{16}$ events/yr~\cite{JUNO:2022hlz,Zhang:2023dha}). A conservative DM bound could be set by requiring that the signal rate is less than this background.  Because the background cannot be predicted precisely, we cannot estimate the DM signal sensitivity by requiring $S_{\rm tot} \gtrsim \sqrt{B_{\rm tot}}$, where those are the total numbers of signal and background events in the observation period, $T$.

However, we can isolate even a small DM signal through the annual modulation caused by Earth's orbital motion~\cite{Drukier:1986tm, Freese:2012xd}. The modulation amplitude, $f_{\text{mod}}$, is determined via the change in the scattering rate that appears via Eq.~(\ref{eq:rate}). The rate as written is used on 2 March and 2 September, while it is modified to take into account the fact that the Earth velocity increases by 15~km/s on 2 June and decreases by 15~km/s on 2 December. We find that $f_{\text{mod}} \sim 0.1$; it is somewhat larger (smaller) for light (heavy) DM, which we take into account. The DM signal is then
\begin{equation}
S(t) = \frac{S_{\text{tot}}}{T} \left[1 + f_{\text{mod}} \cos(\frac{2\pi t}{\text{1 yr}} - \frac{\pi}{2})\right]\,,
\label{eq:sigfunc}
\end{equation}
where the DM signal is increased during the first half year after 2 March and decreased in the second half year.

%%%%%%%%%%%%%%%%%%%%%%%%%%%%%%%%%%%%%%%%%%%%%%
\begin{figure*}
    \begin{subfigure}
    {\includegraphics[width=0.49\textwidth]{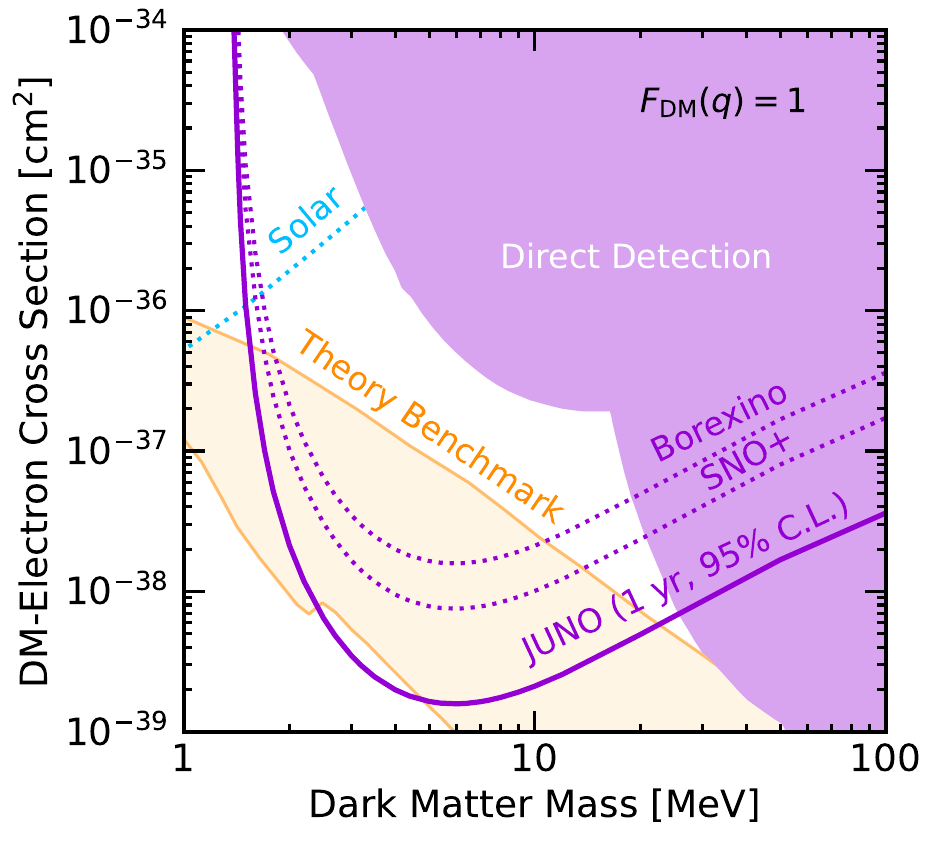}}
    \end{subfigure}
    \begin{subfigure}
    {\includegraphics[width=0.49\textwidth]{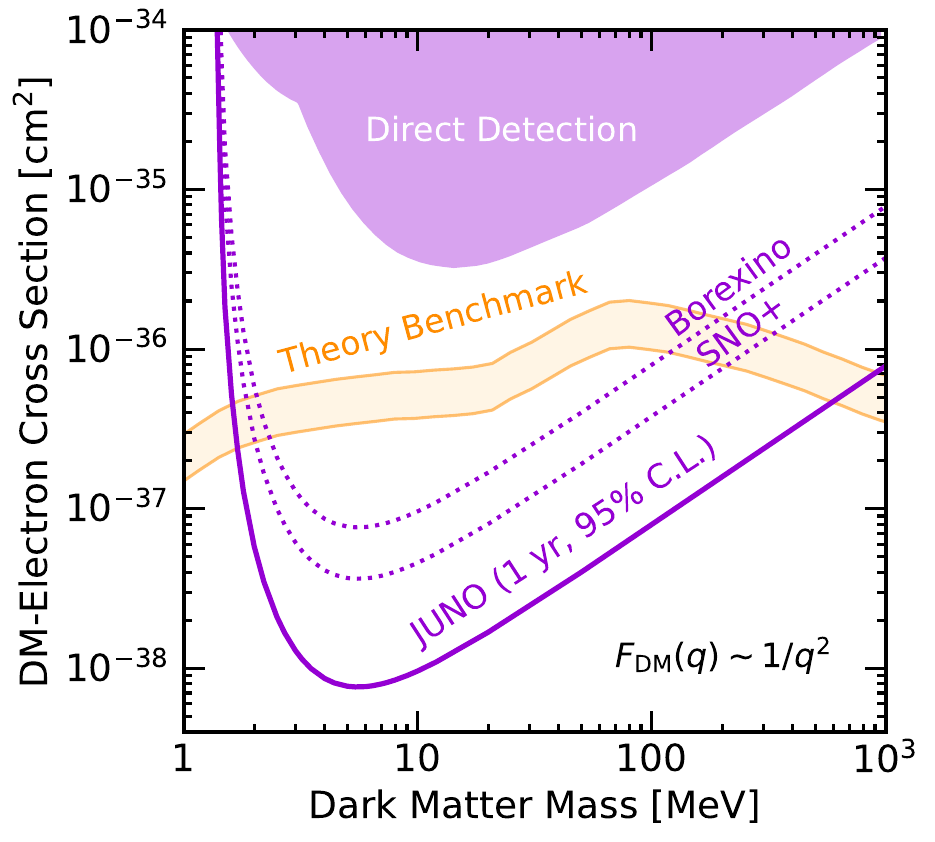}}
    \end{subfigure}
    \caption{JUNO's annual-modulation potential $95\%$~C.L. sensitivity for the DM-electron scattering cross section as a function of DM mass, assuming a nominal one year of exposure (20,000~ton-yr). We also show estimated potential one-year sensitivities for Borexino and SNO+. We show results for a heavy mediator (left panel) and a light mediator (right panel), along with the complementary direct detection~\cite{SuperCDMS:2020ymb, DAMIC:2019dcn, SENSEI:2020dpa, PandaX-II:2021nsg, DarkSide:2022knj, XENON:2019gfn} and solar reflection (``Solar")~\cite{An:2017ojc, Emken:2017hnp} constraints. ``Theory Benchmark" shows parameter spaces for some benchmark particle models~\cite{Essig:2022dfa}. Note differing axes between the two panels.}
    \label{fig:result}
\end{figure*}
%%%%%%%%%%%%%%%%%%%%%%%%%%%%%%%%%%%%%%%%%%%%%%

To isolate a DM signal, one should fit the binned time-series data for $N(t) = B(t) + S(t)$. For the case where the background rate is constant, $B(t) = B_{\text{tot}}/T$, the basic results and uncertainties can be obtained via simple methods~\cite{Eadie:1971qcl, Beacom:1998fj, Lisi:2004jw}; we discuss in the next two paragraphs and Supplemental Material the challenges in real data. The modulated part of the signal counts is $S_{\text{tot}}f_{\text{mod}}$ and the uncertainty on the background counts is $\sqrt{B_{\text{tot}}}$. To set the 95\% C.L. sensitivity, we therefore use $S_{\text{tot}} =2\,\sqrt{B_{\text{tot}}}/f_{\text{mod}} \sim 20 \sqrt{B_{\text{tot}}}$, where we take into account how $f_{\text{mod}}$ varies with mass. There is no trials factor in searching for this DM signal modulation, as its amplitude, period, and phase are all known. Given the background rate of $\sim2 \times 10^{16}$~events/yr, this leads to a minimum DM signal rate of $S_{\rm tot} \sim 3 \times 10^9$ events/yr.

In real experiments, the raw measured dark rates are not constant (see, $e.g.$, Borexino~\cite{Borexino:2013zhu}, Super-K~\cite{Super-Kamiokande:2023jbt}, and IceCube~\cite{IceCube:2011cwc}). There are secular trends as well as short-timescale fluctuations. Careful work by JUNO will be needed to pre-process the data to avoid reaching spurious conclusions~\cite{COSINE-100:2022dvc}. One step should be removing all physics events above $\sim$0.5~MeV and flashers, both of which produce hits across multiple PMTs correlated to one physics event, as well as anomalous activity of individual PMTs. While this cleaning will not appreciably reduce the dark rate, it will greatly reduce its non-thermal-noise components, though residuals can remain~\cite{IceCube:2011cwc}. A second step should be using independent measurements of detector properties to detrend obscuring variations in the data. In JUNO, a powerful tool for this will be the dark rate of the PMTs in the cylindrical, optically isolated outer detector~\cite{Lu:2023ztx}, which is filled with water, for which DM signals are expected to be much smaller. Ultimately, we expect that several years of data will be needed for robust conclusions, though more conservative analyses could be done sooner, including by using subsets of PMTs. Note that we require only the scalar PMT hit rate, $i.e.$, the number of PMT hits per unit time, which can be logged at coarse time resolution ($e.g.$, daily) and is commonly recorded in neutrino detectors for monitoring purposes. This avoids the need for full per-event readout, and imposes negligible bandwidth requirements. JUNO’s existing infrastructure is expected to support this without hardware changes.

Once JUNO obtains a cleaned, detrended total rate, non-statistical fluctuations may remain an obstacle. Those that occur on timescales much less than one year will average out. Of greater concern are components of the background that have annual modulations, some of which are discussed in Refs.~\cite{Blum:2011jf, Davis:2014cja, BOREXINO:2017yyp}. PMT thermal noise depends on the detector temperature, radon and muons depend on atmospheric conditions, solar neutrinos depend on Earth's distance to the Sun, and reactor antineutrinos depend on power cycles. The cleaning and detrending procedures discussed above should greatly reduce the effects of all of these. In addition, independent measurements will be possible. In any case, these modulations are shifted in phase from the DM signal by weeks to months and will have small phase uncertainties. Finally, combining the results of multiple detectors (discussed below) would make the analysis more robust, as the signals depend on the detector properties, while the backgrounds depend on the detector properties and location. Further details are given in Supplemental Material.

Figure~\ref{fig:result} shows the potential sensitivity (i.e., the theoretical sensitivity, which will take experimental work to approach) for the DM-electron scattering cross section as a function of DM mass. We show results for two benchmark scenarios for the DM-electron interaction: via a heavy mediator ($F_\text{DM} = 1$) and via a light mediator ($F_\text{DM}(q) = (\alpha m_e)^2/q^2$). We show results for one year of exposure; over longer periods, additional sensitivity is gained as square root of exposure time. We also show complementary constraints from direct-detection experiments and solar reflection. The relative performance of our neutrino bounds versus direct detection experiments differs between panels due to the distinct energy dependence of the light and heavy mediator interactions, and the differing thresholds of the experiments. The orange shaded regions show parameter spaces for some concrete benchmark models that produce the observed relic DM abundance, though the parameter space of viable  DM models is of course broader, see, $e.g.$, Refs.~\cite{Boehm:2003hm,Boehm:2003ha,McDonald:2001vt,Fayet:2004bw,Sigurdson:2004zp,Strassler:2006im,Sikivie:2006ni,Nussinov:1985xr,Kaplan:1991ah,Zurek:2008qg,Hooper:2008im,Cholis:2008vb,ArkaniHamed:2008qn,Pospelov:2008jd,Feng:2008ya,Feng:2008dz,Pospelov:2008jk,Hall:2009bx,Morrissey:2009ur,Kaplan:2009ag,Essig:2010ye,Cohen:2010kn,Essig:2011nj,Chu:2011be,Falkowski:2011xh,Lin:2011gj,Feng:2011ik,Nelson:2011sf,Arias:2012az,Graham:2012su,Kaplinghat:2013yxa,Hochberg:2014dra,Hochberg:2014kqa,Boddy:2014yra,Boddy:2014qxa,Essig:2015cda,Hochberg:2015vrg,Izaguirre:2015yja,Kuflik:2015isi,Graham:2015rva,Marsh:2015xka,DAgnolo:2015ujb,Pappadopulo:2016pkp,Farina:2016llk,Dror:2016rxc,Feng:2017drg,Kuflik:2017iqs,Choi:2017zww,DAgnolo:2017dbv,Falkowski:2017uya,DAgnolo:2018wcn,Chu:2018qrm,Dvorkin:2019zdi,DAgnolo:2019zkf,Hambye:2019dwd,Heeba:2019jho,Evans:2019vxr,Koren:2019iuv,Chu:2019rok,Chang:2019xva,Darme:2017glc,March-Russell:2020nun,Dodelson:1993je,Shi:1998km,Abazajian:2001nj,Abazajian:2012ys,Abazajian:2017tcc,Boyarsky:2018tvu,Hochberg:2018vdo,Hochberg:2018rjs,Smirnov:2020zwf,Dutta:2019fxn,Dutta:2020enk,Dutta:2022qvn,Fernandez:2021iti,Iles:2024zka}.

Figure~\ref{fig:result} also shows simple estimates of potential one-year sensitivities with some example liquid scintillator neutrino experiments (Borexino and SNO$+$). As an approximation, we assumed the same PMT efficiencies, in-medium spectrum, and propagation as for JUNO, though these inputs will likely be somewhat worse. We also assumed the same form factor as for JUNO, as these are all benzene-based scintillators. We updated the calculation for their specific volumes, dark rates, and PMT coverage. This shows that this technique is promising for multiple detectors. As the cleaned and detrended data are not publicly available, these are only projections.

The position of the minimum near 10~MeV in Fig.~\ref{fig:result} is driven by the low-excitation threshold of LAB being a few eV, as discussed earlier. At higher masses, the sensitivity degrades linearly due to the lower DM number density. At lower masses, it degrades rapidly because fewer DM particles have sufficient momentum to overcome the excitation threshold. Because the modulation fraction grows with decreasing DM mass, the minimum is shifted to slightly lower DM masses (in agreement with, $e.g.$, Ref.~\cite{Essig:2015cda}). Note that near threshold, the modulation of the DM signal might include other detectable harmonics, which could improve signal recovery~\cite{Lee:2013xxa}.\\

%%%%%%%%%%%%%%%%%%%%%%%%%%%%%%%%%%%%%%%%%%%%%%%%%%%%%%%%%%%%%%%%%%%%%%%%%%%%%%%%%%%%%
%%%%%%%%%%%%%%%%%%%%%%%%%%%%%%%%%%%%%%%%%%%%%%%%%%%%%%%%%%%%%%%%%%%%%%%%%%%%%%%%%%%%%

\noindent\textbf{\textit{Conclusions and Prospects.---}}
The nature of DM remains one of the most important open questions in physics. Despite extensive efforts, a wide array of DM experiments have yet to yield definitive signals. This has motivated increasing interest in exploring broader possibilities than GeV-range mass WIMPs and other theoretically motivated DM candidates. For example, DM candidates in the sub-GeV mass range are now the subject of intense focus. However, detecting sub-GeV DM is notoriously difficult, due to the small energy transfers in their scattering interactions.

We introduce a new technique to probe sub-GeV DM and a new use of neutrino observatories. Specifically, we demonstrate that DM-electron scattering in neutrino observatories can produce detectable signals through the excitation of target molecules. While individual scattering events are indistinguishable, the aggregate rate from many independent scatters can be isolated from the PMT dark rate through annual modulation. Focusing on JUNO, we show that this method enables world-leading sensitivity to the DM-electron scattering cross section in some mass ranges, reaching cross sections as low as about $10^{-39}$~cm$^2$ for MeV-scale DM with a nominal one year of data. This power comes from JUNO's huge size and high fluorescence quantum yield.

The goal for this first paper is to show the community that there is an exciting possible way forward but that it will be challenging, requiring substantial work from both theorists and experimentalists. Next steps include refinement of the theoretical and experimental details for this technique in JUNO. This requires a deeper understanding of the detector's background rates, improved modeling of the signal response, and a detailed study of potential systematic uncertainties. With JUNO's data taking starting in 2025, early-stage experimental validation of key inputs would accelerate these steps. While there will be experimental challenges in implementing our new technique, we are optimistic that they can be surmounted, given the importance of the task. 

Beyond JUNO, this new technique can be generalized to other neutrino observatories, both scintillator-based (those noted above plus Daya Bay, KamLAND, and more) and otherwise (Super-K/Hyper-K, IceCube, and DUNE). It would also be interesting to explore the potential of this technique for dedicated DM detectors. Utilizing multiple detectors is important because they will have very different backgrounds and systematics; combining results would be more robust.  Further, if a signal were observed in JUNO, they would extract only a degenerate combination of DM mass and cross section --- a characteristic trajectory in Fig.~\ref{fig:result}, whereas other detectors would have different trajectories, which could cross JUNO's. In future work, we will address these and other topics.

Ultimately, this work opens a new frontier in DM detection. Neutrino observatories, which are designed for completely different physics goals, may hold the key to uncovering sub-GeV DM. The potential is now clear, but realizing it will require dedicated experimental validation, sharper theoretical modeling, and coordinated multi-detector searches. If DM hides in this mass range, we may already have the tools to find it. We just need to start looking, but in a new way.

%%%%%%%%%%%%%%%%%%%%%%%%%%%%%%%%%%%%%%%%%%%%%%%%%%%%%%%%%%%%%%%%%%%%%%%%%%%%%%%%%%%%%
%%%%%%%%%%%%%%%%%%%%%%%%%%%%%%%%%%%%%%%%%%%%%%%%%%%%%%%%%%%%%%%%%%%%%%%%%%%%%%%%%%%%%

\medskip
\noindent\textbf{\textit{Acknowledgments.---}}
We are grateful for helpful discussions with Carlos Blanco, Rouven Essig, Roni Harnik, Noah Kurinsky, Yufeng Li, Bryce Littlejohn, Stephan Meighen-Berger, Bernhard Mistlberger, Lillian Santos-Olmsted, Mark Vagins, and Tien-Tien Yu. RKL was supported by the U.S.\ Department of Energy under Contract DE-AC02-76SF00515. JFB was supported by National Science Foundation
Grant No.\ PHY-2310018. 

%%%%%%%%%%%%%%%%%%%%%%%%%%%%%%%%%%%%%%%%%%%%%%%%%%%%%%%%%%%%%%%%%%%%%%%%%%%%%%%%%%%%%
%%%%%%%%%%%%%%%%%%%%%%%%%%%%%%%%%%%%%%%%%%%%%%%%%%%%%%%%%%%%%%%%%%%%%%%%%%%%%%%%%%%%%

\clearpage
\newpage
\maketitle
\onecolumngrid
\begin{center}
\textbf{\large Sub-GeV Dark Matter Direct Detection with Neutrino Observatories}

\vspace{0.05in}
{ \it \large Supplemental Material}\\ 
\vspace{0.05in}
{Rebecca K. Leane and John F. Beacom}
\end{center}
\onecolumngrid
\setcounter{equation}{0}
\setcounter{figure}{0}
\setcounter{section}{0}
\setcounter{table}{0}
\setcounter{page}{1}
\makeatletter
\renewcommand{\theequation}{S\arabic{equation}}
\renewcommand{\thefigure}{S\arabic{figure}}
\renewcommand{\thetable}{S\arabic{table}}

\tableofcontents

 \section{Signal and Background Statistics}
 \label{sec:cap}

Using a Fisher Information Matrix approach, we analytically estimate the uncertainty on extracting the DM signal in the presence of multiple significant backgrounds.

\subsection{Analytic Setup}

The expected number of counts per bin is
\begin{equation}
    \lambda_i(t)=N \left(1+\delta_{\rm PMT}+\alpha \textrm{cos}(\omega t_i)+\sum_k \beta_k(1+\delta_k)\textrm{cos}(\omega t_i+\phi_k)\right)\,,
\end{equation}
where $N$ is the expected mean number of events per bin (assumed to be dominated by backgrounds), $\delta_{\text{PMT}}$ is the fractional systematic uncertainty on the flat background, with variance $\sigma_{\text{PMT}}$,  $\alpha$ is the relative amplitude of the DM signal (related to the main text notation as $N\, n_{\rm bins}\,\alpha=S_{\rm tot}f_{\rm mod}\approx0.1 \,S_{\rm tot}$), $\beta_k$ is the amplitude of the $k$-th oscillating background with phase $\phi_k$, and $\delta_k$ is the systematic uncertainty on the amplitude of the $k$-th oscillating background, with variance $\sigma_k$.

The likelihood of observing $n_i$ events in each bin, assuming Poisson-distributed counts (though given our infinite statistics, will be in fact Gaussian), is
\begin{equation}
    \log \mathcal{L}_{\text{total}} = \sum_{i=1}^{n_{\text{bins}}} \left[ -\lambda_i + n_i \log \lambda_i - \log(n_i!) \right] - \frac{\delta_{\text{PMT}}^2}{2 \sigma_{\text{PMT}}^2} - \sum_{k} \frac{\delta_k^2}{2 \sigma_k^2}\,.
\end{equation}
To determine the uncertainty on recovered parameters, we use the Fisher Information Matrix elements, which are computed from the expected value of the second derivative of the log-likelihood function, which is
\begin{equation}
    I_{ab} = -\mathbb{E} \left[ \frac{\partial^2 \log \mathcal{L}_{\text{total}}}{\partial \theta_a \, \partial \theta_b} \right]\,,
\end{equation}
where $\theta_a$ and $\theta_b$ are the parameters being estimated. The relevant derivatives are
\begin{align}
\frac{\partial \lambda_i}{\partial \alpha} &= N \cos(\omega t_i) \,,\\
\frac{\partial \lambda_i}{\partial \delta_{\text{PMT}}} &= N \,,\\
\frac{\partial \lambda_i}{\partial \delta_k} &= N \beta_k \cos(\omega t_i + \phi_k)\,.
\end{align}

The Fisher Information Matrix elements are then
\begin{align}
I_{\alpha\alpha} &= N^2 \sum_{i=1}^{n_{\text{bins}}} \frac{ \cos^2(\omega t_i) }{ \lambda_i } \,,\\
I_{\delta_{\text{PMT}} \delta_{\text{PMT}}} &= N^2 \sum_{i=1}^{n_{\text{bins}}} \frac{1}{\lambda_i} + \frac{1}{\sigma_{\text{PMT}}^2}\,, \\
I_{\delta_k \delta_k} &= N^2 \beta_k^2 \sum_{i=1}^{n_{\text{bins}}} \frac{ \cos^2(\omega t_i + \phi_k) }{ \lambda_i } + \frac{1}{\sigma_k^2}\,.
\end{align}
The cross terms (correlations) are
\begin{align}
I_{\alpha \delta_{\text{PMT}}} &= N^2 \sum_{i=1}^{n_{\text{bins}}} \frac{ \cos(\omega t_i) }{ \lambda_i } \,,\\
I_{\alpha \delta_k} &= N^2 \beta_k \sum_{i=1}^{n_{\text{bins}}} \frac{ \cos(\omega t_i) \cos(\omega t_i + \phi_k) }{ \lambda_i } \,,\\
I_{\delta_{\text{PMT}} \delta_k} &= N^2 \beta_k \sum_{i=1}^{n_{\text{bins}}} \frac{ \cos(\omega t_i + \phi_k) }{ \lambda_i }\,.
\end{align}
The marginalized uncertainty on $\alpha$, accounting for systematic uncertainties, is then
\begin{equation}
   \delta \alpha^2 = \left( I_{\alpha\alpha} - \sum_j \frac{I_{\alpha j}^2}{I_{jj}} \right)^{-1} \,,
\end{equation}
where $j$ runs over all nuisance parameters $(\delta_{\text{PMT}}, \delta_k)$. From this, we can recover the uncertainty on $\alpha$ given $N$ background events. 

\subsection{Flat, Dominant Dark Rate Background with Systematic Uncertainty}

We first consider the case that there is only the PMT background, and we assume that the data are sufficiently cleaned and detrended, such that the background is approximated as flat. At first glance, since $\cos(\omega t)$ integrates to zero over a full period, this term appears to vanish. However, $\lambda_i$ depends on $\alpha$,
\begin{equation}
\lambda_i = N \left[ 1 + \delta_{\rm PMT} +\alpha \cos(\omega t_i) \right]\,.
\end{equation}
Expanding $\lambda_i^{-1}$ to first order in $\alpha$ (assuming $\alpha \ll 1$),
\begin{equation}
\frac{1}{\lambda_i} \approx \frac{1}{N} \left[ 1 + \delta_{\rm PMT} - \alpha \cos(\omega t_i) \right]\,.
\end{equation}
Substituting this into the Fisher information expression,
\begin{equation}
I_{\alpha \, \delta_{\text{PMT}}} = N^2 \sum_{i=1}^{n_{\text{bins}}} \left[ \frac{\cos(\omega t_i)}{N} \left( 1 + \delta_{\rm PMT} - \alpha \cos(\omega t_i) \right) \right]\,,
\end{equation}
which simplifies to
\begin{equation}
I_{\alpha \, \delta_{\text{PMT}}} = N \sum_{i=1}^{n_{\text{bins}}} \left[ (1 + \delta_{\rm PMT})\cos(\omega t_i) - \alpha \cos^2(\omega t_i) \right]\,.
\end{equation}
The first term vanishes over a full cycle since $\sum \cos(\omega t_i) = 0$, such that
\begin{equation}
I_{\alpha \, \delta_{\text{PMT}}} = -\alpha N \sum_{i=1}^{n_{\text{bins}}} \cos^2(\omega t_i)\,.
\end{equation}
Over a full cycle
\begin{equation}
\sum_{i=1}^{n_{\text{bins}}} \cos^2(\omega t_i) \approx \frac{n_{\text{bins}}}{2}\,,
\end{equation}
thus,
\begin{equation}
I_{\alpha \, \delta_{\text{PMT}}} \approx -\frac{\alpha N n_{\text{bins}}}{2}\ .
\end{equation}
We can also simplify $I_{\alpha\alpha}$,
\begin{equation}
I_{\alpha\alpha} = N^2 \sum_{i=1}^{n_{\text{bins}}} \frac{ \cos^2(\omega t_i) }{ \lambda_i } \approx N^2 \sum_{i=1}^{n_{\text{bins}}} \frac{ \cos^2(\omega t_i) }{ N } = N \cdot \frac{n_{\text{bins}}}{2}\,.
\end{equation}
Last, we estimate $I_{\delta_{\text{PMT}} \, \delta_{\text{PMT}}}$,
\begin{equation}
I_{\delta_{\text{PMT}} \, \delta_{\text{PMT}}} = N^2 \sum_{i=1}^{n_{\text{bins}}} \frac{1}{\lambda_i} + \frac{1}{\sigma_{\text{PMT}}^2} \approx N n_{\text{bins}} + \frac{1}{\sigma_{\text{PMT}}^2} \approx N n_{\text{bins}} \, ,
\end{equation}
as $N n_{\text{bins}}\gg1/{\sigma_{\text{PMT}}^2}$, recalling that the total number of counts $N_{\text{tot}}=N n_{\text{bins}}\sim10^{16}$ for one year of data.

Substituting the terms into the expression for $\delta \alpha^2$,
\begin{equation}
\delta \alpha^2 = \left[ \frac{N n_{\text{bins}}}{2} - \frac{\alpha^2 N^2 n_{\text{bins}}^2 / 4}{ N n_{\text{bins}} } \right]^{-1} = \left[ \frac{N n_{\text{bins}}}{2} - \frac{\alpha^2 N n_{\text{bins}}}{4} \right]^{-1} = \frac{2}{N n_{\text{bins}}} \left( 1 - \frac{\alpha^2}{2} \right)^{-1}\,.
\end{equation}
Therefore in the limit of $\alpha \ll 1$, the uncertainty simplifies to
\begin{equation}
\delta \alpha \approx \frac{1}{\sqrt{N_{\text{tot}}}}\,.
\end{equation}
Replacing $\alpha$ with its definition, we then have the minimum sensitivity given by
\begin{equation}
    S_{\rm tot} f_{\rm mod} \approx \sqrt{N_{\rm tot}}, 
\end{equation}
$i.e.$,
\begin{equation}
    0.1 S_{\rm tot} \approx \sqrt{B}\,. 
\end{equation}
This shows that the statistical uncertainty dominates, and our significance scales as stated in the main text, where a factor of 2 is added for a 95\,\%\,C.L. sensitivity.

\subsection{Modulating Backgrounds}

We now consider the uncertainty scaling on the DM modulation signal when there are modulating backgrounds present.  For simplicity, we take only one background.  As the most challenging example, we take the muon background, as it has the closest phase to that of the DM signal. In this case, the expected counts per bin is
\begin{equation}
\lambda(t) = N \left[ 1 + \delta_{\text{PMT}} + \alpha \cos(\omega t) + \beta_k (1 + \delta_k) \cos(\omega t + \phi_k) \right]\,,
\end{equation}
where $\beta_k$ is the amplitude of the modulating background, $\delta_k \sim \mathcal{N}(0, \sigma_k^2)$ is the systematic uncertainty in the modulating background amplitude, and $\phi_k$ is the phase shift between the background modulation and the signal.    

Because the muon background modulation is offset by about 3 weeks in a yearly cycle, we have
\begin{equation}
\phi_k = \frac{3}{52} \times 2\pi = \frac{3\pi}{26}.
\end{equation}
We now calculate the Fisher information for the modulating background,
\begin{equation}
I_{\delta_k \delta_k} = N^2 \beta_k^2 \sum_{i=1}^{n_{\text{bins}}} \frac{\cos^2(\omega t_i + \phi_k)}{\lambda_i} + \frac{1}{\sigma_k^2}\approx N \beta_k^2 \sum_{i=1}^{n_{\text{bins}}} \cos^2(\omega t_i + \phi_k) + \frac{1}{\sigma_k^2}\,.
\end{equation}
Using $\sum \cos^2(\omega t_i + \phi_k) \approx n_{\text{bins}}/2$, we then have
\begin{equation}
I_{\delta_k \delta_k} \approx \frac{N n_{\text{bins}} \beta_k^2}{2} + \frac{1}{\sigma_k^2}\,.
\end{equation}
We can estimate the cross term as
\begin{equation}
I_{\alpha \delta_k} = N^2 \beta_k \sum_{i=1}^{n_{\text{bins}}} \frac{\cos(\omega t_i) \cos(\omega t_i + \phi_k)}{\lambda_i}\approx N \beta_k \sum_{i=1}^{n_{\text{bins}}} \frac{1}{2} \left[ \cos(\phi_k) + \cos(2\omega t_i + \phi_k) \right]\,,
\end{equation}
where we used the identity
\begin{equation}
\cos A \cos B = \frac{1}{2} \left[ \cos(A - B) + \cos(A + B) \right].
\end{equation}
Because $\sum_{i=1}^{n_{\text{bins}}} \cos(2\omega t_i + \phi_k) \approx 0$, we get
\begin{equation}
I_{\alpha \delta_k} \approx \frac{N n_{\text{bins}} \beta_k}{2} \cos(\phi_k).
\end{equation}
Finally, substituting $\phi_k = \frac{3\pi}{26}$,
\begin{equation}
\cos(\phi_k) = \cos\left(\frac{3\pi}{26}\right) \approx 0.9,
\end{equation}
and so we arrive at
\begin{equation}
I_{\alpha \delta_k} \approx \frac{N n_{\text{bins}} \beta_k}{2} \times 0.9\,.
\end{equation}
As above, we can now compare the size of the relevant terms in the marginalized uncertainty on $\alpha$,
\begin{equation}
\delta \alpha^2 = \left( I_{\alpha\alpha} - \frac{I_{\alpha \delta_k}^2}{I_{\delta_k \delta_k}} \right)^{-1} = \left( \frac{N n_{\text{bins}}}{2} - \frac{\left( \frac{N n_{\text{bins}} \beta_k}{2} \times 0.9 \right)^2}{\frac{N n_{\text{bins}} \beta_k^2}{2} + \frac{1}{\sigma_k^2}} \right)^{-1}\,.
\end{equation}
We will retain the desired scaling of $\delta \alpha \sim 1/\sqrt{N}$ if the second term remains small compared to $I_{\alpha\alpha} $. This requires
\begin{equation}
\frac{\left( \frac{N n_{\text{bins}} \beta_k}{2} \times 0.9 \right)^2}{\frac{N n_{\text{bins}} \beta_k^2}{2} + \frac{1}{\sigma_k^2}} \ll \frac{N n_{\text{bins}}}{2}\,.
\end{equation}
Solving for $\sigma_k $, this is obeyed if
\begin{equation}
\sigma_k^2 \ll \frac{1}{\cos^2(\phi_k)} \approx \frac{1}{0.9^2} \approx 1.2\,,
\end{equation}
or $\sigma_k \ll 1.1$.  Therefore, as long as the muon background is known better than an order-one systematic uncertainty, the total uncertainty on $\alpha$ will be dominated by statistics, so that $\delta \alpha \propto 1/\sqrt{N}$, $i.e.$, $ S_{\rm tot} f_{\rm mod} \approx \sqrt{B}$ as used in the main text.  Because the muons can be measured so well, we expect this to be the case.  For example, in an IceCube study where muon events were cut, some residual backgrounds due to muons remained, but the fraction could be measured relatively precisely~\cite{IceCube:2011cwc}.  For other time-dependent backgrounds ($e.g.$, due to radon, solar neutrinos, etc.) with precise auxiliary measurements, we also expect $\delta \alpha \propto 1/\sqrt{N}$ as used in the main text.

\subsection{Minimum Uncertainty on the Background Modulation Phase}

We now consider the pessimistic case where the background phase $\phi_k$ is not known, and we wish to recover it. In this scenario we will consider only the background modulation, such that
\begin{equation}
\lambda_i(t) = N \left[ 1 + \beta_k \cos(\omega t_i + \phi_k) \right]\,.
\end{equation}
We will also assume the background normalization is known, and takes the value $\beta_k=0.1$. This corresponds to the measured value in IceCube~\cite{Tilav:2019xmf} (similar values are expected for JUNO); it is around 0.01 in Borexino~\cite{Borexino:2018pev}.
Following cuts to remove muon events, the effective value of $\beta_k$ will be much smaller~\cite{IceCube:2011cwc}.

The Fisher information for the phase $\phi_k$ of the modulating background is
\begin{equation}
I_{\phi_k \phi_k} = \sum_{i=1}^{n_{\text{bins}}} \frac{1}{\lambda_i} \left( \frac{\partial \lambda_i}{\partial \phi_k} \right)^2\,,
\end{equation}
where 
\begin{equation}
\frac{\partial \lambda_i}{\partial \phi_k} = -N \beta_k \sin(\omega t_i + \phi_k)\,.
\end{equation}
Substituting this into the Fisher information expression,
\begin{equation}
I_{\phi_k \phi_k} = N^2 \beta_k^2 \sum_{i=1}^{n_{\text{bins}}} \frac{\sin^2(\omega t_i + \phi_k)}{\lambda_i}\,,
\end{equation}
For large \( N \), $\lambda_i \approx N$, such that
\begin{equation}
I_{\phi_k \phi_k} \approx N \beta_k^2 \sum_{i=1}^{n_{\text{bins}}} \sin^2(\omega t_i + \phi_k).
\end{equation}
Approximating the sum over a full cycle,
\begin{equation}
\sum_{i=1}^{n_{\text{bins}}} \sin^2(\omega t_i + \phi_k) \approx \frac{n_{\text{bins}}}{2},
\end{equation}
we obtain
\begin{equation}
I_{\phi_k \phi_k} \approx \frac{N n_{\text{bins}} \beta_k^2}{2}\,.
\end{equation}
We can now take the Cramér-Rao bound, that gives the minimum possible variance of an unbiased estimator of $\phi_k$, which is
\begin{equation}
\delta \phi_k^2 \geq \frac{1}{I_{\phi_k \phi_k}}\,.
\end{equation}
Substituting our result for $I_{\phi_k \phi_k}$, we have
\begin{equation}
\delta \phi_k^2 = \frac{2}{N n_{\text{bins}} \beta_k^2}\,,
\end{equation}
and so
\begin{equation}
\delta \phi_k = \frac{\sqrt{2}}{\beta_k \sqrt{N_{\text{tot}}}}\,,
\end{equation}
where $N_{\text{tot}} = N \cdot n_{\text{bins}}$ is the total number of background events. Therefore, for a total of about $2\times10^{16}$ background events, we have the minimum uncertainty on the background phase,
\begin{equation}
    \delta \phi_k \geq 1\times10^{-7}\,.
\end{equation}
We can convert this to the minimum time,
\begin{equation}
    t_{\rm min} = \frac{\phi_{\rm min}}{2\pi}\times 365~\text{days} \approx 0.5~\text{seconds}.
\end{equation}
The muon background differs in phase from the DM signal by about three weeks.  Due to the high statistics of muons, their phase will be more than adequately determined.  With cuts to remove muon events from the data, $\beta_k$ will effectively become smaller.  Nominally, if muons were cut by a factor of $\sim10^5$, then it would seem that muon phase uncertainty would increase to $\sim0.5$~days.  But this is incorrect, because the phase can be well determined from the uncut muon data.  For other time-dependent backgrounds ($e.g.$, due to radon, solar neutrinos, etc.) with precise auxiliary measurements, we also expect no problems with their phase uncertainties.

%%%%%%%%%%%%%%%%%%%%%%%%%%%%%%%%%%%%%%%%%%%%%%%%%%%%%%%%%%%%%%%%%%%%%%%%%%%%%%%%%%%%%
%%%%%%%%%%%%%%%%%%%%%%%%%%%%%%%%%%%%%%%%%%%%%%%%%%%%%%%%%%%%%%%%%%%%%%%%%%%%%%%%%%%%%

\bibliography{main}

%apsrev4-2.bst 2019-01-14 (MD) hand-edited version of apsrev4-1.bst
%Control: key (0)
%Control: author (72) initials jnrlst
%Control: editor formatted (1) identically to author
%Control: production of article title (-1) disabled
%Control: page (0) single
%Control: year (1) truncated
%Control: production of eprint (0) enabled
\begin{thebibliography}{190}%
\makeatletter
\providecommand \@ifxundefined [1]{%
 \@ifx{#1\undefined}
}%
\providecommand \@ifnum [1]{%
 \ifnum #1\expandafter \@firstoftwo
 \else \expandafter \@secondoftwo
 \fi
}%
\providecommand \@ifx [1]{%
 \ifx #1\expandafter \@firstoftwo
 \else \expandafter \@secondoftwo
 \fi
}%
\providecommand \natexlab [1]{#1}%
\providecommand \enquote  [1]{``#1''}%
\providecommand \bibnamefont  [1]{#1}%
\providecommand \bibfnamefont [1]{#1}%
\providecommand \citenamefont [1]{#1}%
\providecommand \href@noop [0]{\@secondoftwo}%
\providecommand \href [0]{\begingroup \@sanitize@url \@href}%
\providecommand \@href[1]{\@@startlink{#1}\@@href}%
\providecommand \@@href[1]{\endgroup#1\@@endlink}%
\providecommand \@sanitize@url [0]{\catcode `\\12\catcode `\$12\catcode `\&12\catcode `\#12\catcode `\^12\catcode `\_12\catcode `\%12\relax}%
\providecommand \@@startlink[1]{}%
\providecommand \@@endlink[0]{}%
\providecommand \url  [0]{\begingroup\@sanitize@url \@url }%
\providecommand \@url [1]{\endgroup\@href {#1}{\urlprefix }}%
\providecommand \urlprefix  [0]{URL }%
\providecommand \Eprint [0]{\href }%
\providecommand \doibase [0]{https://doi.org/}%
\providecommand \selectlanguage [0]{\@gobble}%
\providecommand \bibinfo  [0]{\@secondoftwo}%
\providecommand \bibfield  [0]{\@secondoftwo}%
\providecommand \translation [1]{[#1]}%
\providecommand \BibitemOpen [0]{}%
\providecommand \bibitemStop [0]{}%
\providecommand \bibitemNoStop [0]{.\EOS\space}%
\providecommand \EOS [0]{\spacefactor3000\relax}%
\providecommand \BibitemShut  [1]{\csname bibitem#1\endcsname}%
\let\auto@bib@innerbib\@empty
%</preamble>
\bibitem [{\citenamefont {Essig}\ \emph {et~al.}(2022)\citenamefont {Essig}, \citenamefont {Giovanetti}, \citenamefont {Kurinsky}, \citenamefont {McKinsey}, \citenamefont {Ramanathan}, \citenamefont {Stifter},\ and\ \citenamefont {Yu}}]{Essig:2022dfa}%
  \BibitemOpen
  \bibfield  {author} {\bibinfo {author} {\bibfnamefont {R.}~\bibnamefont {Essig}}, \bibinfo {author} {\bibfnamefont {G.~K.}\ \bibnamefont {Giovanetti}}, \bibinfo {author} {\bibfnamefont {N.}~\bibnamefont {Kurinsky}}, \bibinfo {author} {\bibfnamefont {D.}~\bibnamefont {McKinsey}}, \bibinfo {author} {\bibfnamefont {K.}~\bibnamefont {Ramanathan}}, \bibinfo {author} {\bibfnamefont {K.}~\bibnamefont {Stifter}},\ and\ \bibinfo {author} {\bibfnamefont {T.-T.}\ \bibnamefont {Yu}},\ }in\ \href@noop {} {\emph {\bibinfo {booktitle} {{2022 Snowmass Summer Study}}}}\ (\bibinfo {year} {2022})\ \Eprint {https://arxiv.org/abs/2203.08297} {arXiv:2203.08297 [hep-ph]} \BibitemShut {NoStop}%
\bibitem [{\citenamefont {Hochberg}\ \emph {et~al.}(2022{\natexlab{a}})\citenamefont {Hochberg}, \citenamefont {Kahn}, \citenamefont {Leane}, \citenamefont {Rajendran}, \citenamefont {Van~Tilburg}, \citenamefont {Yu},\ and\ \citenamefont {Zurek}}]{Hochberg:2022apz}%
  \BibitemOpen
  \bibfield  {author} {\bibinfo {author} {\bibfnamefont {Y.}~\bibnamefont {Hochberg}}, \bibinfo {author} {\bibfnamefont {Y.~F.}\ \bibnamefont {Kahn}}, \bibinfo {author} {\bibfnamefont {R.~K.}\ \bibnamefont {Leane}}, \bibinfo {author} {\bibfnamefont {S.}~\bibnamefont {Rajendran}}, \bibinfo {author} {\bibfnamefont {K.}~\bibnamefont {Van~Tilburg}}, \bibinfo {author} {\bibfnamefont {T.-T.}\ \bibnamefont {Yu}},\ and\ \bibinfo {author} {\bibfnamefont {K.~M.}\ \bibnamefont {Zurek}},\ }\href {https://doi.org/10.1038/s42254-022-00509-4} {\bibfield  {journal} {\bibinfo  {journal} {Nature Rev. Phys.}\ }\textbf {\bibinfo {volume} {4}},\ \bibinfo {pages} {637} (\bibinfo {year} {2022}{\natexlab{a}})}\BibitemShut {NoStop}%
\bibitem [{\citenamefont {Essig}\ \emph {et~al.}(2012)\citenamefont {Essig}, \citenamefont {Mardon},\ and\ \citenamefont {Volansky}}]{Essig:2011nj}%
  \BibitemOpen
  \bibfield  {author} {\bibinfo {author} {\bibfnamefont {R.}~\bibnamefont {Essig}}, \bibinfo {author} {\bibfnamefont {J.}~\bibnamefont {Mardon}},\ and\ \bibinfo {author} {\bibfnamefont {T.}~\bibnamefont {Volansky}},\ }\href {https://doi.org/10.1103/PhysRevD.85.076007} {\bibfield  {journal} {\bibinfo  {journal} {Phys. Rev. D}\ }\textbf {\bibinfo {volume} {85}},\ \bibinfo {pages} {076007} (\bibinfo {year} {2012})},\ \Eprint {https://arxiv.org/abs/1108.5383} {arXiv:1108.5383 [hep-ph]} \BibitemShut {NoStop}%
\bibitem [{\citenamefont {Graham}\ \emph {et~al.}(2012)\citenamefont {Graham}, \citenamefont {Kaplan}, \citenamefont {Rajendran},\ and\ \citenamefont {Walters}}]{Graham:2012su}%
  \BibitemOpen
  \bibfield  {author} {\bibinfo {author} {\bibfnamefont {P.~W.}\ \bibnamefont {Graham}}, \bibinfo {author} {\bibfnamefont {D.~E.}\ \bibnamefont {Kaplan}}, \bibinfo {author} {\bibfnamefont {S.}~\bibnamefont {Rajendran}},\ and\ \bibinfo {author} {\bibfnamefont {M.~T.}\ \bibnamefont {Walters}},\ }\href {https://doi.org/10.1016/j.dark.2012.09.001} {\bibfield  {journal} {\bibinfo  {journal} {Phys. Dark Univ.}\ }\textbf {\bibinfo {volume} {1}},\ \bibinfo {pages} {32} (\bibinfo {year} {2012})},\ \Eprint {https://arxiv.org/abs/1203.2531} {arXiv:1203.2531 [hep-ph]} \BibitemShut {NoStop}%
\bibitem [{\citenamefont {Essig}\ \emph {et~al.}(2016)\citenamefont {Essig}, \citenamefont {Fernandez-Serra}, \citenamefont {Mardon}, \citenamefont {Soto}, \citenamefont {Volansky},\ and\ \citenamefont {Yu}}]{Essig:2015cda}%
  \BibitemOpen
  \bibfield  {author} {\bibinfo {author} {\bibfnamefont {R.}~\bibnamefont {Essig}}, \bibinfo {author} {\bibfnamefont {M.}~\bibnamefont {Fernandez-Serra}}, \bibinfo {author} {\bibfnamefont {J.}~\bibnamefont {Mardon}}, \bibinfo {author} {\bibfnamefont {A.}~\bibnamefont {Soto}}, \bibinfo {author} {\bibfnamefont {T.}~\bibnamefont {Volansky}},\ and\ \bibinfo {author} {\bibfnamefont {T.-T.}\ \bibnamefont {Yu}},\ }\href {https://doi.org/10.1007/JHEP05(2016)046} {\bibfield  {journal} {\bibinfo  {journal} {JHEP}\ }\textbf {\bibinfo {volume} {05}},\ \bibinfo {pages} {046}},\ \Eprint {https://arxiv.org/abs/1509.01598} {arXiv:1509.01598 [hep-ph]} \BibitemShut {NoStop}%
\bibitem [{\citenamefont {Hochberg}\ \emph {et~al.}(2016{\natexlab{a}})\citenamefont {Hochberg}, \citenamefont {Zhao},\ and\ \citenamefont {Zurek}}]{Hochberg:2015pha}%
  \BibitemOpen
  \bibfield  {author} {\bibinfo {author} {\bibfnamefont {Y.}~\bibnamefont {Hochberg}}, \bibinfo {author} {\bibfnamefont {Y.}~\bibnamefont {Zhao}},\ and\ \bibinfo {author} {\bibfnamefont {K.~M.}\ \bibnamefont {Zurek}},\ }\href {https://doi.org/10.1103/PhysRevLett.116.011301} {\bibfield  {journal} {\bibinfo  {journal} {Phys. Rev. Lett.}\ }\textbf {\bibinfo {volume} {116}},\ \bibinfo {pages} {011301} (\bibinfo {year} {2016}{\natexlab{a}})},\ \Eprint {https://arxiv.org/abs/1504.07237} {arXiv:1504.07237 [hep-ph]} \BibitemShut {NoStop}%
\bibitem [{\citenamefont {Hochberg}\ \emph {et~al.}(2016{\natexlab{b}})\citenamefont {Hochberg}, \citenamefont {Pyle}, \citenamefont {Zhao},\ and\ \citenamefont {Zurek}}]{Hochberg:2015fth}%
  \BibitemOpen
  \bibfield  {author} {\bibinfo {author} {\bibfnamefont {Y.}~\bibnamefont {Hochberg}}, \bibinfo {author} {\bibfnamefont {M.}~\bibnamefont {Pyle}}, \bibinfo {author} {\bibfnamefont {Y.}~\bibnamefont {Zhao}},\ and\ \bibinfo {author} {\bibfnamefont {K.~M.}\ \bibnamefont {Zurek}},\ }\href {https://doi.org/10.1007/JHEP08(2016)057} {\bibfield  {journal} {\bibinfo  {journal} {JHEP}\ }\textbf {\bibinfo {volume} {08}},\ \bibinfo {pages} {057}},\ \Eprint {https://arxiv.org/abs/1512.04533} {arXiv:1512.04533 [hep-ph]} \BibitemShut {NoStop}%
\bibitem [{\citenamefont {Hochberg}\ \emph {et~al.}(2019{\natexlab{a}})\citenamefont {Hochberg}, \citenamefont {Charaev}, \citenamefont {Nam}, \citenamefont {Verma}, \citenamefont {Colangelo},\ and\ \citenamefont {Berggren}}]{Hochberg:2019cyy}%
  \BibitemOpen
  \bibfield  {author} {\bibinfo {author} {\bibfnamefont {Y.}~\bibnamefont {Hochberg}}, \bibinfo {author} {\bibfnamefont {I.}~\bibnamefont {Charaev}}, \bibinfo {author} {\bibfnamefont {S.-W.}\ \bibnamefont {Nam}}, \bibinfo {author} {\bibfnamefont {V.}~\bibnamefont {Verma}}, \bibinfo {author} {\bibfnamefont {M.}~\bibnamefont {Colangelo}},\ and\ \bibinfo {author} {\bibfnamefont {K.~K.}\ \bibnamefont {Berggren}},\ }\href {https://doi.org/10.1103/PhysRevLett.123.151802} {\bibfield  {journal} {\bibinfo  {journal} {Phys. Rev. Lett.}\ }\textbf {\bibinfo {volume} {123}},\ \bibinfo {pages} {151802} (\bibinfo {year} {2019}{\natexlab{a}})},\ \Eprint {https://arxiv.org/abs/1903.05101} {arXiv:1903.05101 [hep-ph]} \BibitemShut {NoStop}%
\bibitem [{\citenamefont {Hochberg}\ \emph {et~al.}(2021)\citenamefont {Hochberg}, \citenamefont {Kramer}, \citenamefont {Kurinsky},\ and\ \citenamefont {Lehmann}}]{Hochberg:2021ymx}%
  \BibitemOpen
  \bibfield  {author} {\bibinfo {author} {\bibfnamefont {Y.}~\bibnamefont {Hochberg}}, \bibinfo {author} {\bibfnamefont {E.~D.}\ \bibnamefont {Kramer}}, \bibinfo {author} {\bibfnamefont {N.}~\bibnamefont {Kurinsky}},\ and\ \bibinfo {author} {\bibfnamefont {B.~V.}\ \bibnamefont {Lehmann}},\ }\href@noop {} {\  (\bibinfo {year} {2021})},\ \Eprint {https://arxiv.org/abs/2109.04473} {arXiv:2109.04473 [hep-ph]} \BibitemShut {NoStop}%
\bibitem [{\citenamefont {Hochberg}\ \emph {et~al.}(2022{\natexlab{b}})\citenamefont {Hochberg}, \citenamefont {Lehmann}, \citenamefont {Charaev}, \citenamefont {Chiles}, \citenamefont {Colangelo}, \citenamefont {Nam},\ and\ \citenamefont {Berggren}}]{Hochberg:2021yud}%
  \BibitemOpen
  \bibfield  {author} {\bibinfo {author} {\bibfnamefont {Y.}~\bibnamefont {Hochberg}}, \bibinfo {author} {\bibfnamefont {B.~V.}\ \bibnamefont {Lehmann}}, \bibinfo {author} {\bibfnamefont {I.}~\bibnamefont {Charaev}}, \bibinfo {author} {\bibfnamefont {J.}~\bibnamefont {Chiles}}, \bibinfo {author} {\bibfnamefont {M.}~\bibnamefont {Colangelo}}, \bibinfo {author} {\bibfnamefont {S.~W.}\ \bibnamefont {Nam}},\ and\ \bibinfo {author} {\bibfnamefont {K.~K.}\ \bibnamefont {Berggren}},\ }\href {https://doi.org/10.1103/PhysRevD.106.112005} {\bibfield  {journal} {\bibinfo  {journal} {Phys. Rev. D}\ }\textbf {\bibinfo {volume} {106}},\ \bibinfo {pages} {112005} (\bibinfo {year} {2022}{\natexlab{b}})},\ \Eprint {https://arxiv.org/abs/2110.01586} {arXiv:2110.01586 [hep-ph]} \BibitemShut {NoStop}%
\bibitem [{\citenamefont {Derenzo}\ \emph {et~al.}(2017)\citenamefont {Derenzo}, \citenamefont {Essig}, \citenamefont {Massari}, \citenamefont {Soto},\ and\ \citenamefont {Yu}}]{Derenzo:2016fse}%
  \BibitemOpen
  \bibfield  {author} {\bibinfo {author} {\bibfnamefont {S.}~\bibnamefont {Derenzo}}, \bibinfo {author} {\bibfnamefont {R.}~\bibnamefont {Essig}}, \bibinfo {author} {\bibfnamefont {A.}~\bibnamefont {Massari}}, \bibinfo {author} {\bibfnamefont {A.}~\bibnamefont {Soto}},\ and\ \bibinfo {author} {\bibfnamefont {T.-T.}\ \bibnamefont {Yu}},\ }\href {https://doi.org/10.1103/PhysRevD.96.016026} {\bibfield  {journal} {\bibinfo  {journal} {Phys. Rev. D}\ }\textbf {\bibinfo {volume} {96}},\ \bibinfo {pages} {016026} (\bibinfo {year} {2017})},\ \Eprint {https://arxiv.org/abs/1607.01009} {arXiv:1607.01009 [hep-ph]} \BibitemShut {NoStop}%
\bibitem [{\citenamefont {Schutz}\ and\ \citenamefont {Zurek}(2016)}]{Schutz:2016tid}%
  \BibitemOpen
  \bibfield  {author} {\bibinfo {author} {\bibfnamefont {K.}~\bibnamefont {Schutz}}\ and\ \bibinfo {author} {\bibfnamefont {K.~M.}\ \bibnamefont {Zurek}},\ }\href {https://doi.org/10.1103/PhysRevLett.117.121302} {\bibfield  {journal} {\bibinfo  {journal} {Phys. Rev. Lett.}\ }\textbf {\bibinfo {volume} {117}},\ \bibinfo {pages} {121302} (\bibinfo {year} {2016})},\ \Eprint {https://arxiv.org/abs/1604.08206} {arXiv:1604.08206 [hep-ph]} \BibitemShut {NoStop}%
\bibitem [{\citenamefont {Hochberg}\ \emph {et~al.}(2017)\citenamefont {Hochberg}, \citenamefont {Kahn}, \citenamefont {Lisanti}, \citenamefont {Tully},\ and\ \citenamefont {Zurek}}]{Hochberg:2016ntt}%
  \BibitemOpen
  \bibfield  {author} {\bibinfo {author} {\bibfnamefont {Y.}~\bibnamefont {Hochberg}}, \bibinfo {author} {\bibfnamefont {Y.}~\bibnamefont {Kahn}}, \bibinfo {author} {\bibfnamefont {M.}~\bibnamefont {Lisanti}}, \bibinfo {author} {\bibfnamefont {C.~G.}\ \bibnamefont {Tully}},\ and\ \bibinfo {author} {\bibfnamefont {K.~M.}\ \bibnamefont {Zurek}},\ }\href {https://doi.org/10.1016/j.physletb.2017.06.051} {\bibfield  {journal} {\bibinfo  {journal} {Phys. Lett. B}\ }\textbf {\bibinfo {volume} {772}},\ \bibinfo {pages} {239} (\bibinfo {year} {2017})},\ \Eprint {https://arxiv.org/abs/1606.08849} {arXiv:1606.08849 [hep-ph]} \BibitemShut {NoStop}%
\bibitem [{\citenamefont {Essig}\ \emph {et~al.}(2017)\citenamefont {Essig}, \citenamefont {Mardon}, \citenamefont {Slone},\ and\ \citenamefont {Volansky}}]{Essig:2016crl}%
  \BibitemOpen
  \bibfield  {author} {\bibinfo {author} {\bibfnamefont {R.}~\bibnamefont {Essig}}, \bibinfo {author} {\bibfnamefont {J.}~\bibnamefont {Mardon}}, \bibinfo {author} {\bibfnamefont {O.}~\bibnamefont {Slone}},\ and\ \bibinfo {author} {\bibfnamefont {T.}~\bibnamefont {Volansky}},\ }\href {https://doi.org/10.1103/PhysRevD.95.056011} {\bibfield  {journal} {\bibinfo  {journal} {Phys. Rev. D}\ }\textbf {\bibinfo {volume} {95}},\ \bibinfo {pages} {056011} (\bibinfo {year} {2017})},\ \Eprint {https://arxiv.org/abs/1608.02940} {arXiv:1608.02940 [hep-ph]} \BibitemShut {NoStop}%
\bibitem [{\citenamefont {Hochberg}\ \emph {et~al.}(2018{\natexlab{a}})\citenamefont {Hochberg}, \citenamefont {Kahn}, \citenamefont {Lisanti}, \citenamefont {Zurek}, \citenamefont {Grushin}, \citenamefont {Ilan}, \citenamefont {Griffin}, \citenamefont {Liu}, \citenamefont {Weber},\ and\ \citenamefont {Neaton}}]{Hochberg:2017wce}%
  \BibitemOpen
  \bibfield  {author} {\bibinfo {author} {\bibfnamefont {Y.}~\bibnamefont {Hochberg}}, \bibinfo {author} {\bibfnamefont {Y.}~\bibnamefont {Kahn}}, \bibinfo {author} {\bibfnamefont {M.}~\bibnamefont {Lisanti}}, \bibinfo {author} {\bibfnamefont {K.~M.}\ \bibnamefont {Zurek}}, \bibinfo {author} {\bibfnamefont {A.~G.}\ \bibnamefont {Grushin}}, \bibinfo {author} {\bibfnamefont {R.}~\bibnamefont {Ilan}}, \bibinfo {author} {\bibfnamefont {S.~M.}\ \bibnamefont {Griffin}}, \bibinfo {author} {\bibfnamefont {Z.-F.}\ \bibnamefont {Liu}}, \bibinfo {author} {\bibfnamefont {S.~F.}\ \bibnamefont {Weber}},\ and\ \bibinfo {author} {\bibfnamefont {J.~B.}\ \bibnamefont {Neaton}},\ }\href {https://doi.org/10.1103/PhysRevD.97.015004} {\bibfield  {journal} {\bibinfo  {journal} {Phys. Rev. D}\ }\textbf {\bibinfo {volume} {97}},\ \bibinfo {pages} {015004} (\bibinfo {year} {2018}{\natexlab{a}})},\ \Eprint {https://arxiv.org/abs/1708.08929} {arXiv:1708.08929 [hep-ph]} \BibitemShut {NoStop}%
\bibitem [{\citenamefont {Cavoto}\ \emph {et~al.}(2018)\citenamefont {Cavoto}, \citenamefont {Luchetta},\ and\ \citenamefont {Polosa}}]{Cavoto:2017otc}%
  \BibitemOpen
  \bibfield  {author} {\bibinfo {author} {\bibfnamefont {G.}~\bibnamefont {Cavoto}}, \bibinfo {author} {\bibfnamefont {F.}~\bibnamefont {Luchetta}},\ and\ \bibinfo {author} {\bibfnamefont {A.~D.}\ \bibnamefont {Polosa}},\ }\href {https://doi.org/10.1016/j.physletb.2017.11.064} {\bibfield  {journal} {\bibinfo  {journal} {Phys. Lett. B}\ }\textbf {\bibinfo {volume} {776}},\ \bibinfo {pages} {338} (\bibinfo {year} {2018})},\ \Eprint {https://arxiv.org/abs/1706.02487} {arXiv:1706.02487 [hep-ph]} \BibitemShut {NoStop}%
\bibitem [{\citenamefont {Griffin}\ \emph {et~al.}(2018)\citenamefont {Griffin}, \citenamefont {Knapen}, \citenamefont {Lin},\ and\ \citenamefont {Zurek}}]{Griffin:2018bjn}%
  \BibitemOpen
  \bibfield  {author} {\bibinfo {author} {\bibfnamefont {S.}~\bibnamefont {Griffin}}, \bibinfo {author} {\bibfnamefont {S.}~\bibnamefont {Knapen}}, \bibinfo {author} {\bibfnamefont {T.}~\bibnamefont {Lin}},\ and\ \bibinfo {author} {\bibfnamefont {K.~M.}\ \bibnamefont {Zurek}},\ }\href {https://doi.org/10.1103/PhysRevD.98.115034} {\bibfield  {journal} {\bibinfo  {journal} {Phys. Rev. D}\ }\textbf {\bibinfo {volume} {98}},\ \bibinfo {pages} {115034} (\bibinfo {year} {2018})},\ \Eprint {https://arxiv.org/abs/1807.10291} {arXiv:1807.10291 [hep-ph]} \BibitemShut {NoStop}%
\bibitem [{\citenamefont {S\'anchez-Mart\'\i{}nez}\ \emph {et~al.}(2019)\citenamefont {S\'anchez-Mart\'\i{}nez}, \citenamefont {Robredo}, \citenamefont {Bidauzarraga}, \citenamefont {Bergara}, \citenamefont {de~Juan}, \citenamefont {Grushin},\ and\ \citenamefont {Vergniory}}]{Sanchez-Martinez:2019bac}%
  \BibitemOpen
  \bibfield  {author} {\bibinfo {author} {\bibfnamefont {M.-A.}\ \bibnamefont {S\'anchez-Mart\'\i{}nez}}, \bibinfo {author} {\bibfnamefont {I.~n.}\ \bibnamefont {Robredo}}, \bibinfo {author} {\bibfnamefont {A.}~\bibnamefont {Bidauzarraga}}, \bibinfo {author} {\bibfnamefont {A.}~\bibnamefont {Bergara}}, \bibinfo {author} {\bibfnamefont {F.}~\bibnamefont {de~Juan}}, \bibinfo {author} {\bibfnamefont {A.~G.}\ \bibnamefont {Grushin}},\ and\ \bibinfo {author} {\bibfnamefont {M.~G.}\ \bibnamefont {Vergniory}},\ }\href {https://doi.org/10.1088/2515-7639/ab3ea2} {\bibfield  {journal} {\bibinfo  {journal} {Materials}\ }\textbf {\bibinfo {volume} {3}},\ \bibinfo {pages} {014001} (\bibinfo {year} {2019})},\ \Eprint {https://arxiv.org/abs/1905.04805} {arXiv:1905.04805 [cond-mat.mtrl-sci]} \BibitemShut {NoStop}%
\bibitem [{\citenamefont {Essig}\ \emph {et~al.}(2020)\citenamefont {Essig}, \citenamefont {Pradler}, \citenamefont {Sholapurkar},\ and\ \citenamefont {Yu}}]{Essig:2019xkx}%
  \BibitemOpen
  \bibfield  {author} {\bibinfo {author} {\bibfnamefont {R.}~\bibnamefont {Essig}}, \bibinfo {author} {\bibfnamefont {J.}~\bibnamefont {Pradler}}, \bibinfo {author} {\bibfnamefont {M.}~\bibnamefont {Sholapurkar}},\ and\ \bibinfo {author} {\bibfnamefont {T.-T.}\ \bibnamefont {Yu}},\ }\href {https://doi.org/10.1103/PhysRevLett.124.021801} {\bibfield  {journal} {\bibinfo  {journal} {Phys. Rev. Lett.}\ }\textbf {\bibinfo {volume} {124}},\ \bibinfo {pages} {021801} (\bibinfo {year} {2020})},\ \Eprint {https://arxiv.org/abs/1908.10881} {arXiv:1908.10881 [hep-ph]} \BibitemShut {NoStop}%
\bibitem [{\citenamefont {Emken}\ \emph {et~al.}(2019)\citenamefont {Emken}, \citenamefont {Essig}, \citenamefont {Kouvaris},\ and\ \citenamefont {Sholapurkar}}]{Emken:2019tni}%
  \BibitemOpen
  \bibfield  {author} {\bibinfo {author} {\bibfnamefont {T.}~\bibnamefont {Emken}}, \bibinfo {author} {\bibfnamefont {R.}~\bibnamefont {Essig}}, \bibinfo {author} {\bibfnamefont {C.}~\bibnamefont {Kouvaris}},\ and\ \bibinfo {author} {\bibfnamefont {M.}~\bibnamefont {Sholapurkar}},\ }\href {https://doi.org/10.1088/1475-7516/2019/09/070} {\bibfield  {journal} {\bibinfo  {journal} {JCAP}\ }\textbf {\bibinfo {volume} {09}},\ \bibinfo {pages} {070}},\ \Eprint {https://arxiv.org/abs/1905.06348} {arXiv:1905.06348 [hep-ph]} \BibitemShut {NoStop}%
\bibitem [{\citenamefont {Kurinsky}\ \emph {et~al.}(2019)\citenamefont {Kurinsky}, \citenamefont {Yu}, \citenamefont {Hochberg},\ and\ \citenamefont {Cabrera}}]{Kurinsky:2019pgb}%
  \BibitemOpen
  \bibfield  {author} {\bibinfo {author} {\bibfnamefont {N.~A.}\ \bibnamefont {Kurinsky}}, \bibinfo {author} {\bibfnamefont {T.~C.}\ \bibnamefont {Yu}}, \bibinfo {author} {\bibfnamefont {Y.}~\bibnamefont {Hochberg}},\ and\ \bibinfo {author} {\bibfnamefont {B.}~\bibnamefont {Cabrera}},\ }\href {https://doi.org/10.1103/PhysRevD.99.123005} {\bibfield  {journal} {\bibinfo  {journal} {Phys. Rev. D}\ }\textbf {\bibinfo {volume} {99}},\ \bibinfo {pages} {123005} (\bibinfo {year} {2019})},\ \Eprint {https://arxiv.org/abs/1901.07569} {arXiv:1901.07569 [hep-ex]} \BibitemShut {NoStop}%
\bibitem [{\citenamefont {Blanco}\ \emph {et~al.}(2020)\citenamefont {Blanco}, \citenamefont {Collar}, \citenamefont {Kahn},\ and\ \citenamefont {Lillard}}]{Blanco:2019lrf}%
  \BibitemOpen
  \bibfield  {author} {\bibinfo {author} {\bibfnamefont {C.}~\bibnamefont {Blanco}}, \bibinfo {author} {\bibfnamefont {J.~I.}\ \bibnamefont {Collar}}, \bibinfo {author} {\bibfnamefont {Y.}~\bibnamefont {Kahn}},\ and\ \bibinfo {author} {\bibfnamefont {B.}~\bibnamefont {Lillard}},\ }\href {https://doi.org/10.1103/PhysRevD.101.056001} {\bibfield  {journal} {\bibinfo  {journal} {Phys. Rev. D}\ }\textbf {\bibinfo {volume} {101}},\ \bibinfo {pages} {056001} (\bibinfo {year} {2020})},\ \Eprint {https://arxiv.org/abs/1912.02822} {arXiv:1912.02822 [hep-ph]} \BibitemShut {NoStop}%
\bibitem [{\citenamefont {Griffin}\ \emph {et~al.}(2021)\citenamefont {Griffin}, \citenamefont {Hochberg}, \citenamefont {Inzani}, \citenamefont {Kurinsky}, \citenamefont {Lin},\ and\ \citenamefont {Chin}}]{Griffin:2020lgd}%
  \BibitemOpen
  \bibfield  {author} {\bibinfo {author} {\bibfnamefont {S.~M.}\ \bibnamefont {Griffin}}, \bibinfo {author} {\bibfnamefont {Y.}~\bibnamefont {Hochberg}}, \bibinfo {author} {\bibfnamefont {K.}~\bibnamefont {Inzani}}, \bibinfo {author} {\bibfnamefont {N.}~\bibnamefont {Kurinsky}}, \bibinfo {author} {\bibfnamefont {T.}~\bibnamefont {Lin}},\ and\ \bibinfo {author} {\bibfnamefont {T.}~\bibnamefont {Chin}},\ }\href {https://doi.org/10.1103/PhysRevD.103.075002} {\bibfield  {journal} {\bibinfo  {journal} {Phys. Rev. D}\ }\textbf {\bibinfo {volume} {103}},\ \bibinfo {pages} {075002} (\bibinfo {year} {2021})},\ \Eprint {https://arxiv.org/abs/2008.08560} {arXiv:2008.08560 [hep-ph]} \BibitemShut {NoStop}%
\bibitem [{\citenamefont {Blanco}\ \emph {et~al.}(2021)\citenamefont {Blanco}, \citenamefont {Kahn}, \citenamefont {Lillard},\ and\ \citenamefont {McDermott}}]{Blanco:2021hlm}%
  \BibitemOpen
  \bibfield  {author} {\bibinfo {author} {\bibfnamefont {C.}~\bibnamefont {Blanco}}, \bibinfo {author} {\bibfnamefont {Y.}~\bibnamefont {Kahn}}, \bibinfo {author} {\bibfnamefont {B.}~\bibnamefont {Lillard}},\ and\ \bibinfo {author} {\bibfnamefont {S.~D.}\ \bibnamefont {McDermott}},\ }\href {https://doi.org/10.1103/PhysRevD.104.036011} {\bibfield  {journal} {\bibinfo  {journal} {Phys. Rev. D}\ }\textbf {\bibinfo {volume} {104}},\ \bibinfo {pages} {036011} (\bibinfo {year} {2021})},\ \Eprint {https://arxiv.org/abs/2103.08601} {arXiv:2103.08601 [hep-ph]} \BibitemShut {NoStop}%
\bibitem [{\citenamefont {Baxter}\ \emph {et~al.}(2020)\citenamefont {Baxter}, \citenamefont {Kahn},\ and\ \citenamefont {Krnjaic}}]{Baxter:2019pnz}%
  \BibitemOpen
  \bibfield  {author} {\bibinfo {author} {\bibfnamefont {D.}~\bibnamefont {Baxter}}, \bibinfo {author} {\bibfnamefont {Y.}~\bibnamefont {Kahn}},\ and\ \bibinfo {author} {\bibfnamefont {G.}~\bibnamefont {Krnjaic}},\ }\href {https://doi.org/10.1103/PhysRevD.101.076014} {\bibfield  {journal} {\bibinfo  {journal} {Phys. Rev. D}\ }\textbf {\bibinfo {volume} {101}},\ \bibinfo {pages} {076014} (\bibinfo {year} {2020})},\ \Eprint {https://arxiv.org/abs/1908.00012} {arXiv:1908.00012 [hep-ph]} \BibitemShut {NoStop}%
\bibitem [{\citenamefont {Kurinsky}\ \emph {et~al.}(2020)\citenamefont {Kurinsky}, \citenamefont {Baxter}, \citenamefont {Kahn},\ and\ \citenamefont {Krnjaic}}]{Kurinsky:2020dpb}%
  \BibitemOpen
  \bibfield  {author} {\bibinfo {author} {\bibfnamefont {N.}~\bibnamefont {Kurinsky}}, \bibinfo {author} {\bibfnamefont {D.}~\bibnamefont {Baxter}}, \bibinfo {author} {\bibfnamefont {Y.}~\bibnamefont {Kahn}},\ and\ \bibinfo {author} {\bibfnamefont {G.}~\bibnamefont {Krnjaic}},\ }\href {https://doi.org/10.1103/PhysRevD.102.015017} {\bibfield  {journal} {\bibinfo  {journal} {Phys. Rev. D}\ }\textbf {\bibinfo {volume} {102}},\ \bibinfo {pages} {015017} (\bibinfo {year} {2020})},\ \Eprint {https://arxiv.org/abs/2002.06937} {arXiv:2002.06937 [hep-ph]} \BibitemShut {NoStop}%
\bibitem [{\citenamefont {Geilhufe}\ \emph {et~al.}(2020)\citenamefont {Geilhufe}, \citenamefont {Kahlhoefer},\ and\ \citenamefont {Winkler}}]{Geilhufe:2019ndy}%
  \BibitemOpen
  \bibfield  {author} {\bibinfo {author} {\bibfnamefont {R.~M.}\ \bibnamefont {Geilhufe}}, \bibinfo {author} {\bibfnamefont {F.}~\bibnamefont {Kahlhoefer}},\ and\ \bibinfo {author} {\bibfnamefont {M.~W.}\ \bibnamefont {Winkler}},\ }\href {https://doi.org/10.1103/PhysRevD.101.055005} {\bibfield  {journal} {\bibinfo  {journal} {Phys. Rev. D}\ }\textbf {\bibinfo {volume} {101}},\ \bibinfo {pages} {055005} (\bibinfo {year} {2020})},\ \Eprint {https://arxiv.org/abs/1910.02091} {arXiv:1910.02091 [hep-ph]} \BibitemShut {NoStop}%
\bibitem [{\citenamefont {Emken}\ \emph {et~al.}(2017)\citenamefont {Emken}, \citenamefont {Kouvaris},\ and\ \citenamefont {Shoemaker}}]{Emken:2017erx}%
  \BibitemOpen
  \bibfield  {author} {\bibinfo {author} {\bibfnamefont {T.}~\bibnamefont {Emken}}, \bibinfo {author} {\bibfnamefont {C.}~\bibnamefont {Kouvaris}},\ and\ \bibinfo {author} {\bibfnamefont {I.~M.}\ \bibnamefont {Shoemaker}},\ }\href {https://doi.org/10.1103/PhysRevD.96.015018} {\bibfield  {journal} {\bibinfo  {journal} {Phys. Rev. D}\ }\textbf {\bibinfo {volume} {96}},\ \bibinfo {pages} {015018} (\bibinfo {year} {2017})},\ \Eprint {https://arxiv.org/abs/1702.07750} {arXiv:1702.07750 [hep-ph]} \BibitemShut {NoStop}%
\bibitem [{\citenamefont {Emken}\ and\ \citenamefont {Kouvaris}(2017)}]{Emken:2017qmp}%
  \BibitemOpen
  \bibfield  {author} {\bibinfo {author} {\bibfnamefont {T.}~\bibnamefont {Emken}}\ and\ \bibinfo {author} {\bibfnamefont {C.}~\bibnamefont {Kouvaris}},\ }\href {https://doi.org/10.1088/1475-7516/2017/10/031} {\bibfield  {journal} {\bibinfo  {journal} {JCAP}\ }\textbf {\bibinfo {volume} {10}},\ \bibinfo {pages} {031}},\ \Eprint {https://arxiv.org/abs/1706.02249} {arXiv:1706.02249 [hep-ph]} \BibitemShut {NoStop}%
\bibitem [{\citenamefont {Catena}\ \emph {et~al.}(2020)\citenamefont {Catena}, \citenamefont {Emken}, \citenamefont {Spaldin},\ and\ \citenamefont {Tarantino}}]{Catena:2019gfa}%
  \BibitemOpen
  \bibfield  {author} {\bibinfo {author} {\bibfnamefont {R.}~\bibnamefont {Catena}}, \bibinfo {author} {\bibfnamefont {T.}~\bibnamefont {Emken}}, \bibinfo {author} {\bibfnamefont {N.~A.}\ \bibnamefont {Spaldin}},\ and\ \bibinfo {author} {\bibfnamefont {W.}~\bibnamefont {Tarantino}},\ }\href {https://doi.org/10.1103/PhysRevResearch.2.033195} {\bibfield  {journal} {\bibinfo  {journal} {Phys. Rev. Res.}\ }\textbf {\bibinfo {volume} {2}},\ \bibinfo {pages} {033195} (\bibinfo {year} {2020})},\ \bibinfo {note} {[Erratum: Phys.Rev.Res. 7, 019001 (2025)]},\ \Eprint {https://arxiv.org/abs/1912.08204} {arXiv:1912.08204 [hep-ph]} \BibitemShut {NoStop}%
\bibitem [{\citenamefont {Radick}\ \emph {et~al.}(2021)\citenamefont {Radick}, \citenamefont {Taki},\ and\ \citenamefont {Yu}}]{Radick:2020qip}%
  \BibitemOpen
  \bibfield  {author} {\bibinfo {author} {\bibfnamefont {A.}~\bibnamefont {Radick}}, \bibinfo {author} {\bibfnamefont {A.-M.}\ \bibnamefont {Taki}},\ and\ \bibinfo {author} {\bibfnamefont {T.-T.}\ \bibnamefont {Yu}},\ }\href {https://doi.org/10.1088/1475-7516/2021/02/004} {\bibfield  {journal} {\bibinfo  {journal} {JCAP}\ }\textbf {\bibinfo {volume} {02}},\ \bibinfo {pages} {004}},\ \Eprint {https://arxiv.org/abs/2011.02493} {arXiv:2011.02493 [hep-ph]} \BibitemShut {NoStop}%
\bibitem [{\citenamefont {Gelmini}\ \emph {et~al.}(2020)\citenamefont {Gelmini}, \citenamefont {Takhistov},\ and\ \citenamefont {Vitagliano}}]{Gelmini:2020xir}%
  \BibitemOpen
  \bibfield  {author} {\bibinfo {author} {\bibfnamefont {G.~B.}\ \bibnamefont {Gelmini}}, \bibinfo {author} {\bibfnamefont {V.}~\bibnamefont {Takhistov}},\ and\ \bibinfo {author} {\bibfnamefont {E.}~\bibnamefont {Vitagliano}},\ }\href {https://doi.org/10.1016/j.physletb.2020.135779} {\bibfield  {journal} {\bibinfo  {journal} {Phys. Lett. B}\ }\textbf {\bibinfo {volume} {809}},\ \bibinfo {pages} {135779} (\bibinfo {year} {2020})},\ \Eprint {https://arxiv.org/abs/2006.13909} {arXiv:2006.13909 [hep-ph]} \BibitemShut {NoStop}%
\bibitem [{\citenamefont {Trickle}\ \emph {et~al.}(2022)\citenamefont {Trickle}, \citenamefont {Zhang},\ and\ \citenamefont {Zurek}}]{Trickle:2020oki}%
  \BibitemOpen
  \bibfield  {author} {\bibinfo {author} {\bibfnamefont {T.}~\bibnamefont {Trickle}}, \bibinfo {author} {\bibfnamefont {Z.}~\bibnamefont {Zhang}},\ and\ \bibinfo {author} {\bibfnamefont {K.~M.}\ \bibnamefont {Zurek}},\ }\href {https://doi.org/10.1103/PhysRevD.105.015001} {\bibfield  {journal} {\bibinfo  {journal} {Phys. Rev. D}\ }\textbf {\bibinfo {volume} {105}},\ \bibinfo {pages} {015001} (\bibinfo {year} {2022})},\ \Eprint {https://arxiv.org/abs/2009.13534} {arXiv:2009.13534 [hep-ph]} \BibitemShut {NoStop}%
\bibitem [{\citenamefont {Knapen}\ \emph {et~al.}(2021)\citenamefont {Knapen}, \citenamefont {Kozaczuk},\ and\ \citenamefont {Lin}}]{Knapen:2021run}%
  \BibitemOpen
  \bibfield  {author} {\bibinfo {author} {\bibfnamefont {S.}~\bibnamefont {Knapen}}, \bibinfo {author} {\bibfnamefont {J.}~\bibnamefont {Kozaczuk}},\ and\ \bibinfo {author} {\bibfnamefont {T.}~\bibnamefont {Lin}},\ }\href {https://doi.org/10.1103/PhysRevD.104.015031} {\bibfield  {journal} {\bibinfo  {journal} {Phys. Rev. D}\ }\textbf {\bibinfo {volume} {104}},\ \bibinfo {pages} {015031} (\bibinfo {year} {2021})},\ \Eprint {https://arxiv.org/abs/2101.08275} {arXiv:2101.08275 [hep-ph]} \BibitemShut {NoStop}%
\bibitem [{\citenamefont {Cook}\ \emph {et~al.}(2024)\citenamefont {Cook}, \citenamefont {Blanco},\ and\ \citenamefont {Smirnov}}]{Cook:2024cgm}%
  \BibitemOpen
  \bibfield  {author} {\bibinfo {author} {\bibfnamefont {C.}~\bibnamefont {Cook}}, \bibinfo {author} {\bibfnamefont {C.}~\bibnamefont {Blanco}},\ and\ \bibinfo {author} {\bibfnamefont {J.}~\bibnamefont {Smirnov}},\ }\href@noop {} {\  (\bibinfo {year} {2024})},\ \Eprint {https://arxiv.org/abs/2501.00091} {arXiv:2501.00091 [hep-ph]} \BibitemShut {NoStop}%
\bibitem [{\citenamefont {Simchony}\ \emph {et~al.}(2024)\citenamefont {Simchony} \emph {et~al.}}]{Simchony:2024kcn}%
  \BibitemOpen
  \bibfield  {author} {\bibinfo {author} {\bibfnamefont {A.}~\bibnamefont {Simchony}} \emph {et~al.},\ }\href {https://doi.org/10.1007/s10909-024-03148-4} {\bibfield  {journal} {\bibinfo  {journal} {J. Low Temp. Phys.}\ }\textbf {\bibinfo {volume} {216}},\ \bibinfo {pages} {363} (\bibinfo {year} {2024})}\BibitemShut {NoStop}%
\bibitem [{\citenamefont {Das}\ \emph {et~al.}(2024{\natexlab{a}})\citenamefont {Das}, \citenamefont {Kurinsky},\ and\ \citenamefont {Leane}}]{Das:2022srn}%
  \BibitemOpen
  \bibfield  {author} {\bibinfo {author} {\bibfnamefont {A.}~\bibnamefont {Das}}, \bibinfo {author} {\bibfnamefont {N.}~\bibnamefont {Kurinsky}},\ and\ \bibinfo {author} {\bibfnamefont {R.~K.}\ \bibnamefont {Leane}},\ }\href {https://doi.org/10.1103/PhysRevLett.132.121801} {\bibfield  {journal} {\bibinfo  {journal} {Phys. Rev. Lett.}\ }\textbf {\bibinfo {volume} {132}},\ \bibinfo {pages} {121801} (\bibinfo {year} {2024}{\natexlab{a}})},\ \Eprint {https://arxiv.org/abs/2210.09313} {arXiv:2210.09313 [hep-ph]} \BibitemShut {NoStop}%
\bibitem [{\citenamefont {Das}\ \emph {et~al.}(2024{\natexlab{b}})\citenamefont {Das}, \citenamefont {Kurinsky},\ and\ \citenamefont {Leane}}]{Das:2024jdz}%
  \BibitemOpen
  \bibfield  {author} {\bibinfo {author} {\bibfnamefont {A.}~\bibnamefont {Das}}, \bibinfo {author} {\bibfnamefont {N.}~\bibnamefont {Kurinsky}},\ and\ \bibinfo {author} {\bibfnamefont {R.~K.}\ \bibnamefont {Leane}},\ }\href {https://doi.org/10.1007/JHEP07(2024)233} {\bibfield  {journal} {\bibinfo  {journal} {JHEP}\ }\textbf {\bibinfo {volume} {07}},\ \bibinfo {pages} {233}},\ \Eprint {https://arxiv.org/abs/2405.00112} {arXiv:2405.00112 [hep-ph]} \BibitemShut {NoStop}%
\bibitem [{\citenamefont {Griffin}\ \emph {et~al.}(2024)\citenamefont {Griffin}, \citenamefont {Hadas}, \citenamefont {Hochberg}, \citenamefont {Inzani},\ and\ \citenamefont {Lehmann}}]{Griffin:2024cew}%
  \BibitemOpen
  \bibfield  {author} {\bibinfo {author} {\bibfnamefont {S.~M.}\ \bibnamefont {Griffin}}, \bibinfo {author} {\bibfnamefont {G.~D.}\ \bibnamefont {Hadas}}, \bibinfo {author} {\bibfnamefont {Y.}~\bibnamefont {Hochberg}}, \bibinfo {author} {\bibfnamefont {K.}~\bibnamefont {Inzani}},\ and\ \bibinfo {author} {\bibfnamefont {B.~V.}\ \bibnamefont {Lehmann}},\ }\href@noop {} {\  (\bibinfo {year} {2024})},\ \Eprint {https://arxiv.org/abs/2412.16283} {arXiv:2412.16283 [hep-ph]} \BibitemShut {NoStop}%
\bibitem [{\citenamefont {Baudis}\ \emph {et~al.}(2024)\citenamefont {Baudis} \emph {et~al.}}]{QROCODILE:2024zmg}%
  \BibitemOpen
  \bibfield  {author} {\bibinfo {author} {\bibfnamefont {L.}~\bibnamefont {Baudis}} \emph {et~al.} (\bibinfo {collaboration} {QROCODILE}),\ }\href@noop {} {\  (\bibinfo {year} {2024})},\ \Eprint {https://arxiv.org/abs/2412.16279} {arXiv:2412.16279 [hep-ph]} \BibitemShut {NoStop}%
\bibitem [{\citenamefont {Marocco}\ and\ \citenamefont {Wheater}(2025)}]{Marocco:2025eqw}%
  \BibitemOpen
  \bibfield  {author} {\bibinfo {author} {\bibfnamefont {G.}~\bibnamefont {Marocco}}\ and\ \bibinfo {author} {\bibfnamefont {J.}~\bibnamefont {Wheater}},\ }\href@noop {} {\  (\bibinfo {year} {2025})},\ \Eprint {https://arxiv.org/abs/2501.18120} {arXiv:2501.18120 [hep-ph]} \BibitemShut {NoStop}%
\bibitem [{\citenamefont {Abusleme}\ \emph {et~al.}(2021{\natexlab{a}})\citenamefont {Abusleme} \emph {et~al.}}]{JUNO:2020xtj}%
  \BibitemOpen
  \bibfield  {author} {\bibinfo {author} {\bibfnamefont {A.}~\bibnamefont {Abusleme}} \emph {et~al.} (\bibinfo {collaboration} {JUNO}),\ }\href {https://doi.org/10.1007/JHEP03(2021)004} {\bibfield  {journal} {\bibinfo  {journal} {JHEP}\ }\textbf {\bibinfo {volume} {03}},\ \bibinfo {pages} {004}},\ \Eprint {https://arxiv.org/abs/2011.06405} {arXiv:2011.06405 [physics.ins-det]} \BibitemShut {NoStop}%
\bibitem [{\citenamefont {Abusleme}\ \emph {et~al.}(2022{\natexlab{a}})\citenamefont {Abusleme} \emph {et~al.}}]{JUNO:2021vlw}%
  \BibitemOpen
  \bibfield  {author} {\bibinfo {author} {\bibfnamefont {A.}~\bibnamefont {Abusleme}} \emph {et~al.} (\bibinfo {collaboration} {JUNO}),\ }\href {https://doi.org/10.1016/j.ppnp.2021.103927} {\bibfield  {journal} {\bibinfo  {journal} {Prog. Part. Nucl. Phys.}\ }\textbf {\bibinfo {volume} {123}},\ \bibinfo {pages} {103927} (\bibinfo {year} {2022}{\natexlab{a}})},\ \Eprint {https://arxiv.org/abs/2104.02565} {arXiv:2104.02565 [hep-ex]} \BibitemShut {NoStop}%
\bibitem [{\citenamefont {{Press}}\ and\ \citenamefont {{Spergel}}(1985)}]{1985ApJ296679P}%
  \BibitemOpen
  \bibfield  {author} {\bibinfo {author} {\bibfnamefont {W.~H.}\ \bibnamefont {{Press}}}\ and\ \bibinfo {author} {\bibfnamefont {D.~N.}\ \bibnamefont {{Spergel}}},\ }\href {https://doi.org/10.1086/163485} {\bibfield  {journal} {\bibinfo  {journal} {\apj}\ }\textbf {\bibinfo {volume} {296}},\ \bibinfo {pages} {679} (\bibinfo {year} {1985})}\BibitemShut {NoStop}%
\bibitem [{\citenamefont {Srednicki}\ \emph {et~al.}(1987)\citenamefont {Srednicki}, \citenamefont {Olive},\ and\ \citenamefont {Silk}}]{Srednicki:1986vj}%
  \BibitemOpen
  \bibfield  {author} {\bibinfo {author} {\bibfnamefont {M.}~\bibnamefont {Srednicki}}, \bibinfo {author} {\bibfnamefont {K.~A.}\ \bibnamefont {Olive}},\ and\ \bibinfo {author} {\bibfnamefont {J.}~\bibnamefont {Silk}},\ }\href {https://doi.org/10.1016/0550-3213(87)90020-4} {\bibfield  {journal} {\bibinfo  {journal} {Nucl. Phys. B}\ }\textbf {\bibinfo {volume} {279}},\ \bibinfo {pages} {804} (\bibinfo {year} {1987})}\BibitemShut {NoStop}%
\bibitem [{\citenamefont {Ritz}\ and\ \citenamefont {Seckel}(1988)}]{Ritz:1987mh}%
  \BibitemOpen
  \bibfield  {author} {\bibinfo {author} {\bibfnamefont {S.}~\bibnamefont {Ritz}}\ and\ \bibinfo {author} {\bibfnamefont {D.}~\bibnamefont {Seckel}},\ }\href {https://doi.org/10.1016/0550-3213(88)90660-8} {\bibfield  {journal} {\bibinfo  {journal} {Nucl. Phys. B}\ }\textbf {\bibinfo {volume} {304}},\ \bibinfo {pages} {877} (\bibinfo {year} {1988})}\BibitemShut {NoStop}%
\bibitem [{\citenamefont {Meade}\ \emph {et~al.}(2010)\citenamefont {Meade}, \citenamefont {Nussinov}, \citenamefont {Papucci},\ and\ \citenamefont {Volansky}}]{Meade:2009mu}%
  \BibitemOpen
  \bibfield  {author} {\bibinfo {author} {\bibfnamefont {P.}~\bibnamefont {Meade}}, \bibinfo {author} {\bibfnamefont {S.}~\bibnamefont {Nussinov}}, \bibinfo {author} {\bibfnamefont {M.}~\bibnamefont {Papucci}},\ and\ \bibinfo {author} {\bibfnamefont {T.}~\bibnamefont {Volansky}},\ }\href {https://doi.org/10.1007/JHEP06(2010)029} {\bibfield  {journal} {\bibinfo  {journal} {JHEP}\ }\textbf {\bibinfo {volume} {06}},\ \bibinfo {pages} {029}},\ \Eprint {https://arxiv.org/abs/0910.4160} {arXiv:0910.4160 [hep-ph]} \BibitemShut {NoStop}%
\bibitem [{\citenamefont {Bell}\ and\ \citenamefont {Petraki}(2011)}]{Bell:2011sn}%
  \BibitemOpen
  \bibfield  {author} {\bibinfo {author} {\bibfnamefont {N.~F.}\ \bibnamefont {Bell}}\ and\ \bibinfo {author} {\bibfnamefont {K.}~\bibnamefont {Petraki}},\ }\href {https://doi.org/10.1088/1475-7516/2011/04/003} {\bibfield  {journal} {\bibinfo  {journal} {JCAP}\ }\textbf {\bibinfo {volume} {04}},\ \bibinfo {pages} {003}},\ \Eprint {https://arxiv.org/abs/1102.2958} {arXiv:1102.2958 [hep-ph]} \BibitemShut {NoStop}%
\bibitem [{\citenamefont {Feng}\ \emph {et~al.}(2016)\citenamefont {Feng}, \citenamefont {Smolinsky},\ and\ \citenamefont {Tanedo}}]{Feng:2016ijc}%
  \BibitemOpen
  \bibfield  {author} {\bibinfo {author} {\bibfnamefont {J.~L.}\ \bibnamefont {Feng}}, \bibinfo {author} {\bibfnamefont {J.}~\bibnamefont {Smolinsky}},\ and\ \bibinfo {author} {\bibfnamefont {P.}~\bibnamefont {Tanedo}},\ }\href {https://doi.org/10.1103/PhysRevD.93.115036} {\bibfield  {journal} {\bibinfo  {journal} {Phys. Rev. D}\ }\textbf {\bibinfo {volume} {93}},\ \bibinfo {pages} {115036} (\bibinfo {year} {2016})},\ \bibinfo {note} {[Erratum: Phys.Rev.D 96, 099903 (2017)]},\ \Eprint {https://arxiv.org/abs/1602.01465} {arXiv:1602.01465 [hep-ph]} \BibitemShut {NoStop}%
\bibitem [{\citenamefont {Leane}\ \emph {et~al.}(2017)\citenamefont {Leane}, \citenamefont {Ng},\ and\ \citenamefont {Beacom}}]{Leane:2017vag}%
  \BibitemOpen
  \bibfield  {author} {\bibinfo {author} {\bibfnamefont {R.~K.}\ \bibnamefont {Leane}}, \bibinfo {author} {\bibfnamefont {K.~C.~Y.}\ \bibnamefont {Ng}},\ and\ \bibinfo {author} {\bibfnamefont {J.~F.}\ \bibnamefont {Beacom}},\ }\href {https://doi.org/10.1103/PhysRevD.95.123016} {\bibfield  {journal} {\bibinfo  {journal} {Phys. Rev. D}\ }\textbf {\bibinfo {volume} {95}},\ \bibinfo {pages} {123016} (\bibinfo {year} {2017})},\ \Eprint {https://arxiv.org/abs/1703.04629} {arXiv:1703.04629 [astro-ph.HE]} \BibitemShut {NoStop}%
\bibitem [{\citenamefont {Arina}\ \emph {et~al.}(2017)\citenamefont {Arina}, \citenamefont {Backovi\'c}, \citenamefont {Heisig},\ and\ \citenamefont {Lucente}}]{Arina:2017sng}%
  \BibitemOpen
  \bibfield  {author} {\bibinfo {author} {\bibfnamefont {C.}~\bibnamefont {Arina}}, \bibinfo {author} {\bibfnamefont {M.}~\bibnamefont {Backovi\'c}}, \bibinfo {author} {\bibfnamefont {J.}~\bibnamefont {Heisig}},\ and\ \bibinfo {author} {\bibfnamefont {M.}~\bibnamefont {Lucente}},\ }\href {https://doi.org/10.1103/PhysRevD.96.063010} {\bibfield  {journal} {\bibinfo  {journal} {Phys. Rev. D}\ }\textbf {\bibinfo {volume} {96}},\ \bibinfo {pages} {063010} (\bibinfo {year} {2017})},\ \Eprint {https://arxiv.org/abs/1703.08087} {arXiv:1703.08087 [astro-ph.HE]} \BibitemShut {NoStop}%
\bibitem [{\citenamefont {Mazziotta}\ \emph {et~al.}(2020)\citenamefont {Mazziotta}, \citenamefont {Loparco}, \citenamefont {Serini}, \citenamefont {Cuoco}, \citenamefont {De~La Torre~Luque}, \citenamefont {Gargano},\ and\ \citenamefont {Gustafsson}}]{Mazziotta:2020foa}%
  \BibitemOpen
  \bibfield  {author} {\bibinfo {author} {\bibfnamefont {M.}~\bibnamefont {Mazziotta}}, \bibinfo {author} {\bibfnamefont {F.}~\bibnamefont {Loparco}}, \bibinfo {author} {\bibfnamefont {D.}~\bibnamefont {Serini}}, \bibinfo {author} {\bibfnamefont {A.}~\bibnamefont {Cuoco}}, \bibinfo {author} {\bibfnamefont {P.}~\bibnamefont {De~La Torre~Luque}}, \bibinfo {author} {\bibfnamefont {F.}~\bibnamefont {Gargano}},\ and\ \bibinfo {author} {\bibfnamefont {M.}~\bibnamefont {Gustafsson}},\ }\href {https://doi.org/10.1103/PhysRevD.102.022003} {\bibfield  {journal} {\bibinfo  {journal} {Phys. Rev. D}\ }\textbf {\bibinfo {volume} {102}},\ \bibinfo {pages} {022003} (\bibinfo {year} {2020})},\ \Eprint {https://arxiv.org/abs/2006.04114} {arXiv:2006.04114 [astro-ph.HE]} \BibitemShut {NoStop}%
\bibitem [{\citenamefont {Choi}\ \emph {et~al.}(2015)\citenamefont {Choi} \emph {et~al.}}]{Super-Kamiokande:2015xms}%
  \BibitemOpen
  \bibfield  {author} {\bibinfo {author} {\bibfnamefont {K.}~\bibnamefont {Choi}} \emph {et~al.} (\bibinfo {collaboration} {Super-Kamiokande}),\ }\href {https://doi.org/10.1103/PhysRevLett.114.141301} {\bibfield  {journal} {\bibinfo  {journal} {Phys. Rev. Lett.}\ }\textbf {\bibinfo {volume} {114}},\ \bibinfo {pages} {141301} (\bibinfo {year} {2015})},\ \Eprint {https://arxiv.org/abs/1503.04858} {arXiv:1503.04858 [hep-ex]} \BibitemShut {NoStop}%
\bibitem [{\citenamefont {Abe}\ \emph {et~al.}(2011)\citenamefont {Abe} \emph {et~al.}}]{Abe:2011ts}%
  \BibitemOpen
  \bibfield  {author} {\bibinfo {author} {\bibfnamefont {K.}~\bibnamefont {Abe}} \emph {et~al.},\ }\href@noop {} {\  (\bibinfo {year} {2011})},\ \Eprint {https://arxiv.org/abs/1109.3262} {arXiv:1109.3262 [hep-ex]} \BibitemShut {NoStop}%
\bibitem [{\citenamefont {Adrian-Martinez}\ \emph {et~al.}(2016)\citenamefont {Adrian-Martinez} \emph {et~al.}}]{ANTARES:2016xuh}%
  \BibitemOpen
  \bibfield  {author} {\bibinfo {author} {\bibfnamefont {S.}~\bibnamefont {Adrian-Martinez}} \emph {et~al.} (\bibinfo {collaboration} {ANTARES}),\ }\href {https://doi.org/10.1016/j.physletb.2016.05.019} {\bibfield  {journal} {\bibinfo  {journal} {Phys. Lett. B}\ }\textbf {\bibinfo {volume} {759}},\ \bibinfo {pages} {69} (\bibinfo {year} {2016})},\ \Eprint {https://arxiv.org/abs/1603.02228} {arXiv:1603.02228 [astro-ph.HE]} \BibitemShut {NoStop}%
\bibitem [{\citenamefont {Albert}\ \emph {et~al.}(2017)\citenamefont {Albert} \emph {et~al.}}]{ANTARES:2016bxz}%
  \BibitemOpen
  \bibfield  {author} {\bibinfo {author} {\bibfnamefont {A.}~\bibnamefont {Albert}} \emph {et~al.} (\bibinfo {collaboration} {ANTARES}),\ }\href {https://doi.org/10.1016/j.dark.2017.04.005} {\bibfield  {journal} {\bibinfo  {journal} {Phys. Dark Univ.}\ }\textbf {\bibinfo {volume} {16}},\ \bibinfo {pages} {41} (\bibinfo {year} {2017})},\ \Eprint {https://arxiv.org/abs/1612.06792} {arXiv:1612.06792 [hep-ex]} \BibitemShut {NoStop}%
\bibitem [{\citenamefont {Aartsen}\ \emph {et~al.}(2016)\citenamefont {Aartsen} \emph {et~al.}}]{IceCube:2016yoy}%
  \BibitemOpen
  \bibfield  {author} {\bibinfo {author} {\bibfnamefont {M.~G.}\ \bibnamefont {Aartsen}} \emph {et~al.} (\bibinfo {collaboration} {IceCube}),\ }\href {https://doi.org/10.1088/1475-7516/2016/04/022} {\bibfield  {journal} {\bibinfo  {journal} {JCAP}\ }\textbf {\bibinfo {volume} {04}},\ \bibinfo {pages} {022}},\ \Eprint {https://arxiv.org/abs/1601.00653} {arXiv:1601.00653 [hep-ph]} \BibitemShut {NoStop}%
\bibitem [{\citenamefont {Abbasi}\ \emph {et~al.}(2020)\citenamefont {Abbasi} \emph {et~al.}}]{IceCube:2020wxa}%
  \BibitemOpen
  \bibfield  {author} {\bibinfo {author} {\bibfnamefont {R.}~\bibnamefont {Abbasi}} \emph {et~al.} (\bibinfo {collaboration} {IceCube}),\ }\href {https://doi.org/10.22323/1.358.0541} {\bibfield  {journal} {\bibinfo  {journal} {PoS}\ }\textbf {\bibinfo {volume} {ICRC2019}},\ \bibinfo {pages} {541} (\bibinfo {year} {2020})},\ \Eprint {https://arxiv.org/abs/1908.07255} {arXiv:1908.07255 [astro-ph.HE]} \BibitemShut {NoStop}%
\bibitem [{\citenamefont {Bell}\ \emph {et~al.}(2021)\citenamefont {Bell}, \citenamefont {Dolan},\ and\ \citenamefont {Robles}}]{Bell:2021esh}%
  \BibitemOpen
  \bibfield  {author} {\bibinfo {author} {\bibfnamefont {N.~F.}\ \bibnamefont {Bell}}, \bibinfo {author} {\bibfnamefont {M.~J.}\ \bibnamefont {Dolan}},\ and\ \bibinfo {author} {\bibfnamefont {S.}~\bibnamefont {Robles}},\ }\href {https://doi.org/10.1088/1475-7516/2021/11/004} {\bibfield  {journal} {\bibinfo  {journal} {JCAP}\ }\textbf {\bibinfo {volume} {11}},\ \bibinfo {pages} {004}},\ \Eprint {https://arxiv.org/abs/2107.04216} {arXiv:2107.04216 [hep-ph]} \BibitemShut {NoStop}%
\bibitem [{\citenamefont {Leane}\ and\ \citenamefont {Tong}(2024)}]{Leane:2024bvh}%
  \BibitemOpen
  \bibfield  {author} {\bibinfo {author} {\bibfnamefont {R.~K.}\ \bibnamefont {Leane}}\ and\ \bibinfo {author} {\bibfnamefont {J.}~\bibnamefont {Tong}},\ }\href {https://doi.org/10.1088/1475-7516/2024/12/031} {\bibfield  {journal} {\bibinfo  {journal} {JCAP}\ }\textbf {\bibinfo {volume} {12}},\ \bibinfo {pages} {031}},\ \Eprint {https://arxiv.org/abs/2405.05312} {arXiv:2405.05312 [hep-ph]} \BibitemShut {NoStop}%
\bibitem [{\citenamefont {Nguyen}\ \emph {et~al.}(2025)\citenamefont {Nguyen}, \citenamefont {Linden}, \citenamefont {Carenza},\ and\ \citenamefont {Widmark}}]{Nguyen:2025ygc}%
  \BibitemOpen
  \bibfield  {author} {\bibinfo {author} {\bibfnamefont {T.~T.~Q.}\ \bibnamefont {Nguyen}}, \bibinfo {author} {\bibfnamefont {T.}~\bibnamefont {Linden}}, \bibinfo {author} {\bibfnamefont {P.}~\bibnamefont {Carenza}},\ and\ \bibinfo {author} {\bibfnamefont {A.}~\bibnamefont {Widmark}},\ }\href@noop {} {\  (\bibinfo {year} {2025})},\ \Eprint {https://arxiv.org/abs/2501.14864} {arXiv:2501.14864 [astro-ph.HE]} \BibitemShut {NoStop}%
\bibitem [{\citenamefont {Yin}(2019)}]{Yin:2018yjn}%
  \BibitemOpen
  \bibfield  {author} {\bibinfo {author} {\bibfnamefont {W.}~\bibnamefont {Yin}},\ }\href {https://doi.org/10.1051/epjconf/201920804003} {\bibfield  {journal} {\bibinfo  {journal} {EPJ Web Conf.}\ }\textbf {\bibinfo {volume} {208}},\ \bibinfo {pages} {04003} (\bibinfo {year} {2019})},\ \Eprint {https://arxiv.org/abs/1809.08610} {arXiv:1809.08610 [hep-ph]} \BibitemShut {NoStop}%
\bibitem [{\citenamefont {Ema}\ \emph {et~al.}(2019)\citenamefont {Ema}, \citenamefont {Sala},\ and\ \citenamefont {Sato}}]{Ema:2018bih}%
  \BibitemOpen
  \bibfield  {author} {\bibinfo {author} {\bibfnamefont {Y.}~\bibnamefont {Ema}}, \bibinfo {author} {\bibfnamefont {F.}~\bibnamefont {Sala}},\ and\ \bibinfo {author} {\bibfnamefont {R.}~\bibnamefont {Sato}},\ }\href {https://doi.org/10.1103/PhysRevLett.122.181802} {\bibfield  {journal} {\bibinfo  {journal} {Phys. Rev. Lett.}\ }\textbf {\bibinfo {volume} {122}},\ \bibinfo {pages} {181802} (\bibinfo {year} {2019})},\ \Eprint {https://arxiv.org/abs/1811.00520} {arXiv:1811.00520 [hep-ph]} \BibitemShut {NoStop}%
\bibitem [{\citenamefont {Bringmann}\ and\ \citenamefont {Pospelov}(2019)}]{Bringmann:2018cvk}%
  \BibitemOpen
  \bibfield  {author} {\bibinfo {author} {\bibfnamefont {T.}~\bibnamefont {Bringmann}}\ and\ \bibinfo {author} {\bibfnamefont {M.}~\bibnamefont {Pospelov}},\ }\href {https://doi.org/10.1103/PhysRevLett.122.171801} {\bibfield  {journal} {\bibinfo  {journal} {Phys. Rev. Lett.}\ }\textbf {\bibinfo {volume} {122}},\ \bibinfo {pages} {171801} (\bibinfo {year} {2019})},\ \Eprint {https://arxiv.org/abs/1810.10543} {arXiv:1810.10543 [hep-ph]} \BibitemShut {NoStop}%
\bibitem [{\citenamefont {Cappiello}\ and\ \citenamefont {Beacom}(2019)}]{Cappiello:2019qsw}%
  \BibitemOpen
  \bibfield  {author} {\bibinfo {author} {\bibfnamefont {C.~V.}\ \bibnamefont {Cappiello}}\ and\ \bibinfo {author} {\bibfnamefont {J.~F.}\ \bibnamefont {Beacom}},\ }\href {https://doi.org/10.1103/PhysRevD.104.069901} {\bibfield  {journal} {\bibinfo  {journal} {Phys. Rev. D}\ }\textbf {\bibinfo {volume} {100}},\ \bibinfo {pages} {103011} (\bibinfo {year} {2019})},\ \bibinfo {note} {[Erratum: Phys.Rev.D 104, 069901 (2021)]},\ \Eprint {https://arxiv.org/abs/1906.11283} {arXiv:1906.11283 [hep-ph]} \BibitemShut {NoStop}%
\bibitem [{\citenamefont {Kahn}\ \emph {et~al.}(2015)\citenamefont {Kahn}, \citenamefont {Krnjaic}, \citenamefont {Thaler},\ and\ \citenamefont {Toups}}]{Kahn:2014sra}%
  \BibitemOpen
  \bibfield  {author} {\bibinfo {author} {\bibfnamefont {Y.}~\bibnamefont {Kahn}}, \bibinfo {author} {\bibfnamefont {G.}~\bibnamefont {Krnjaic}}, \bibinfo {author} {\bibfnamefont {J.}~\bibnamefont {Thaler}},\ and\ \bibinfo {author} {\bibfnamefont {M.}~\bibnamefont {Toups}},\ }\href {https://doi.org/10.1103/PhysRevD.91.055006} {\bibfield  {journal} {\bibinfo  {journal} {Phys. Rev. D}\ }\textbf {\bibinfo {volume} {91}},\ \bibinfo {pages} {055006} (\bibinfo {year} {2015})},\ \Eprint {https://arxiv.org/abs/1411.1055} {arXiv:1411.1055 [hep-ph]} \BibitemShut {NoStop}%
\bibitem [{\citenamefont {Agashe}\ \emph {et~al.}(2014)\citenamefont {Agashe}, \citenamefont {Cui}, \citenamefont {Necib},\ and\ \citenamefont {Thaler}}]{Agashe:2014yua}%
  \BibitemOpen
  \bibfield  {author} {\bibinfo {author} {\bibfnamefont {K.}~\bibnamefont {Agashe}}, \bibinfo {author} {\bibfnamefont {Y.}~\bibnamefont {Cui}}, \bibinfo {author} {\bibfnamefont {L.}~\bibnamefont {Necib}},\ and\ \bibinfo {author} {\bibfnamefont {J.}~\bibnamefont {Thaler}},\ }\href {https://doi.org/10.1088/1475-7516/2014/10/062} {\bibfield  {journal} {\bibinfo  {journal} {JCAP}\ }\textbf {\bibinfo {volume} {10}},\ \bibinfo {pages} {062}},\ \Eprint {https://arxiv.org/abs/1405.7370} {arXiv:1405.7370 [hep-ph]} \BibitemShut {NoStop}%
\bibitem [{\citenamefont {Kopp}\ \emph {et~al.}(2015)\citenamefont {Kopp}, \citenamefont {Liu},\ and\ \citenamefont {Wang}}]{Kopp:2015bfa}%
  \BibitemOpen
  \bibfield  {author} {\bibinfo {author} {\bibfnamefont {J.}~\bibnamefont {Kopp}}, \bibinfo {author} {\bibfnamefont {J.}~\bibnamefont {Liu}},\ and\ \bibinfo {author} {\bibfnamefont {X.-P.}\ \bibnamefont {Wang}},\ }\href {https://doi.org/10.1007/JHEP04(2015)105} {\bibfield  {journal} {\bibinfo  {journal} {JHEP}\ }\textbf {\bibinfo {volume} {04}},\ \bibinfo {pages} {105}},\ \Eprint {https://arxiv.org/abs/1503.02669} {arXiv:1503.02669 [hep-ph]} \BibitemShut {NoStop}%
\bibitem [{\citenamefont {Lei}\ \emph {et~al.}(2022)\citenamefont {Lei}, \citenamefont {Tang},\ and\ \citenamefont {Zhang}}]{Lei:2020mii}%
  \BibitemOpen
  \bibfield  {author} {\bibinfo {author} {\bibfnamefont {Z.-H.}\ \bibnamefont {Lei}}, \bibinfo {author} {\bibfnamefont {J.}~\bibnamefont {Tang}},\ and\ \bibinfo {author} {\bibfnamefont {B.-L.}\ \bibnamefont {Zhang}},\ }\href {https://doi.org/10.1088/1674-1137/ac68da} {\bibfield  {journal} {\bibinfo  {journal} {Chin. Phys. C}\ }\textbf {\bibinfo {volume} {46}},\ \bibinfo {pages} {085103} (\bibinfo {year} {2022})},\ \Eprint {https://arxiv.org/abs/2008.07116} {arXiv:2008.07116 [hep-ph]} \BibitemShut {NoStop}%
\bibitem [{\citenamefont {An}\ \emph {et~al.}(2021)\citenamefont {An}, \citenamefont {Nie}, \citenamefont {Pospelov}, \citenamefont {Pradler},\ and\ \citenamefont {Ritz}}]{An:2021qdl}%
  \BibitemOpen
  \bibfield  {author} {\bibinfo {author} {\bibfnamefont {H.}~\bibnamefont {An}}, \bibinfo {author} {\bibfnamefont {H.}~\bibnamefont {Nie}}, \bibinfo {author} {\bibfnamefont {M.}~\bibnamefont {Pospelov}}, \bibinfo {author} {\bibfnamefont {J.}~\bibnamefont {Pradler}},\ and\ \bibinfo {author} {\bibfnamefont {A.}~\bibnamefont {Ritz}},\ }\href {https://doi.org/10.1103/PhysRevD.104.103026} {\bibfield  {journal} {\bibinfo  {journal} {Phys. Rev. D}\ }\textbf {\bibinfo {volume} {104}},\ \bibinfo {pages} {103026} (\bibinfo {year} {2021})},\ \Eprint {https://arxiv.org/abs/2108.10332} {arXiv:2108.10332 [hep-ph]} \BibitemShut {NoStop}%
\bibitem [{\citenamefont {Wang}\ \emph {et~al.}(2022)\citenamefont {Wang}, \citenamefont {Granelli},\ and\ \citenamefont {Ullio}}]{Wang:2021jic}%
  \BibitemOpen
  \bibfield  {author} {\bibinfo {author} {\bibfnamefont {J.-W.}\ \bibnamefont {Wang}}, \bibinfo {author} {\bibfnamefont {A.}~\bibnamefont {Granelli}},\ and\ \bibinfo {author} {\bibfnamefont {P.}~\bibnamefont {Ullio}},\ }\href {https://doi.org/10.1103/PhysRevLett.128.221104} {\bibfield  {journal} {\bibinfo  {journal} {Phys. Rev. Lett.}\ }\textbf {\bibinfo {volume} {128}},\ \bibinfo {pages} {221104} (\bibinfo {year} {2022})},\ \Eprint {https://arxiv.org/abs/2111.13644} {arXiv:2111.13644 [astro-ph.HE]} \BibitemShut {NoStop}%
\bibitem [{\citenamefont {Toma}(2022)}]{Toma:2021vlw}%
  \BibitemOpen
  \bibfield  {author} {\bibinfo {author} {\bibfnamefont {T.}~\bibnamefont {Toma}},\ }\href {https://doi.org/10.1103/PhysRevD.105.043007} {\bibfield  {journal} {\bibinfo  {journal} {Phys. Rev. D}\ }\textbf {\bibinfo {volume} {105}},\ \bibinfo {pages} {043007} (\bibinfo {year} {2022})},\ \Eprint {https://arxiv.org/abs/2109.05911} {arXiv:2109.05911 [hep-ph]} \BibitemShut {NoStop}%
\bibitem [{\citenamefont {Aoki}\ and\ \citenamefont {Toma}(2024)}]{Aoki:2023tlb}%
  \BibitemOpen
  \bibfield  {author} {\bibinfo {author} {\bibfnamefont {M.}~\bibnamefont {Aoki}}\ and\ \bibinfo {author} {\bibfnamefont {T.}~\bibnamefont {Toma}},\ }\href {https://doi.org/10.1088/1475-7516/2024/02/033} {\bibfield  {journal} {\bibinfo  {journal} {JCAP}\ }\textbf {\bibinfo {volume} {02}},\ \bibinfo {pages} {033}},\ \Eprint {https://arxiv.org/abs/2309.00395} {arXiv:2309.00395 [hep-ph]} \BibitemShut {NoStop}%
\bibitem [{\citenamefont {Das}\ and\ \citenamefont {Sen}(2021)}]{Das:2021lcr}%
  \BibitemOpen
  \bibfield  {author} {\bibinfo {author} {\bibfnamefont {A.}~\bibnamefont {Das}}\ and\ \bibinfo {author} {\bibfnamefont {M.}~\bibnamefont {Sen}},\ }\href {https://doi.org/10.1103/PhysRevD.104.075029} {\bibfield  {journal} {\bibinfo  {journal} {Phys. Rev. D}\ }\textbf {\bibinfo {volume} {104}},\ \bibinfo {pages} {075029} (\bibinfo {year} {2021})},\ \Eprint {https://arxiv.org/abs/2104.00027} {arXiv:2104.00027 [hep-ph]} \BibitemShut {NoStop}%
\bibitem [{\citenamefont {Zhang}\ \emph {et~al.}(2024{\natexlab{a}})\citenamefont {Zhang} \emph {et~al.}}]{CDEX:2023wfz}%
  \BibitemOpen
  \bibfield  {author} {\bibinfo {author} {\bibfnamefont {Z.~Y.}\ \bibnamefont {Zhang}} \emph {et~al.} (\bibinfo {collaboration} {CDEX}),\ }\href {https://doi.org/10.1103/PhysRevLett.132.171001} {\bibfield  {journal} {\bibinfo  {journal} {Phys. Rev. Lett.}\ }\textbf {\bibinfo {volume} {132}},\ \bibinfo {pages} {171001} (\bibinfo {year} {2024}{\natexlab{a}})},\ \Eprint {https://arxiv.org/abs/2309.14982} {arXiv:2309.14982 [hep-ex]} \BibitemShut {NoStop}%
\bibitem [{\citenamefont {Bell}\ \emph {et~al.}(2024)\citenamefont {Bell}, \citenamefont {Newstead},\ and\ \citenamefont {Shaukat-Ali}}]{Bell:2023sdq}%
  \BibitemOpen
  \bibfield  {author} {\bibinfo {author} {\bibfnamefont {N.~F.}\ \bibnamefont {Bell}}, \bibinfo {author} {\bibfnamefont {J.~L.}\ \bibnamefont {Newstead}},\ and\ \bibinfo {author} {\bibfnamefont {I.}~\bibnamefont {Shaukat-Ali}},\ }\href {https://doi.org/10.1103/PhysRevD.109.063034} {\bibfield  {journal} {\bibinfo  {journal} {Phys. Rev. D}\ }\textbf {\bibinfo {volume} {109}},\ \bibinfo {pages} {063034} (\bibinfo {year} {2024})},\ \Eprint {https://arxiv.org/abs/2309.11003} {arXiv:2309.11003 [hep-ph]} \BibitemShut {NoStop}%
\bibitem [{\citenamefont {Liang}\ \emph {et~al.}(2024)\citenamefont {Liang}, \citenamefont {Su}, \citenamefont {Wu},\ and\ \citenamefont {Zhu}}]{Liang:2024xcx}%
  \BibitemOpen
  \bibfield  {author} {\bibinfo {author} {\bibfnamefont {Z.-L.}\ \bibnamefont {Liang}}, \bibinfo {author} {\bibfnamefont {L.}~\bibnamefont {Su}}, \bibinfo {author} {\bibfnamefont {L.}~\bibnamefont {Wu}},\ and\ \bibinfo {author} {\bibfnamefont {B.}~\bibnamefont {Zhu}},\ }\href@noop {} {\  (\bibinfo {year} {2024})},\ \Eprint {https://arxiv.org/abs/2401.11971} {arXiv:2401.11971 [hep-ph]} \BibitemShut {NoStop}%
\bibitem [{\citenamefont {Dutta}\ \emph {et~al.}(2024)\citenamefont {Dutta}, \citenamefont {Huang}, \citenamefont {Kim}, \citenamefont {Newstead}, \citenamefont {Park},\ and\ \citenamefont {Ali}}]{Dutta:2024kuj}%
  \BibitemOpen
  \bibfield  {author} {\bibinfo {author} {\bibfnamefont {B.}~\bibnamefont {Dutta}}, \bibinfo {author} {\bibfnamefont {W.-C.}\ \bibnamefont {Huang}}, \bibinfo {author} {\bibfnamefont {D.}~\bibnamefont {Kim}}, \bibinfo {author} {\bibfnamefont {J.~L.}\ \bibnamefont {Newstead}}, \bibinfo {author} {\bibfnamefont {J.-C.}\ \bibnamefont {Park}},\ and\ \bibinfo {author} {\bibfnamefont {I.~S.}\ \bibnamefont {Ali}},\ }\href {https://doi.org/10.1103/PhysRevLett.133.161801} {\bibfield  {journal} {\bibinfo  {journal} {Phys. Rev. Lett.}\ }\textbf {\bibinfo {volume} {133}},\ \bibinfo {pages} {161801} (\bibinfo {year} {2024})},\ \Eprint {https://arxiv.org/abs/2402.04184} {arXiv:2402.04184 [hep-ph]} \BibitemShut {NoStop}%
\bibitem [{\citenamefont {Bramante}\ \emph {et~al.}(2019{\natexlab{a}})\citenamefont {Bramante}, \citenamefont {Broerman}, \citenamefont {Kumar}, \citenamefont {Lang}, \citenamefont {Pospelov},\ and\ \citenamefont {Raj}}]{Bramante:2018tos}%
  \BibitemOpen
  \bibfield  {author} {\bibinfo {author} {\bibfnamefont {J.}~\bibnamefont {Bramante}}, \bibinfo {author} {\bibfnamefont {B.}~\bibnamefont {Broerman}}, \bibinfo {author} {\bibfnamefont {J.}~\bibnamefont {Kumar}}, \bibinfo {author} {\bibfnamefont {R.~F.}\ \bibnamefont {Lang}}, \bibinfo {author} {\bibfnamefont {M.}~\bibnamefont {Pospelov}},\ and\ \bibinfo {author} {\bibfnamefont {N.}~\bibnamefont {Raj}},\ }\href {https://doi.org/10.1103/PhysRevD.99.083010} {\bibfield  {journal} {\bibinfo  {journal} {Phys. Rev. D}\ }\textbf {\bibinfo {volume} {99}},\ \bibinfo {pages} {083010} (\bibinfo {year} {2019}{\natexlab{a}})},\ \Eprint {https://arxiv.org/abs/1812.09325} {arXiv:1812.09325 [hep-ph]} \BibitemShut {NoStop}%
\bibitem [{\citenamefont {Bai}\ \emph {et~al.}(2022)\citenamefont {Bai}, \citenamefont {Berger},\ and\ \citenamefont {Korwar}}]{Bai:2022nsv}%
  \BibitemOpen
  \bibfield  {author} {\bibinfo {author} {\bibfnamefont {Y.}~\bibnamefont {Bai}}, \bibinfo {author} {\bibfnamefont {J.}~\bibnamefont {Berger}},\ and\ \bibinfo {author} {\bibfnamefont {M.}~\bibnamefont {Korwar}},\ }\href {https://doi.org/10.1007/JHEP11(2022)079} {\bibfield  {journal} {\bibinfo  {journal} {JHEP}\ }\textbf {\bibinfo {volume} {11}},\ \bibinfo {pages} {079}},\ \Eprint {https://arxiv.org/abs/2206.07928} {arXiv:2206.07928 [hep-ph]} \BibitemShut {NoStop}%
\bibitem [{\citenamefont {Eby}\ \emph {et~al.}(2019)\citenamefont {Eby}, \citenamefont {Fox}, \citenamefont {Harnik},\ and\ \citenamefont {Kribs}}]{Eby:2019mgs}%
  \BibitemOpen
  \bibfield  {author} {\bibinfo {author} {\bibfnamefont {J.}~\bibnamefont {Eby}}, \bibinfo {author} {\bibfnamefont {P.~J.}\ \bibnamefont {Fox}}, \bibinfo {author} {\bibfnamefont {R.}~\bibnamefont {Harnik}},\ and\ \bibinfo {author} {\bibfnamefont {G.~D.}\ \bibnamefont {Kribs}},\ }\href {https://doi.org/10.1007/JHEP09(2019)115} {\bibfield  {journal} {\bibinfo  {journal} {JHEP}\ }\textbf {\bibinfo {volume} {09}},\ \bibinfo {pages} {115}},\ \Eprint {https://arxiv.org/abs/1904.09994} {arXiv:1904.09994 [hep-ph]} \BibitemShut {NoStop}%
\bibitem [{\citenamefont {Aggarwal}\ and\ \citenamefont {Raj}(2024)}]{Aggarwal:2024ngx}%
  \BibitemOpen
  \bibfield  {author} {\bibinfo {author} {\bibfnamefont {H.}~\bibnamefont {Aggarwal}}\ and\ \bibinfo {author} {\bibfnamefont {N.}~\bibnamefont {Raj}},\ }\href@noop {} {\  (\bibinfo {year} {2024})},\ \Eprint {https://arxiv.org/abs/2410.22168} {arXiv:2410.22168 [hep-ph]} \BibitemShut {NoStop}%
\bibitem [{\citenamefont {Bramante}\ \emph {et~al.}(2019{\natexlab{b}})\citenamefont {Bramante}, \citenamefont {Kumar},\ and\ \citenamefont {Raj}}]{Bramante:2019yss}%
  \BibitemOpen
  \bibfield  {author} {\bibinfo {author} {\bibfnamefont {J.}~\bibnamefont {Bramante}}, \bibinfo {author} {\bibfnamefont {J.}~\bibnamefont {Kumar}},\ and\ \bibinfo {author} {\bibfnamefont {N.}~\bibnamefont {Raj}},\ }\href {https://doi.org/10.1103/PhysRevD.100.123016} {\bibfield  {journal} {\bibinfo  {journal} {Phys. Rev. D}\ }\textbf {\bibinfo {volume} {100}},\ \bibinfo {pages} {123016} (\bibinfo {year} {2019}{\natexlab{b}})},\ \Eprint {https://arxiv.org/abs/1910.05380} {arXiv:1910.05380 [hep-ph]} \BibitemShut {NoStop}%
\bibitem [{\citenamefont {Pont\'on}\ \emph {et~al.}(2019)\citenamefont {Pont\'on}, \citenamefont {Bai},\ and\ \citenamefont {Jain}}]{Ponton:2019hux}%
  \BibitemOpen
  \bibfield  {author} {\bibinfo {author} {\bibfnamefont {E.}~\bibnamefont {Pont\'on}}, \bibinfo {author} {\bibfnamefont {Y.}~\bibnamefont {Bai}},\ and\ \bibinfo {author} {\bibfnamefont {B.}~\bibnamefont {Jain}},\ }\href {https://doi.org/10.1007/s13130-019-11194-5} {\bibfield  {journal} {\bibinfo  {journal} {JHEP}\ }\textbf {\bibinfo {volume} {09}},\ \bibinfo {pages} {011}},\ \Eprint {https://arxiv.org/abs/1906.10739} {arXiv:1906.10739 [hep-ph]} \BibitemShut {NoStop}%
\bibitem [{\citenamefont {Church}\ \emph {et~al.}(2020)\citenamefont {Church}, \citenamefont {Jackson},\ and\ \citenamefont {Saldanha}}]{Church:2020env}%
  \BibitemOpen
  \bibfield  {author} {\bibinfo {author} {\bibfnamefont {E.}~\bibnamefont {Church}}, \bibinfo {author} {\bibfnamefont {C.~M.}\ \bibnamefont {Jackson}},\ and\ \bibinfo {author} {\bibfnamefont {R.}~\bibnamefont {Saldanha}},\ }\href {https://doi.org/10.1088/1748-0221/15/09/P09026} {\bibfield  {journal} {\bibinfo  {journal} {JINST}\ }\textbf {\bibinfo {volume} {15}}\bibfield  {number} {\bibinfo  {number} { (09)},\ \bibinfo {pages} {P09026}},\ }\Eprint {https://arxiv.org/abs/2005.04824} {arXiv:2005.04824 [physics.ins-det]} \BibitemShut {NoStop}%
\bibitem [{\citenamefont {Abusleme}\ \emph {et~al.}(2021{\natexlab{b}})\citenamefont {Abusleme} \emph {et~al.}}]{JUNO:2020bcl}%
  \BibitemOpen
  \bibfield  {author} {\bibinfo {author} {\bibfnamefont {A.}~\bibnamefont {Abusleme}} \emph {et~al.} (\bibinfo {collaboration} {JUNO, Daya Bay}),\ }\href {https://doi.org/10.1016/j.nima.2020.164823} {\bibfield  {journal} {\bibinfo  {journal} {Nucl. Instrum. Meth. A}\ }\textbf {\bibinfo {volume} {988}},\ \bibinfo {pages} {164823} (\bibinfo {year} {2021}{\natexlab{b}})},\ \Eprint {https://arxiv.org/abs/2007.00314} {arXiv:2007.00314 [physics.ins-det]} \BibitemShut {NoStop}%
\bibitem [{\citenamefont {Abusleme}\ \emph {et~al.}(2022{\natexlab{b}})\citenamefont {Abusleme} \emph {et~al.}}]{JUNO:2022hlz}%
  \BibitemOpen
  \bibfield  {author} {\bibinfo {author} {\bibfnamefont {A.}~\bibnamefont {Abusleme}} \emph {et~al.} (\bibinfo {collaboration} {JUNO}),\ }\href {https://doi.org/10.1140/epjc/s10052-022-11002-8} {\bibfield  {journal} {\bibinfo  {journal} {Eur. Phys. J. C}\ }\textbf {\bibinfo {volume} {82}},\ \bibinfo {pages} {1168} (\bibinfo {year} {2022}{\natexlab{b}})},\ \Eprint {https://arxiv.org/abs/2205.08629} {arXiv:2205.08629 [physics.ins-det]} \BibitemShut {NoStop}%
\bibitem [{\citenamefont {Zhang}\ \emph {et~al.}(2024{\natexlab{b}})\citenamefont {Zhang}, \citenamefont {Wang}, \citenamefont {Li}, \citenamefont {Liu}, \citenamefont {Rodphai}, \citenamefont {Zhang}, \citenamefont {Xu}, \citenamefont {Yang},\ and\ \citenamefont {Heng}}]{Zhang:2023dha}%
  \BibitemOpen
  \bibfield  {author} {\bibinfo {author} {\bibfnamefont {Y.}~\bibnamefont {Zhang}}, \bibinfo {author} {\bibfnamefont {Z.}~\bibnamefont {Wang}}, \bibinfo {author} {\bibfnamefont {M.}~\bibnamefont {Li}}, \bibinfo {author} {\bibfnamefont {C.}~\bibnamefont {Liu}}, \bibinfo {author} {\bibfnamefont {N.}~\bibnamefont {Rodphai}}, \bibinfo {author} {\bibfnamefont {Y.}~\bibnamefont {Zhang}}, \bibinfo {author} {\bibfnamefont {J.}~\bibnamefont {Xu}}, \bibinfo {author} {\bibfnamefont {C.}~\bibnamefont {Yang}},\ and\ \bibinfo {author} {\bibfnamefont {Y.}~\bibnamefont {Heng}},\ }\href {https://doi.org/10.1088/1748-0221/19/02/P02026} {\bibfield  {journal} {\bibinfo  {journal} {JINST}\ }\textbf {\bibinfo {volume} {19}}\bibfield  {number} {\bibinfo  {number} { (02)},\ \bibinfo {pages} {P02026}},\ }\Eprint {https://arxiv.org/abs/2307.15104} {arXiv:2307.15104 [physics.ins-det]} \BibitemShut {NoStop}%
\bibitem [{\citenamefont {Lombardi}\ \emph {et~al.}(2019)\citenamefont {Lombardi} \emph {et~al.}}]{Lombardi:2019epz}%
  \BibitemOpen
  \bibfield  {author} {\bibinfo {author} {\bibfnamefont {P.}~\bibnamefont {Lombardi}} \emph {et~al.},\ }\href {https://doi.org/10.1016/j.nima.2019.01.071} {\bibfield  {journal} {\bibinfo  {journal} {Nucl. Instrum. Meth. A}\ }\textbf {\bibinfo {volume} {925}},\ \bibinfo {pages} {6} (\bibinfo {year} {2019})},\ \Eprint {https://arxiv.org/abs/1902.05288} {arXiv:1902.05288 [physics.ins-det]} \BibitemShut {NoStop}%
\bibitem [{\citenamefont {{L.~S.~B.}}(1972)}]{1972JMoSt..13..140L}%
  \BibitemOpen
  \bibfield  {author} {\bibinfo {author} {\bibnamefont {{L.~S.~B.}}},\ }\href {https://doi.org/10.1016/0022-2860(72)87041-8} {\bibfield  {journal} {\bibinfo  {journal} {Journal of Molecular Structure}\ }\textbf {\bibinfo {volume} {13}},\ \bibinfo {pages} {140} (\bibinfo {year} {1972})}\BibitemShut {NoStop}%
\bibitem [{\citenamefont {Beretta}\ \emph {et~al.}(2025)\citenamefont {Beretta} \emph {et~al.}}]{Beretta:2025rwj}%
  \BibitemOpen
  \bibfield  {author} {\bibinfo {author} {\bibfnamefont {M.}~\bibnamefont {Beretta}} \emph {et~al.},\ }\href@noop {} {\  (\bibinfo {year} {2025})},\ \Eprint {https://arxiv.org/abs/2501.09988} {arXiv:2501.09988 [physics.ins-det]} \BibitemShut {NoStop}%
\bibitem [{\citenamefont {Zhang}\ \emph {et~al.}(2020)\citenamefont {Zhang}, \citenamefont {Yu}, \citenamefont {Li}, \citenamefont {Deng},\ and\ \citenamefont {Wen}}]{Zhang:2020mqz}%
  \BibitemOpen
  \bibfield  {author} {\bibinfo {author} {\bibfnamefont {Y.}~\bibnamefont {Zhang}}, \bibinfo {author} {\bibfnamefont {Z.-Y.}\ \bibnamefont {Yu}}, \bibinfo {author} {\bibfnamefont {X.-Y.}\ \bibnamefont {Li}}, \bibinfo {author} {\bibfnamefont {Z.-Y.}\ \bibnamefont {Deng}},\ and\ \bibinfo {author} {\bibfnamefont {L.-J.}\ \bibnamefont {Wen}},\ }\href {https://doi.org/10.1016/j.nima.2020.163860} {\bibfield  {journal} {\bibinfo  {journal} {Nucl. Instrum. Meth. A}\ }\textbf {\bibinfo {volume} {967}},\ \bibinfo {pages} {163860} (\bibinfo {year} {2020})},\ \Eprint {https://arxiv.org/abs/2003.12212} {arXiv:2003.12212 [physics.ins-det]} \BibitemShut {NoStop}%
\bibitem [{\citenamefont {Li}\ \emph {et~al.}(2016)\citenamefont {Li}, \citenamefont {Guo}, \citenamefont {Yeh}, \citenamefont {Wang},\ and\ \citenamefont {Chen}}]{Li:2015phc}%
  \BibitemOpen
  \bibfield  {author} {\bibinfo {author} {\bibfnamefont {M.}~\bibnamefont {Li}}, \bibinfo {author} {\bibfnamefont {Z.}~\bibnamefont {Guo}}, \bibinfo {author} {\bibfnamefont {M.}~\bibnamefont {Yeh}}, \bibinfo {author} {\bibfnamefont {Z.}~\bibnamefont {Wang}},\ and\ \bibinfo {author} {\bibfnamefont {S.}~\bibnamefont {Chen}},\ }\href {https://doi.org/10.1016/j.nima.2016.05.132} {\bibfield  {journal} {\bibinfo  {journal} {Nucl. Instrum. Meth. A}\ }\textbf {\bibinfo {volume} {830}},\ \bibinfo {pages} {303} (\bibinfo {year} {2016})},\ \Eprint {https://arxiv.org/abs/1511.09339} {arXiv:1511.09339 [physics.ins-det]} \BibitemShut {NoStop}%
\bibitem [{\citenamefont {Drukier}\ \emph {et~al.}(1986)\citenamefont {Drukier}, \citenamefont {Freese},\ and\ \citenamefont {Spergel}}]{Drukier:1986tm}%
  \BibitemOpen
  \bibfield  {author} {\bibinfo {author} {\bibfnamefont {A.~K.}\ \bibnamefont {Drukier}}, \bibinfo {author} {\bibfnamefont {K.}~\bibnamefont {Freese}},\ and\ \bibinfo {author} {\bibfnamefont {D.~N.}\ \bibnamefont {Spergel}},\ }\href {https://doi.org/10.1103/PhysRevD.33.3495} {\bibfield  {journal} {\bibinfo  {journal} {Phys. Rev. D}\ }\textbf {\bibinfo {volume} {33}},\ \bibinfo {pages} {3495} (\bibinfo {year} {1986})}\BibitemShut {NoStop}%
\bibitem [{\citenamefont {Freese}\ \emph {et~al.}(2013)\citenamefont {Freese}, \citenamefont {Lisanti},\ and\ \citenamefont {Savage}}]{Freese:2012xd}%
  \BibitemOpen
  \bibfield  {author} {\bibinfo {author} {\bibfnamefont {K.}~\bibnamefont {Freese}}, \bibinfo {author} {\bibfnamefont {M.}~\bibnamefont {Lisanti}},\ and\ \bibinfo {author} {\bibfnamefont {C.}~\bibnamefont {Savage}},\ }\href {https://doi.org/10.1103/RevModPhys.85.1561} {\bibfield  {journal} {\bibinfo  {journal} {Rev. Mod. Phys.}\ }\textbf {\bibinfo {volume} {85}},\ \bibinfo {pages} {1561} (\bibinfo {year} {2013})},\ \Eprint {https://arxiv.org/abs/1209.3339} {arXiv:1209.3339 [astro-ph.CO]} \BibitemShut {NoStop}%
\bibitem [{\citenamefont {Amaral}\ \emph {et~al.}(2020)\citenamefont {Amaral} \emph {et~al.}}]{SuperCDMS:2020ymb}%
  \BibitemOpen
  \bibfield  {author} {\bibinfo {author} {\bibfnamefont {D.~W.}\ \bibnamefont {Amaral}} \emph {et~al.} (\bibinfo {collaboration} {SuperCDMS}),\ }\href {https://doi.org/10.1103/PhysRevD.102.091101} {\bibfield  {journal} {\bibinfo  {journal} {Phys. Rev. D}\ }\textbf {\bibinfo {volume} {102}},\ \bibinfo {pages} {091101} (\bibinfo {year} {2020})},\ \Eprint {https://arxiv.org/abs/2005.14067} {arXiv:2005.14067 [hep-ex]} \BibitemShut {NoStop}%
\bibitem [{\citenamefont {Aguilar-Arevalo}\ \emph {et~al.}(2019)\citenamefont {Aguilar-Arevalo} \emph {et~al.}}]{DAMIC:2019dcn}%
  \BibitemOpen
  \bibfield  {author} {\bibinfo {author} {\bibfnamefont {A.}~\bibnamefont {Aguilar-Arevalo}} \emph {et~al.} (\bibinfo {collaboration} {DAMIC}),\ }\href {https://doi.org/10.1103/PhysRevLett.123.181802} {\bibfield  {journal} {\bibinfo  {journal} {Phys. Rev. Lett.}\ }\textbf {\bibinfo {volume} {123}},\ \bibinfo {pages} {181802} (\bibinfo {year} {2019})},\ \Eprint {https://arxiv.org/abs/1907.12628} {arXiv:1907.12628 [astro-ph.CO]} \BibitemShut {NoStop}%
\bibitem [{\citenamefont {Barak}\ \emph {et~al.}(2020)\citenamefont {Barak} \emph {et~al.}}]{SENSEI:2020dpa}%
  \BibitemOpen
  \bibfield  {author} {\bibinfo {author} {\bibfnamefont {L.}~\bibnamefont {Barak}} \emph {et~al.} (\bibinfo {collaboration} {SENSEI}),\ }\href {https://doi.org/10.1103/PhysRevLett.125.171802} {\bibfield  {journal} {\bibinfo  {journal} {Phys. Rev. Lett.}\ }\textbf {\bibinfo {volume} {125}},\ \bibinfo {pages} {171802} (\bibinfo {year} {2020})},\ \Eprint {https://arxiv.org/abs/2004.11378} {arXiv:2004.11378 [astro-ph.CO]} \BibitemShut {NoStop}%
\bibitem [{\citenamefont {Cheng}\ \emph {et~al.}(2021)\citenamefont {Cheng} \emph {et~al.}}]{PandaX-II:2021nsg}%
  \BibitemOpen
  \bibfield  {author} {\bibinfo {author} {\bibfnamefont {C.}~\bibnamefont {Cheng}} \emph {et~al.} (\bibinfo {collaboration} {PandaX-II}),\ }\href {https://doi.org/10.1103/PhysRevLett.126.211803} {\bibfield  {journal} {\bibinfo  {journal} {Phys. Rev. Lett.}\ }\textbf {\bibinfo {volume} {126}},\ \bibinfo {pages} {211803} (\bibinfo {year} {2021})},\ \Eprint {https://arxiv.org/abs/2101.07479} {arXiv:2101.07479 [hep-ex]} \BibitemShut {NoStop}%
\bibitem [{\citenamefont {Agnes}\ \emph {et~al.}(2023)\citenamefont {Agnes} \emph {et~al.}}]{DarkSide:2022knj}%
  \BibitemOpen
  \bibfield  {author} {\bibinfo {author} {\bibfnamefont {P.}~\bibnamefont {Agnes}} \emph {et~al.} (\bibinfo {collaboration} {DarkSide}),\ }\href {https://doi.org/10.1103/PhysRevLett.130.101002} {\bibfield  {journal} {\bibinfo  {journal} {Phys. Rev. Lett.}\ }\textbf {\bibinfo {volume} {130}},\ \bibinfo {pages} {101002} (\bibinfo {year} {2023})},\ \Eprint {https://arxiv.org/abs/2207.11968} {arXiv:2207.11968 [hep-ex]} \BibitemShut {NoStop}%
\bibitem [{\citenamefont {Aprile}\ \emph {et~al.}(2019)\citenamefont {Aprile} \emph {et~al.}}]{XENON:2019gfn}%
  \BibitemOpen
  \bibfield  {author} {\bibinfo {author} {\bibfnamefont {E.}~\bibnamefont {Aprile}} \emph {et~al.} (\bibinfo {collaboration} {XENON}),\ }\href {https://doi.org/10.1103/PhysRevLett.123.251801} {\bibfield  {journal} {\bibinfo  {journal} {Phys. Rev. Lett.}\ }\textbf {\bibinfo {volume} {123}},\ \bibinfo {pages} {251801} (\bibinfo {year} {2019})},\ \Eprint {https://arxiv.org/abs/1907.11485} {arXiv:1907.11485 [hep-ex]} \BibitemShut {NoStop}%
\bibitem [{\citenamefont {An}\ \emph {et~al.}(2018)\citenamefont {An}, \citenamefont {Pospelov}, \citenamefont {Pradler},\ and\ \citenamefont {Ritz}}]{An:2017ojc}%
  \BibitemOpen
  \bibfield  {author} {\bibinfo {author} {\bibfnamefont {H.}~\bibnamefont {An}}, \bibinfo {author} {\bibfnamefont {M.}~\bibnamefont {Pospelov}}, \bibinfo {author} {\bibfnamefont {J.}~\bibnamefont {Pradler}},\ and\ \bibinfo {author} {\bibfnamefont {A.}~\bibnamefont {Ritz}},\ }\href {https://doi.org/10.1103/PhysRevLett.120.141801} {\bibfield  {journal} {\bibinfo  {journal} {Phys. Rev. Lett.}\ }\textbf {\bibinfo {volume} {120}},\ \bibinfo {pages} {141801} (\bibinfo {year} {2018})},\ \bibinfo {note} {[Erratum: Phys.Rev.Lett. 121, 259903 (2018)]},\ \Eprint {https://arxiv.org/abs/1708.03642} {arXiv:1708.03642 [hep-ph]} \BibitemShut {NoStop}%
\bibitem [{\citenamefont {Emken}\ \emph {et~al.}(2018)\citenamefont {Emken}, \citenamefont {Kouvaris},\ and\ \citenamefont {Nielsen}}]{Emken:2017hnp}%
  \BibitemOpen
  \bibfield  {author} {\bibinfo {author} {\bibfnamefont {T.}~\bibnamefont {Emken}}, \bibinfo {author} {\bibfnamefont {C.}~\bibnamefont {Kouvaris}},\ and\ \bibinfo {author} {\bibfnamefont {N.~G.}\ \bibnamefont {Nielsen}},\ }\href {https://doi.org/10.1103/PhysRevD.97.063007} {\bibfield  {journal} {\bibinfo  {journal} {Phys. Rev. D}\ }\textbf {\bibinfo {volume} {97}},\ \bibinfo {pages} {063007} (\bibinfo {year} {2018})},\ \Eprint {https://arxiv.org/abs/1709.06573} {arXiv:1709.06573 [hep-ph]} \BibitemShut {NoStop}%
\bibitem [{\citenamefont {Eadie}\ \emph {et~al.}(1971)\citenamefont {Eadie}, \citenamefont {Drijard}, \citenamefont {James}, \citenamefont {Roos}, \citenamefont {James},\ and\ \citenamefont {Sadoulet}}]{Eadie:1971qcl}%
  \BibitemOpen
  \bibfield  {author} {\bibinfo {author} {\bibfnamefont {W.~T.}\ \bibnamefont {Eadie}}, \bibinfo {author} {\bibfnamefont {D.}~\bibnamefont {Drijard}}, \bibinfo {author} {\bibfnamefont {F.~E.}\ \bibnamefont {James}}, \bibinfo {author} {\bibfnamefont {M.}~\bibnamefont {Roos}}, \bibinfo {author} {\bibfnamefont {F.~E.}\ \bibnamefont {James}},\ and\ \bibinfo {author} {\bibfnamefont {B.}~\bibnamefont {Sadoulet}},\ }\href@noop {} {\emph {\bibinfo {title} {{Statistical Methods in Experimental Physics}}}}\ (\bibinfo  {publisher} {North-Holland Publishing Company},\ \bibinfo {address} {Netherlands},\ \bibinfo {year} {1971})\BibitemShut {NoStop}%
\bibitem [{\citenamefont {Beacom}\ and\ \citenamefont {Vogel}(1999)}]{Beacom:1998fj}%
  \BibitemOpen
  \bibfield  {author} {\bibinfo {author} {\bibfnamefont {J.~F.}\ \bibnamefont {Beacom}}\ and\ \bibinfo {author} {\bibfnamefont {P.}~\bibnamefont {Vogel}},\ }\href {https://doi.org/10.1103/PhysRevD.60.033007} {\bibfield  {journal} {\bibinfo  {journal} {Phys. Rev. D}\ }\textbf {\bibinfo {volume} {60}},\ \bibinfo {pages} {033007} (\bibinfo {year} {1999})},\ \Eprint {https://arxiv.org/abs/astro-ph/9811350} {arXiv:astro-ph/9811350} \BibitemShut {NoStop}%
\bibitem [{\citenamefont {Lisi}\ \emph {et~al.}(2004)\citenamefont {Lisi}, \citenamefont {Palazzo},\ and\ \citenamefont {Rotunno}}]{Lisi:2004jw}%
  \BibitemOpen
  \bibfield  {author} {\bibinfo {author} {\bibfnamefont {E.}~\bibnamefont {Lisi}}, \bibinfo {author} {\bibfnamefont {A.}~\bibnamefont {Palazzo}},\ and\ \bibinfo {author} {\bibfnamefont {A.~M.}\ \bibnamefont {Rotunno}},\ }\href {https://doi.org/10.1016/j.astropartphys.2004.03.005} {\bibfield  {journal} {\bibinfo  {journal} {Astropart. Phys.}\ }\textbf {\bibinfo {volume} {21}},\ \bibinfo {pages} {511} (\bibinfo {year} {2004})},\ \Eprint {https://arxiv.org/abs/hep-ph/0403036} {arXiv:hep-ph/0403036} \BibitemShut {NoStop}%
\bibitem [{\citenamefont {Bellini}\ \emph {et~al.}(2014)\citenamefont {Bellini} \emph {et~al.}}]{Borexino:2013zhu}%
  \BibitemOpen
  \bibfield  {author} {\bibinfo {author} {\bibfnamefont {G.}~\bibnamefont {Bellini}} \emph {et~al.} (\bibinfo {collaboration} {Borexino}),\ }\href {https://doi.org/10.1103/PhysRevD.89.112007} {\bibfield  {journal} {\bibinfo  {journal} {Phys. Rev. D}\ }\textbf {\bibinfo {volume} {89}},\ \bibinfo {pages} {112007} (\bibinfo {year} {2014})},\ \Eprint {https://arxiv.org/abs/1308.0443} {arXiv:1308.0443 [hep-ex]} \BibitemShut {NoStop}%
\bibitem [{\citenamefont {Abe}\ \emph {et~al.}(2024)\citenamefont {Abe} \emph {et~al.}}]{Super-Kamiokande:2023jbt}%
  \BibitemOpen
  \bibfield  {author} {\bibinfo {author} {\bibfnamefont {K.}~\bibnamefont {Abe}} \emph {et~al.} (\bibinfo {collaboration} {Super-Kamiokande}),\ }\href {https://doi.org/10.1103/PhysRevD.109.092001} {\bibfield  {journal} {\bibinfo  {journal} {Phys. Rev. D}\ }\textbf {\bibinfo {volume} {109}},\ \bibinfo {pages} {092001} (\bibinfo {year} {2024})},\ \Eprint {https://arxiv.org/abs/2312.12907} {arXiv:2312.12907 [hep-ex]} \BibitemShut {NoStop}%
\bibitem [{\citenamefont {Abbasi}\ \emph {et~al.}(2011)\citenamefont {Abbasi} \emph {et~al.}}]{IceCube:2011cwc}%
  \BibitemOpen
  \bibfield  {author} {\bibinfo {author} {\bibfnamefont {R.}~\bibnamefont {Abbasi}} \emph {et~al.} (\bibinfo {collaboration} {IceCube}),\ }\href {https://doi.org/10.1051/0004-6361/201117810e} {\bibfield  {journal} {\bibinfo  {journal} {Astron. Astrophys.}\ }\textbf {\bibinfo {volume} {535}},\ \bibinfo {pages} {A109} (\bibinfo {year} {2011})},\ \bibinfo {note} {[Erratum: Astron.Astrophys. 563, C1 (2014)]},\ \Eprint {https://arxiv.org/abs/1108.0171} {arXiv:1108.0171 [astro-ph.HE]} \BibitemShut {NoStop}%
\bibitem [{\citenamefont {Adhikari}\ \emph {et~al.}(2023)\citenamefont {Adhikari} \emph {et~al.}}]{COSINE-100:2022dvc}%
  \BibitemOpen
  \bibfield  {author} {\bibinfo {author} {\bibfnamefont {G.}~\bibnamefont {Adhikari}} \emph {et~al.} (\bibinfo {collaboration} {COSINE-100}),\ }\href {https://doi.org/10.1038/s41598-023-31688-4} {\bibfield  {journal} {\bibinfo  {journal} {Sci. Rep.}\ }\textbf {\bibinfo {volume} {13}},\ \bibinfo {pages} {4676} (\bibinfo {year} {2023})},\ \Eprint {https://arxiv.org/abs/2208.05158} {arXiv:2208.05158 [hep-ex]} \BibitemShut {NoStop}%
\bibitem [{\citenamefont {Lu}(2023)}]{Lu:2023ztx}%
  \BibitemOpen
  \bibfield  {author} {\bibinfo {author} {\bibfnamefont {H.}~\bibnamefont {Lu}} (\bibinfo {collaboration} {JUNO}),\ }\href {https://doi.org/10.1016/j.nima.2023.168623} {\bibfield  {journal} {\bibinfo  {journal} {Nucl. Instrum. Meth. A}\ }\textbf {\bibinfo {volume} {1056}},\ \bibinfo {pages} {168623} (\bibinfo {year} {2023})}\BibitemShut {NoStop}%
\bibitem [{\citenamefont {Blum}(2011)}]{Blum:2011jf}%
  \BibitemOpen
  \bibfield  {author} {\bibinfo {author} {\bibfnamefont {K.}~\bibnamefont {Blum}},\ }\href@noop {} {\  (\bibinfo {year} {2011})},\ \Eprint {https://arxiv.org/abs/1110.0857} {arXiv:1110.0857 [astro-ph.HE]} \BibitemShut {NoStop}%
\bibitem [{\citenamefont {Davis}(2014)}]{Davis:2014cja}%
  \BibitemOpen
  \bibfield  {author} {\bibinfo {author} {\bibfnamefont {J.~H.}\ \bibnamefont {Davis}},\ }\href {https://doi.org/10.1103/PhysRevLett.113.081302} {\bibfield  {journal} {\bibinfo  {journal} {Phys. Rev. Lett.}\ }\textbf {\bibinfo {volume} {113}},\ \bibinfo {pages} {081302} (\bibinfo {year} {2014})},\ \Eprint {https://arxiv.org/abs/1407.1052} {arXiv:1407.1052 [hep-ph]} \BibitemShut {NoStop}%
\bibitem [{\citenamefont {Agostini}\ \emph {et~al.}(2017)\citenamefont {Agostini} \emph {et~al.}}]{BOREXINO:2017yyp}%
  \BibitemOpen
  \bibfield  {author} {\bibinfo {author} {\bibfnamefont {M.}~\bibnamefont {Agostini}} \emph {et~al.} (\bibinfo {collaboration} {BOREXINO}),\ }\href {https://doi.org/10.1016/j.astropartphys.2017.04.004} {\bibfield  {journal} {\bibinfo  {journal} {Astropart. Phys.}\ }\textbf {\bibinfo {volume} {92}},\ \bibinfo {pages} {21} (\bibinfo {year} {2017})},\ \Eprint {https://arxiv.org/abs/1701.07970} {arXiv:1701.07970 [hep-ex]} \BibitemShut {NoStop}%
\bibitem [{\citenamefont {Boehm}\ and\ \citenamefont {Fayet}(2004)}]{Boehm:2003hm}%
  \BibitemOpen
  \bibfield  {author} {\bibinfo {author} {\bibfnamefont {C.}~\bibnamefont {Boehm}}\ and\ \bibinfo {author} {\bibfnamefont {P.}~\bibnamefont {Fayet}},\ }\href {https://doi.org/10.1016/j.nuclphysb.2004.01.015} {\bibfield  {journal} {\bibinfo  {journal} {Nucl. Phys. B}\ }\textbf {\bibinfo {volume} {683}},\ \bibinfo {pages} {219} (\bibinfo {year} {2004})},\ \Eprint {https://arxiv.org/abs/hep-ph/0305261} {arXiv:hep-ph/0305261} \BibitemShut {NoStop}%
\bibitem [{\citenamefont {Boehm}\ \emph {et~al.}(2004)\citenamefont {Boehm}, \citenamefont {Fayet},\ and\ \citenamefont {Silk}}]{Boehm:2003ha}%
  \BibitemOpen
  \bibfield  {author} {\bibinfo {author} {\bibfnamefont {C.}~\bibnamefont {Boehm}}, \bibinfo {author} {\bibfnamefont {P.}~\bibnamefont {Fayet}},\ and\ \bibinfo {author} {\bibfnamefont {J.}~\bibnamefont {Silk}},\ }\href {https://doi.org/10.1103/PhysRevD.69.101302} {\bibfield  {journal} {\bibinfo  {journal} {Phys. Rev. D}\ }\textbf {\bibinfo {volume} {69}},\ \bibinfo {pages} {101302} (\bibinfo {year} {2004})},\ \Eprint {https://arxiv.org/abs/hep-ph/0311143} {arXiv:hep-ph/0311143} \BibitemShut {NoStop}%
\bibitem [{\citenamefont {McDonald}(2002)}]{McDonald:2001vt}%
  \BibitemOpen
  \bibfield  {author} {\bibinfo {author} {\bibfnamefont {J.}~\bibnamefont {McDonald}},\ }\href {https://doi.org/10.1103/PhysRevLett.88.091304} {\bibfield  {journal} {\bibinfo  {journal} {Phys. Rev. Lett.}\ }\textbf {\bibinfo {volume} {88}},\ \bibinfo {pages} {091304} (\bibinfo {year} {2002})},\ \Eprint {https://arxiv.org/abs/hep-ph/0106249} {arXiv:hep-ph/0106249} \BibitemShut {NoStop}%
\bibitem [{\citenamefont {Fayet}(2004)}]{Fayet:2004bw}%
  \BibitemOpen
  \bibfield  {author} {\bibinfo {author} {\bibfnamefont {P.}~\bibnamefont {Fayet}},\ }\href {https://doi.org/10.1103/PhysRevD.70.023514} {\bibfield  {journal} {\bibinfo  {journal} {Phys. Rev. D}\ }\textbf {\bibinfo {volume} {70}},\ \bibinfo {pages} {023514} (\bibinfo {year} {2004})},\ \Eprint {https://arxiv.org/abs/hep-ph/0403226} {arXiv:hep-ph/0403226} \BibitemShut {NoStop}%
\bibitem [{\citenamefont {Sigurdson}\ \emph {et~al.}(2004)\citenamefont {Sigurdson}, \citenamefont {Doran}, \citenamefont {Kurylov}, \citenamefont {Caldwell},\ and\ \citenamefont {Kamionkowski}}]{Sigurdson:2004zp}%
  \BibitemOpen
  \bibfield  {author} {\bibinfo {author} {\bibfnamefont {K.}~\bibnamefont {Sigurdson}}, \bibinfo {author} {\bibfnamefont {M.}~\bibnamefont {Doran}}, \bibinfo {author} {\bibfnamefont {A.}~\bibnamefont {Kurylov}}, \bibinfo {author} {\bibfnamefont {R.~R.}\ \bibnamefont {Caldwell}},\ and\ \bibinfo {author} {\bibfnamefont {M.}~\bibnamefont {Kamionkowski}},\ }\href {https://doi.org/10.1103/PhysRevD.70.083501} {\bibfield  {journal} {\bibinfo  {journal} {Phys. Rev. D}\ }\textbf {\bibinfo {volume} {70}},\ \bibinfo {pages} {083501} (\bibinfo {year} {2004})},\ \bibinfo {note} {[Erratum: Phys.Rev.D 73, 089903 (2006)]},\ \Eprint {https://arxiv.org/abs/astro-ph/0406355} {arXiv:astro-ph/0406355} \BibitemShut {NoStop}%
\bibitem [{\citenamefont {Strassler}\ and\ \citenamefont {Zurek}(2007)}]{Strassler:2006im}%
  \BibitemOpen
  \bibfield  {author} {\bibinfo {author} {\bibfnamefont {M.~J.}\ \bibnamefont {Strassler}}\ and\ \bibinfo {author} {\bibfnamefont {K.~M.}\ \bibnamefont {Zurek}},\ }\href {https://doi.org/10.1016/j.physletb.2007.06.055} {\bibfield  {journal} {\bibinfo  {journal} {Phys. Lett. B}\ }\textbf {\bibinfo {volume} {651}},\ \bibinfo {pages} {374} (\bibinfo {year} {2007})},\ \Eprint {https://arxiv.org/abs/hep-ph/0604261} {arXiv:hep-ph/0604261} \BibitemShut {NoStop}%
\bibitem [{\citenamefont {Sikivie}(2008)}]{Sikivie:2006ni}%
  \BibitemOpen
  \bibfield  {author} {\bibinfo {author} {\bibfnamefont {P.}~\bibnamefont {Sikivie}},\ }\href {https://doi.org/10.1007/978-3-540-73518-2_2} {\bibfield  {journal} {\bibinfo  {journal} {Lect. Notes Phys.}\ }\textbf {\bibinfo {volume} {741}},\ \bibinfo {pages} {19} (\bibinfo {year} {2008})},\ \Eprint {https://arxiv.org/abs/astro-ph/0610440} {arXiv:astro-ph/0610440} \BibitemShut {NoStop}%
\bibitem [{\citenamefont {Nussinov}(1985)}]{Nussinov:1985xr}%
  \BibitemOpen
  \bibfield  {author} {\bibinfo {author} {\bibfnamefont {S.}~\bibnamefont {Nussinov}},\ }\href {https://doi.org/10.1016/0370-2693(85)90689-6} {\bibfield  {journal} {\bibinfo  {journal} {Phys. Lett. B}\ }\textbf {\bibinfo {volume} {165}},\ \bibinfo {pages} {55} (\bibinfo {year} {1985})}\BibitemShut {NoStop}%
\bibitem [{\citenamefont {Kaplan}(1992)}]{Kaplan:1991ah}%
  \BibitemOpen
  \bibfield  {author} {\bibinfo {author} {\bibfnamefont {D.~B.}\ \bibnamefont {Kaplan}},\ }\href {https://doi.org/10.1103/PhysRevLett.68.741} {\bibfield  {journal} {\bibinfo  {journal} {Phys. Rev. Lett.}\ }\textbf {\bibinfo {volume} {68}},\ \bibinfo {pages} {741} (\bibinfo {year} {1992})}\BibitemShut {NoStop}%
\bibitem [{\citenamefont {Zurek}(2009)}]{Zurek:2008qg}%
  \BibitemOpen
  \bibfield  {author} {\bibinfo {author} {\bibfnamefont {K.~M.}\ \bibnamefont {Zurek}},\ }\href {https://doi.org/10.1103/PhysRevD.79.115002} {\bibfield  {journal} {\bibinfo  {journal} {Phys. Rev. D}\ }\textbf {\bibinfo {volume} {79}},\ \bibinfo {pages} {115002} (\bibinfo {year} {2009})},\ \Eprint {https://arxiv.org/abs/0811.4429} {arXiv:0811.4429 [hep-ph]} \BibitemShut {NoStop}%
\bibitem [{\citenamefont {Hooper}\ and\ \citenamefont {Zurek}(2008)}]{Hooper:2008im}%
  \BibitemOpen
  \bibfield  {author} {\bibinfo {author} {\bibfnamefont {D.}~\bibnamefont {Hooper}}\ and\ \bibinfo {author} {\bibfnamefont {K.~M.}\ \bibnamefont {Zurek}},\ }\href {https://doi.org/10.1103/PhysRevD.77.087302} {\bibfield  {journal} {\bibinfo  {journal} {Phys. Rev. D}\ }\textbf {\bibinfo {volume} {77}},\ \bibinfo {pages} {087302} (\bibinfo {year} {2008})},\ \Eprint {https://arxiv.org/abs/0801.3686} {arXiv:0801.3686 [hep-ph]} \BibitemShut {NoStop}%
\bibitem [{\citenamefont {Cholis}\ \emph {et~al.}(2009)\citenamefont {Cholis}, \citenamefont {Goodenough},\ and\ \citenamefont {Weiner}}]{Cholis:2008vb}%
  \BibitemOpen
  \bibfield  {author} {\bibinfo {author} {\bibfnamefont {I.}~\bibnamefont {Cholis}}, \bibinfo {author} {\bibfnamefont {L.}~\bibnamefont {Goodenough}},\ and\ \bibinfo {author} {\bibfnamefont {N.}~\bibnamefont {Weiner}},\ }\href {https://doi.org/10.1103/PhysRevD.79.123505} {\bibfield  {journal} {\bibinfo  {journal} {Phys. Rev. D}\ }\textbf {\bibinfo {volume} {79}},\ \bibinfo {pages} {123505} (\bibinfo {year} {2009})},\ \Eprint {https://arxiv.org/abs/0802.2922} {arXiv:0802.2922 [astro-ph]} \BibitemShut {NoStop}%
\bibitem [{\citenamefont {Arkani-Hamed}\ \emph {et~al.}(2009)\citenamefont {Arkani-Hamed}, \citenamefont {Finkbeiner}, \citenamefont {Slatyer},\ and\ \citenamefont {Weiner}}]{ArkaniHamed:2008qn}%
  \BibitemOpen
  \bibfield  {author} {\bibinfo {author} {\bibfnamefont {N.}~\bibnamefont {Arkani-Hamed}}, \bibinfo {author} {\bibfnamefont {D.~P.}\ \bibnamefont {Finkbeiner}}, \bibinfo {author} {\bibfnamefont {T.~R.}\ \bibnamefont {Slatyer}},\ and\ \bibinfo {author} {\bibfnamefont {N.}~\bibnamefont {Weiner}},\ }\href {https://doi.org/10.1103/PhysRevD.79.015014} {\bibfield  {journal} {\bibinfo  {journal} {Phys. Rev. D}\ }\textbf {\bibinfo {volume} {79}},\ \bibinfo {pages} {015014} (\bibinfo {year} {2009})},\ \Eprint {https://arxiv.org/abs/0810.0713} {arXiv:0810.0713 [hep-ph]} \BibitemShut {NoStop}%
\bibitem [{\citenamefont {Pospelov}\ and\ \citenamefont {Ritz}(2009)}]{Pospelov:2008jd}%
  \BibitemOpen
  \bibfield  {author} {\bibinfo {author} {\bibfnamefont {M.}~\bibnamefont {Pospelov}}\ and\ \bibinfo {author} {\bibfnamefont {A.}~\bibnamefont {Ritz}},\ }\href {https://doi.org/10.1016/j.physletb.2008.12.012} {\bibfield  {journal} {\bibinfo  {journal} {Phys. Lett. B}\ }\textbf {\bibinfo {volume} {671}},\ \bibinfo {pages} {391} (\bibinfo {year} {2009})},\ \Eprint {https://arxiv.org/abs/0810.1502} {arXiv:0810.1502 [hep-ph]} \BibitemShut {NoStop}%
\bibitem [{\citenamefont {Feng}\ and\ \citenamefont {Kumar}(2008)}]{Feng:2008ya}%
  \BibitemOpen
  \bibfield  {author} {\bibinfo {author} {\bibfnamefont {J.~L.}\ \bibnamefont {Feng}}\ and\ \bibinfo {author} {\bibfnamefont {J.}~\bibnamefont {Kumar}},\ }\href {https://doi.org/10.1103/PhysRevLett.101.231301} {\bibfield  {journal} {\bibinfo  {journal} {Phys. Rev. Lett.}\ }\textbf {\bibinfo {volume} {101}},\ \bibinfo {pages} {231301} (\bibinfo {year} {2008})},\ \Eprint {https://arxiv.org/abs/0803.4196} {arXiv:0803.4196 [hep-ph]} \BibitemShut {NoStop}%
\bibitem [{\citenamefont {Feng}\ \emph {et~al.}(2008)\citenamefont {Feng}, \citenamefont {Kumar},\ and\ \citenamefont {Strigari}}]{Feng:2008dz}%
  \BibitemOpen
  \bibfield  {author} {\bibinfo {author} {\bibfnamefont {J.~L.}\ \bibnamefont {Feng}}, \bibinfo {author} {\bibfnamefont {J.}~\bibnamefont {Kumar}},\ and\ \bibinfo {author} {\bibfnamefont {L.~E.}\ \bibnamefont {Strigari}},\ }\href {https://doi.org/10.1016/j.physletb.2008.10.038} {\bibfield  {journal} {\bibinfo  {journal} {Phys. Lett. B}\ }\textbf {\bibinfo {volume} {670}},\ \bibinfo {pages} {37} (\bibinfo {year} {2008})},\ \Eprint {https://arxiv.org/abs/0806.3746} {arXiv:0806.3746 [hep-ph]} \BibitemShut {NoStop}%
\bibitem [{\citenamefont {Pospelov}\ \emph {et~al.}(2008)\citenamefont {Pospelov}, \citenamefont {Ritz},\ and\ \citenamefont {Voloshin}}]{Pospelov:2008jk}%
  \BibitemOpen
  \bibfield  {author} {\bibinfo {author} {\bibfnamefont {M.}~\bibnamefont {Pospelov}}, \bibinfo {author} {\bibfnamefont {A.}~\bibnamefont {Ritz}},\ and\ \bibinfo {author} {\bibfnamefont {M.~B.}\ \bibnamefont {Voloshin}},\ }\href {https://doi.org/10.1103/PhysRevD.78.115012} {\bibfield  {journal} {\bibinfo  {journal} {Phys. Rev. D}\ }\textbf {\bibinfo {volume} {78}},\ \bibinfo {pages} {115012} (\bibinfo {year} {2008})},\ \Eprint {https://arxiv.org/abs/0807.3279} {arXiv:0807.3279 [hep-ph]} \BibitemShut {NoStop}%
\bibitem [{\citenamefont {Hall}\ \emph {et~al.}(2010)\citenamefont {Hall}, \citenamefont {Jedamzik}, \citenamefont {March-Russell},\ and\ \citenamefont {West}}]{Hall:2009bx}%
  \BibitemOpen
  \bibfield  {author} {\bibinfo {author} {\bibfnamefont {L.~J.}\ \bibnamefont {Hall}}, \bibinfo {author} {\bibfnamefont {K.}~\bibnamefont {Jedamzik}}, \bibinfo {author} {\bibfnamefont {J.}~\bibnamefont {March-Russell}},\ and\ \bibinfo {author} {\bibfnamefont {S.~M.}\ \bibnamefont {West}},\ }\href {https://doi.org/10.1007/JHEP03(2010)080} {\bibfield  {journal} {\bibinfo  {journal} {JHEP}\ }\textbf {\bibinfo {volume} {03}},\ \bibinfo {pages} {080}},\ \Eprint {https://arxiv.org/abs/0911.1120} {arXiv:0911.1120 [hep-ph]} \BibitemShut {NoStop}%
\bibitem [{\citenamefont {Morrissey}\ \emph {et~al.}(2009)\citenamefont {Morrissey}, \citenamefont {Poland},\ and\ \citenamefont {Zurek}}]{Morrissey:2009ur}%
  \BibitemOpen
  \bibfield  {author} {\bibinfo {author} {\bibfnamefont {D.~E.}\ \bibnamefont {Morrissey}}, \bibinfo {author} {\bibfnamefont {D.}~\bibnamefont {Poland}},\ and\ \bibinfo {author} {\bibfnamefont {K.~M.}\ \bibnamefont {Zurek}},\ }\href {https://doi.org/10.1088/1126-6708/2009/07/050} {\bibfield  {journal} {\bibinfo  {journal} {JHEP}\ }\textbf {\bibinfo {volume} {07}},\ \bibinfo {pages} {050}},\ \Eprint {https://arxiv.org/abs/0904.2567} {arXiv:0904.2567 [hep-ph]} \BibitemShut {NoStop}%
\bibitem [{\citenamefont {Kaplan}\ \emph {et~al.}(2009)\citenamefont {Kaplan}, \citenamefont {Luty},\ and\ \citenamefont {Zurek}}]{Kaplan:2009ag}%
  \BibitemOpen
  \bibfield  {author} {\bibinfo {author} {\bibfnamefont {D.~E.}\ \bibnamefont {Kaplan}}, \bibinfo {author} {\bibfnamefont {M.~A.}\ \bibnamefont {Luty}},\ and\ \bibinfo {author} {\bibfnamefont {K.~M.}\ \bibnamefont {Zurek}},\ }\href {https://doi.org/10.1103/PhysRevD.79.115016} {\bibfield  {journal} {\bibinfo  {journal} {Phys. Rev. D}\ }\textbf {\bibinfo {volume} {79}},\ \bibinfo {pages} {115016} (\bibinfo {year} {2009})},\ \Eprint {https://arxiv.org/abs/0901.4117} {arXiv:0901.4117 [hep-ph]} \BibitemShut {NoStop}%
\bibitem [{\citenamefont {Essig}\ \emph {et~al.}(2010)\citenamefont {Essig}, \citenamefont {Kaplan}, \citenamefont {Schuster},\ and\ \citenamefont {Toro}}]{Essig:2010ye}%
  \BibitemOpen
  \bibfield  {author} {\bibinfo {author} {\bibfnamefont {R.}~\bibnamefont {Essig}}, \bibinfo {author} {\bibfnamefont {J.}~\bibnamefont {Kaplan}}, \bibinfo {author} {\bibfnamefont {P.}~\bibnamefont {Schuster}},\ and\ \bibinfo {author} {\bibfnamefont {N.}~\bibnamefont {Toro}},\ }\href@noop {} {\  (\bibinfo {year} {2010})},\ \Eprint {https://arxiv.org/abs/1004.0691} {arXiv:1004.0691 [hep-ph]} \BibitemShut {NoStop}%
\bibitem [{\citenamefont {Cohen}\ \emph {et~al.}(2010)\citenamefont {Cohen}, \citenamefont {Phalen}, \citenamefont {Pierce},\ and\ \citenamefont {Zurek}}]{Cohen:2010kn}%
  \BibitemOpen
  \bibfield  {author} {\bibinfo {author} {\bibfnamefont {T.}~\bibnamefont {Cohen}}, \bibinfo {author} {\bibfnamefont {D.~J.}\ \bibnamefont {Phalen}}, \bibinfo {author} {\bibfnamefont {A.}~\bibnamefont {Pierce}},\ and\ \bibinfo {author} {\bibfnamefont {K.~M.}\ \bibnamefont {Zurek}},\ }\href {https://doi.org/10.1103/PhysRevD.82.056001} {\bibfield  {journal} {\bibinfo  {journal} {Phys. Rev. D}\ }\textbf {\bibinfo {volume} {82}},\ \bibinfo {pages} {056001} (\bibinfo {year} {2010})},\ \Eprint {https://arxiv.org/abs/1005.1655} {arXiv:1005.1655 [hep-ph]} \BibitemShut {NoStop}%
\bibitem [{\citenamefont {Chu}\ \emph {et~al.}(2012)\citenamefont {Chu}, \citenamefont {Hambye},\ and\ \citenamefont {Tytgat}}]{Chu:2011be}%
  \BibitemOpen
  \bibfield  {author} {\bibinfo {author} {\bibfnamefont {X.}~\bibnamefont {Chu}}, \bibinfo {author} {\bibfnamefont {T.}~\bibnamefont {Hambye}},\ and\ \bibinfo {author} {\bibfnamefont {M.~H.~G.}\ \bibnamefont {Tytgat}},\ }\href {https://doi.org/10.1088/1475-7516/2012/05/034} {\bibfield  {journal} {\bibinfo  {journal} {JCAP}\ }\textbf {\bibinfo {volume} {05}},\ \bibinfo {pages} {034}},\ \Eprint {https://arxiv.org/abs/1112.0493} {arXiv:1112.0493 [hep-ph]} \BibitemShut {NoStop}%
\bibitem [{\citenamefont {Falkowski}\ \emph {et~al.}(2011)\citenamefont {Falkowski}, \citenamefont {Ruderman},\ and\ \citenamefont {Volansky}}]{Falkowski:2011xh}%
  \BibitemOpen
  \bibfield  {author} {\bibinfo {author} {\bibfnamefont {A.}~\bibnamefont {Falkowski}}, \bibinfo {author} {\bibfnamefont {J.~T.}\ \bibnamefont {Ruderman}},\ and\ \bibinfo {author} {\bibfnamefont {T.}~\bibnamefont {Volansky}},\ }\href {https://doi.org/10.1007/JHEP05(2011)106} {\bibfield  {journal} {\bibinfo  {journal} {JHEP}\ }\textbf {\bibinfo {volume} {05}},\ \bibinfo {pages} {106}},\ \Eprint {https://arxiv.org/abs/1101.4936} {arXiv:1101.4936 [hep-ph]} \BibitemShut {NoStop}%
\bibitem [{\citenamefont {Lin}\ \emph {et~al.}(2012)\citenamefont {Lin}, \citenamefont {Yu},\ and\ \citenamefont {Zurek}}]{Lin:2011gj}%
  \BibitemOpen
  \bibfield  {author} {\bibinfo {author} {\bibfnamefont {T.}~\bibnamefont {Lin}}, \bibinfo {author} {\bibfnamefont {H.-B.}\ \bibnamefont {Yu}},\ and\ \bibinfo {author} {\bibfnamefont {K.~M.}\ \bibnamefont {Zurek}},\ }\href {https://doi.org/10.1103/PhysRevD.85.063503} {\bibfield  {journal} {\bibinfo  {journal} {Phys. Rev. D}\ }\textbf {\bibinfo {volume} {85}},\ \bibinfo {pages} {063503} (\bibinfo {year} {2012})},\ \Eprint {https://arxiv.org/abs/1111.0293} {arXiv:1111.0293 [hep-ph]} \BibitemShut {NoStop}%
\bibitem [{\citenamefont {Feng}\ and\ \citenamefont {Shadmi}(2011)}]{Feng:2011ik}%
  \BibitemOpen
  \bibfield  {author} {\bibinfo {author} {\bibfnamefont {J.~L.}\ \bibnamefont {Feng}}\ and\ \bibinfo {author} {\bibfnamefont {Y.}~\bibnamefont {Shadmi}},\ }\href {https://doi.org/10.1103/PhysRevD.83.095011} {\bibfield  {journal} {\bibinfo  {journal} {Phys. Rev. D}\ }\textbf {\bibinfo {volume} {83}},\ \bibinfo {pages} {095011} (\bibinfo {year} {2011})},\ \Eprint {https://arxiv.org/abs/1102.0282} {arXiv:1102.0282 [hep-ph]} \BibitemShut {NoStop}%
\bibitem [{\citenamefont {Nelson}\ and\ \citenamefont {Scholtz}(2011)}]{Nelson:2011sf}%
  \BibitemOpen
  \bibfield  {author} {\bibinfo {author} {\bibfnamefont {A.~E.}\ \bibnamefont {Nelson}}\ and\ \bibinfo {author} {\bibfnamefont {J.}~\bibnamefont {Scholtz}},\ }\href {https://doi.org/10.1103/PhysRevD.84.103501} {\bibfield  {journal} {\bibinfo  {journal} {Phys. Rev. D}\ }\textbf {\bibinfo {volume} {84}},\ \bibinfo {pages} {103501} (\bibinfo {year} {2011})},\ \Eprint {https://arxiv.org/abs/1105.2812} {arXiv:1105.2812 [hep-ph]} \BibitemShut {NoStop}%
\bibitem [{\citenamefont {Arias}\ \emph {et~al.}(2012)\citenamefont {Arias}, \citenamefont {Cadamuro}, \citenamefont {Goodsell}, \citenamefont {Jaeckel}, \citenamefont {Redondo},\ and\ \citenamefont {Ringwald}}]{Arias:2012az}%
  \BibitemOpen
  \bibfield  {author} {\bibinfo {author} {\bibfnamefont {P.}~\bibnamefont {Arias}}, \bibinfo {author} {\bibfnamefont {D.}~\bibnamefont {Cadamuro}}, \bibinfo {author} {\bibfnamefont {M.}~\bibnamefont {Goodsell}}, \bibinfo {author} {\bibfnamefont {J.}~\bibnamefont {Jaeckel}}, \bibinfo {author} {\bibfnamefont {J.}~\bibnamefont {Redondo}},\ and\ \bibinfo {author} {\bibfnamefont {A.}~\bibnamefont {Ringwald}},\ }\href {https://doi.org/10.1088/1475-7516/2012/06/013} {\bibfield  {journal} {\bibinfo  {journal} {JCAP}\ }\textbf {\bibinfo {volume} {06}},\ \bibinfo {pages} {013}},\ \Eprint {https://arxiv.org/abs/1201.5902} {arXiv:1201.5902 [hep-ph]} \BibitemShut {NoStop}%
\bibitem [{\citenamefont {Kaplinghat}\ \emph {et~al.}(2014)\citenamefont {Kaplinghat}, \citenamefont {Tulin},\ and\ \citenamefont {Yu}}]{Kaplinghat:2013yxa}%
  \BibitemOpen
  \bibfield  {author} {\bibinfo {author} {\bibfnamefont {M.}~\bibnamefont {Kaplinghat}}, \bibinfo {author} {\bibfnamefont {S.}~\bibnamefont {Tulin}},\ and\ \bibinfo {author} {\bibfnamefont {H.-B.}\ \bibnamefont {Yu}},\ }\href {https://doi.org/10.1103/PhysRevD.89.035009} {\bibfield  {journal} {\bibinfo  {journal} {Phys. Rev. D}\ }\textbf {\bibinfo {volume} {89}},\ \bibinfo {pages} {035009} (\bibinfo {year} {2014})},\ \Eprint {https://arxiv.org/abs/1310.7945} {arXiv:1310.7945 [hep-ph]} \BibitemShut {NoStop}%
\bibitem [{\citenamefont {Hochberg}\ \emph {et~al.}(2014)\citenamefont {Hochberg}, \citenamefont {Kuflik}, \citenamefont {Volansky},\ and\ \citenamefont {Wacker}}]{Hochberg:2014dra}%
  \BibitemOpen
  \bibfield  {author} {\bibinfo {author} {\bibfnamefont {Y.}~\bibnamefont {Hochberg}}, \bibinfo {author} {\bibfnamefont {E.}~\bibnamefont {Kuflik}}, \bibinfo {author} {\bibfnamefont {T.}~\bibnamefont {Volansky}},\ and\ \bibinfo {author} {\bibfnamefont {J.~G.}\ \bibnamefont {Wacker}},\ }\href {https://doi.org/10.1103/PhysRevLett.113.171301} {\bibfield  {journal} {\bibinfo  {journal} {Phys. Rev. Lett.}\ }\textbf {\bibinfo {volume} {113}},\ \bibinfo {pages} {171301} (\bibinfo {year} {2014})},\ \Eprint {https://arxiv.org/abs/1402.5143} {arXiv:1402.5143 [hep-ph]} \BibitemShut {NoStop}%
\bibitem [{\citenamefont {Hochberg}\ \emph {et~al.}(2015)\citenamefont {Hochberg}, \citenamefont {Kuflik}, \citenamefont {Murayama}, \citenamefont {Volansky},\ and\ \citenamefont {Wacker}}]{Hochberg:2014kqa}%
  \BibitemOpen
  \bibfield  {author} {\bibinfo {author} {\bibfnamefont {Y.}~\bibnamefont {Hochberg}}, \bibinfo {author} {\bibfnamefont {E.}~\bibnamefont {Kuflik}}, \bibinfo {author} {\bibfnamefont {H.}~\bibnamefont {Murayama}}, \bibinfo {author} {\bibfnamefont {T.}~\bibnamefont {Volansky}},\ and\ \bibinfo {author} {\bibfnamefont {J.~G.}\ \bibnamefont {Wacker}},\ }\href {https://doi.org/10.1103/PhysRevLett.115.021301} {\bibfield  {journal} {\bibinfo  {journal} {Phys. Rev. Lett.}\ }\textbf {\bibinfo {volume} {115}},\ \bibinfo {pages} {021301} (\bibinfo {year} {2015})},\ \Eprint {https://arxiv.org/abs/1411.3727} {arXiv:1411.3727 [hep-ph]} \BibitemShut {NoStop}%
\bibitem [{\citenamefont {Boddy}\ \emph {et~al.}(2014{\natexlab{a}})\citenamefont {Boddy}, \citenamefont {Feng}, \citenamefont {Kaplinghat},\ and\ \citenamefont {Tait}}]{Boddy:2014yra}%
  \BibitemOpen
  \bibfield  {author} {\bibinfo {author} {\bibfnamefont {K.~K.}\ \bibnamefont {Boddy}}, \bibinfo {author} {\bibfnamefont {J.~L.}\ \bibnamefont {Feng}}, \bibinfo {author} {\bibfnamefont {M.}~\bibnamefont {Kaplinghat}},\ and\ \bibinfo {author} {\bibfnamefont {T.~M.~P.}\ \bibnamefont {Tait}},\ }\href {https://doi.org/10.1103/PhysRevD.89.115017} {\bibfield  {journal} {\bibinfo  {journal} {Phys. Rev. D}\ }\textbf {\bibinfo {volume} {89}},\ \bibinfo {pages} {115017} (\bibinfo {year} {2014}{\natexlab{a}})},\ \Eprint {https://arxiv.org/abs/1402.3629} {arXiv:1402.3629 [hep-ph]} \BibitemShut {NoStop}%
\bibitem [{\citenamefont {Boddy}\ \emph {et~al.}(2014{\natexlab{b}})\citenamefont {Boddy}, \citenamefont {Feng}, \citenamefont {Kaplinghat}, \citenamefont {Shadmi},\ and\ \citenamefont {Tait}}]{Boddy:2014qxa}%
  \BibitemOpen
  \bibfield  {author} {\bibinfo {author} {\bibfnamefont {K.~K.}\ \bibnamefont {Boddy}}, \bibinfo {author} {\bibfnamefont {J.~L.}\ \bibnamefont {Feng}}, \bibinfo {author} {\bibfnamefont {M.}~\bibnamefont {Kaplinghat}}, \bibinfo {author} {\bibfnamefont {Y.}~\bibnamefont {Shadmi}},\ and\ \bibinfo {author} {\bibfnamefont {T.~M.~P.}\ \bibnamefont {Tait}},\ }\href {https://doi.org/10.1103/PhysRevD.90.095016} {\bibfield  {journal} {\bibinfo  {journal} {Phys. Rev. D}\ }\textbf {\bibinfo {volume} {90}},\ \bibinfo {pages} {095016} (\bibinfo {year} {2014}{\natexlab{b}})},\ \Eprint {https://arxiv.org/abs/1408.6532} {arXiv:1408.6532 [hep-ph]} \BibitemShut {NoStop}%
\bibitem [{\citenamefont {Hochberg}\ \emph {et~al.}(2016{\natexlab{c}})\citenamefont {Hochberg}, \citenamefont {Kuflik},\ and\ \citenamefont {Murayama}}]{Hochberg:2015vrg}%
  \BibitemOpen
  \bibfield  {author} {\bibinfo {author} {\bibfnamefont {Y.}~\bibnamefont {Hochberg}}, \bibinfo {author} {\bibfnamefont {E.}~\bibnamefont {Kuflik}},\ and\ \bibinfo {author} {\bibfnamefont {H.}~\bibnamefont {Murayama}},\ }\href {https://doi.org/10.1007/JHEP05(2016)090} {\bibfield  {journal} {\bibinfo  {journal} {JHEP}\ }\textbf {\bibinfo {volume} {05}},\ \bibinfo {pages} {090}},\ \Eprint {https://arxiv.org/abs/1512.07917} {arXiv:1512.07917 [hep-ph]} \BibitemShut {NoStop}%
\bibitem [{\citenamefont {Izaguirre}\ \emph {et~al.}(2015)\citenamefont {Izaguirre}, \citenamefont {Krnjaic}, \citenamefont {Schuster},\ and\ \citenamefont {Toro}}]{Izaguirre:2015yja}%
  \BibitemOpen
  \bibfield  {author} {\bibinfo {author} {\bibfnamefont {E.}~\bibnamefont {Izaguirre}}, \bibinfo {author} {\bibfnamefont {G.}~\bibnamefont {Krnjaic}}, \bibinfo {author} {\bibfnamefont {P.}~\bibnamefont {Schuster}},\ and\ \bibinfo {author} {\bibfnamefont {N.}~\bibnamefont {Toro}},\ }\href {https://doi.org/10.1103/PhysRevLett.115.251301} {\bibfield  {journal} {\bibinfo  {journal} {Phys. Rev. Lett.}\ }\textbf {\bibinfo {volume} {115}},\ \bibinfo {pages} {251301} (\bibinfo {year} {2015})},\ \Eprint {https://arxiv.org/abs/1505.00011} {arXiv:1505.00011 [hep-ph]} \BibitemShut {NoStop}%
\bibitem [{\citenamefont {Kuflik}\ \emph {et~al.}(2016)\citenamefont {Kuflik}, \citenamefont {Perelstein}, \citenamefont {Lorier},\ and\ \citenamefont {Tsai}}]{Kuflik:2015isi}%
  \BibitemOpen
  \bibfield  {author} {\bibinfo {author} {\bibfnamefont {E.}~\bibnamefont {Kuflik}}, \bibinfo {author} {\bibfnamefont {M.}~\bibnamefont {Perelstein}}, \bibinfo {author} {\bibfnamefont {N.~R.-L.}\ \bibnamefont {Lorier}},\ and\ \bibinfo {author} {\bibfnamefont {Y.-D.}\ \bibnamefont {Tsai}},\ }\href {https://doi.org/10.1103/PhysRevLett.116.221302} {\bibfield  {journal} {\bibinfo  {journal} {Phys. Rev. Lett.}\ }\textbf {\bibinfo {volume} {116}},\ \bibinfo {pages} {221302} (\bibinfo {year} {2016})},\ \Eprint {https://arxiv.org/abs/1512.04545} {arXiv:1512.04545 [hep-ph]} \BibitemShut {NoStop}%
\bibitem [{\citenamefont {Graham}\ \emph {et~al.}(2016)\citenamefont {Graham}, \citenamefont {Mardon},\ and\ \citenamefont {Rajendran}}]{Graham:2015rva}%
  \BibitemOpen
  \bibfield  {author} {\bibinfo {author} {\bibfnamefont {P.~W.}\ \bibnamefont {Graham}}, \bibinfo {author} {\bibfnamefont {J.}~\bibnamefont {Mardon}},\ and\ \bibinfo {author} {\bibfnamefont {S.}~\bibnamefont {Rajendran}},\ }\href {https://doi.org/10.1103/PhysRevD.93.103520} {\bibfield  {journal} {\bibinfo  {journal} {Phys. Rev. D}\ }\textbf {\bibinfo {volume} {93}},\ \bibinfo {pages} {103520} (\bibinfo {year} {2016})},\ \Eprint {https://arxiv.org/abs/1504.02102} {arXiv:1504.02102 [hep-ph]} \BibitemShut {NoStop}%
\bibitem [{\citenamefont {Marsh}(2016)}]{Marsh:2015xka}%
  \BibitemOpen
  \bibfield  {author} {\bibinfo {author} {\bibfnamefont {D.~J.~E.}\ \bibnamefont {Marsh}},\ }\href {https://doi.org/10.1016/j.physrep.2016.06.005} {\bibfield  {journal} {\bibinfo  {journal} {Phys. Rept.}\ }\textbf {\bibinfo {volume} {643}},\ \bibinfo {pages} {1} (\bibinfo {year} {2016})},\ \Eprint {https://arxiv.org/abs/1510.07633} {arXiv:1510.07633 [astro-ph.CO]} \BibitemShut {NoStop}%
\bibitem [{\citenamefont {D'Agnolo}\ and\ \citenamefont {Ruderman}(2015)}]{DAgnolo:2015ujb}%
  \BibitemOpen
  \bibfield  {author} {\bibinfo {author} {\bibfnamefont {R.~T.}\ \bibnamefont {D'Agnolo}}\ and\ \bibinfo {author} {\bibfnamefont {J.~T.}\ \bibnamefont {Ruderman}},\ }\href {https://doi.org/10.1103/PhysRevLett.115.061301} {\bibfield  {journal} {\bibinfo  {journal} {Phys. Rev. Lett.}\ }\textbf {\bibinfo {volume} {115}},\ \bibinfo {pages} {061301} (\bibinfo {year} {2015})},\ \Eprint {https://arxiv.org/abs/1505.07107} {arXiv:1505.07107 [hep-ph]} \BibitemShut {NoStop}%
\bibitem [{\citenamefont {Pappadopulo}\ \emph {et~al.}(2016)\citenamefont {Pappadopulo}, \citenamefont {Ruderman},\ and\ \citenamefont {Trevisan}}]{Pappadopulo:2016pkp}%
  \BibitemOpen
  \bibfield  {author} {\bibinfo {author} {\bibfnamefont {D.}~\bibnamefont {Pappadopulo}}, \bibinfo {author} {\bibfnamefont {J.~T.}\ \bibnamefont {Ruderman}},\ and\ \bibinfo {author} {\bibfnamefont {G.}~\bibnamefont {Trevisan}},\ }\href {https://doi.org/10.1103/PhysRevD.94.035005} {\bibfield  {journal} {\bibinfo  {journal} {Phys. Rev. D}\ }\textbf {\bibinfo {volume} {94}},\ \bibinfo {pages} {035005} (\bibinfo {year} {2016})},\ \Eprint {https://arxiv.org/abs/1602.04219} {arXiv:1602.04219 [hep-ph]} \BibitemShut {NoStop}%
\bibitem [{\citenamefont {Farina}\ \emph {et~al.}(2016)\citenamefont {Farina}, \citenamefont {Pappadopulo}, \citenamefont {Ruderman},\ and\ \citenamefont {Trevisan}}]{Farina:2016llk}%
  \BibitemOpen
  \bibfield  {author} {\bibinfo {author} {\bibfnamefont {M.}~\bibnamefont {Farina}}, \bibinfo {author} {\bibfnamefont {D.}~\bibnamefont {Pappadopulo}}, \bibinfo {author} {\bibfnamefont {J.~T.}\ \bibnamefont {Ruderman}},\ and\ \bibinfo {author} {\bibfnamefont {G.}~\bibnamefont {Trevisan}},\ }\href {https://doi.org/10.1007/JHEP12(2016)039} {\bibfield  {journal} {\bibinfo  {journal} {JHEP}\ }\textbf {\bibinfo {volume} {12}},\ \bibinfo {pages} {039}},\ \Eprint {https://arxiv.org/abs/1607.03108} {arXiv:1607.03108 [hep-ph]} \BibitemShut {NoStop}%
\bibitem [{\citenamefont {Dror}\ \emph {et~al.}(2016)\citenamefont {Dror}, \citenamefont {Kuflik},\ and\ \citenamefont {Ng}}]{Dror:2016rxc}%
  \BibitemOpen
  \bibfield  {author} {\bibinfo {author} {\bibfnamefont {J.~A.}\ \bibnamefont {Dror}}, \bibinfo {author} {\bibfnamefont {E.}~\bibnamefont {Kuflik}},\ and\ \bibinfo {author} {\bibfnamefont {W.~H.}\ \bibnamefont {Ng}},\ }\href {https://doi.org/10.1103/PhysRevLett.117.211801} {\bibfield  {journal} {\bibinfo  {journal} {Phys. Rev. Lett.}\ }\textbf {\bibinfo {volume} {117}},\ \bibinfo {pages} {211801} (\bibinfo {year} {2016})},\ \Eprint {https://arxiv.org/abs/1607.03110} {arXiv:1607.03110 [hep-ph]} \BibitemShut {NoStop}%
\bibitem [{\citenamefont {Feng}\ and\ \citenamefont {Smolinsky}(2017)}]{Feng:2017drg}%
  \BibitemOpen
  \bibfield  {author} {\bibinfo {author} {\bibfnamefont {J.~L.}\ \bibnamefont {Feng}}\ and\ \bibinfo {author} {\bibfnamefont {J.}~\bibnamefont {Smolinsky}},\ }\href {https://doi.org/10.1103/PhysRevD.96.095022} {\bibfield  {journal} {\bibinfo  {journal} {Phys. Rev. D}\ }\textbf {\bibinfo {volume} {96}},\ \bibinfo {pages} {095022} (\bibinfo {year} {2017})},\ \Eprint {https://arxiv.org/abs/1707.03835} {arXiv:1707.03835 [hep-ph]} \BibitemShut {NoStop}%
\bibitem [{\citenamefont {Kuflik}\ \emph {et~al.}(2017)\citenamefont {Kuflik}, \citenamefont {Perelstein}, \citenamefont {Lorier},\ and\ \citenamefont {Tsai}}]{Kuflik:2017iqs}%
  \BibitemOpen
  \bibfield  {author} {\bibinfo {author} {\bibfnamefont {E.}~\bibnamefont {Kuflik}}, \bibinfo {author} {\bibfnamefont {M.}~\bibnamefont {Perelstein}}, \bibinfo {author} {\bibfnamefont {N.~R.-L.}\ \bibnamefont {Lorier}},\ and\ \bibinfo {author} {\bibfnamefont {Y.-D.}\ \bibnamefont {Tsai}},\ }\href {https://doi.org/10.1007/JHEP08(2017)078} {\bibfield  {journal} {\bibinfo  {journal} {JHEP}\ }\textbf {\bibinfo {volume} {08}},\ \bibinfo {pages} {078}},\ \Eprint {https://arxiv.org/abs/1706.05381} {arXiv:1706.05381 [hep-ph]} \BibitemShut {NoStop}%
\bibitem [{\citenamefont {Choi}\ \emph {et~al.}(2017)\citenamefont {Choi}, \citenamefont {Hochberg}, \citenamefont {Kuflik}, \citenamefont {Lee}, \citenamefont {Mambrini}, \citenamefont {Murayama},\ and\ \citenamefont {Pierre}}]{Choi:2017zww}%
  \BibitemOpen
  \bibfield  {author} {\bibinfo {author} {\bibfnamefont {S.-M.}\ \bibnamefont {Choi}}, \bibinfo {author} {\bibfnamefont {Y.}~\bibnamefont {Hochberg}}, \bibinfo {author} {\bibfnamefont {E.}~\bibnamefont {Kuflik}}, \bibinfo {author} {\bibfnamefont {H.~M.}\ \bibnamefont {Lee}}, \bibinfo {author} {\bibfnamefont {Y.}~\bibnamefont {Mambrini}}, \bibinfo {author} {\bibfnamefont {H.}~\bibnamefont {Murayama}},\ and\ \bibinfo {author} {\bibfnamefont {M.}~\bibnamefont {Pierre}},\ }\href {https://doi.org/10.1007/JHEP10(2017)162} {\bibfield  {journal} {\bibinfo  {journal} {JHEP}\ }\textbf {\bibinfo {volume} {10}},\ \bibinfo {pages} {162}},\ \Eprint {https://arxiv.org/abs/1707.01434} {arXiv:1707.01434 [hep-ph]} \BibitemShut {NoStop}%
\bibitem [{\citenamefont {D'Agnolo}\ \emph {et~al.}(2017)\citenamefont {D'Agnolo}, \citenamefont {Pappadopulo},\ and\ \citenamefont {Ruderman}}]{DAgnolo:2017dbv}%
  \BibitemOpen
  \bibfield  {author} {\bibinfo {author} {\bibfnamefont {R.~T.}\ \bibnamefont {D'Agnolo}}, \bibinfo {author} {\bibfnamefont {D.}~\bibnamefont {Pappadopulo}},\ and\ \bibinfo {author} {\bibfnamefont {J.~T.}\ \bibnamefont {Ruderman}},\ }\href {https://doi.org/10.1103/PhysRevLett.119.061102} {\bibfield  {journal} {\bibinfo  {journal} {Phys. Rev. Lett.}\ }\textbf {\bibinfo {volume} {119}},\ \bibinfo {pages} {061102} (\bibinfo {year} {2017})},\ \Eprint {https://arxiv.org/abs/1705.08450} {arXiv:1705.08450 [hep-ph]} \BibitemShut {NoStop}%
\bibitem [{\citenamefont {Falkowski}\ \emph {et~al.}(2019)\citenamefont {Falkowski}, \citenamefont {Kuflik}, \citenamefont {Levi},\ and\ \citenamefont {Volansky}}]{Falkowski:2017uya}%
  \BibitemOpen
  \bibfield  {author} {\bibinfo {author} {\bibfnamefont {A.}~\bibnamefont {Falkowski}}, \bibinfo {author} {\bibfnamefont {E.}~\bibnamefont {Kuflik}}, \bibinfo {author} {\bibfnamefont {N.}~\bibnamefont {Levi}},\ and\ \bibinfo {author} {\bibfnamefont {T.}~\bibnamefont {Volansky}},\ }\href {https://doi.org/10.1103/PhysRevD.99.015022} {\bibfield  {journal} {\bibinfo  {journal} {Phys. Rev. D}\ }\textbf {\bibinfo {volume} {99}},\ \bibinfo {pages} {015022} (\bibinfo {year} {2019})},\ \Eprint {https://arxiv.org/abs/1712.07652} {arXiv:1712.07652 [hep-ph]} \BibitemShut {NoStop}%
\bibitem [{\citenamefont {D'Agnolo}\ \emph {et~al.}(2018)\citenamefont {D'Agnolo}, \citenamefont {Mondino}, \citenamefont {Ruderman},\ and\ \citenamefont {Wang}}]{DAgnolo:2018wcn}%
  \BibitemOpen
  \bibfield  {author} {\bibinfo {author} {\bibfnamefont {R.~T.}\ \bibnamefont {D'Agnolo}}, \bibinfo {author} {\bibfnamefont {C.}~\bibnamefont {Mondino}}, \bibinfo {author} {\bibfnamefont {J.~T.}\ \bibnamefont {Ruderman}},\ and\ \bibinfo {author} {\bibfnamefont {P.-J.}\ \bibnamefont {Wang}},\ }\href {https://doi.org/10.1007/JHEP08(2018)079} {\bibfield  {journal} {\bibinfo  {journal} {JHEP}\ }\textbf {\bibinfo {volume} {08}},\ \bibinfo {pages} {079}},\ \Eprint {https://arxiv.org/abs/1803.02901} {arXiv:1803.02901 [hep-ph]} \BibitemShut {NoStop}%
\bibitem [{\citenamefont {Chu}\ \emph {et~al.}(2019{\natexlab{a}})\citenamefont {Chu}, \citenamefont {Pradler},\ and\ \citenamefont {Semmelrock}}]{Chu:2018qrm}%
  \BibitemOpen
  \bibfield  {author} {\bibinfo {author} {\bibfnamefont {X.}~\bibnamefont {Chu}}, \bibinfo {author} {\bibfnamefont {J.}~\bibnamefont {Pradler}},\ and\ \bibinfo {author} {\bibfnamefont {L.}~\bibnamefont {Semmelrock}},\ }\href {https://doi.org/10.1103/PhysRevD.99.015040} {\bibfield  {journal} {\bibinfo  {journal} {Phys. Rev. D}\ }\textbf {\bibinfo {volume} {99}},\ \bibinfo {pages} {015040} (\bibinfo {year} {2019}{\natexlab{a}})},\ \Eprint {https://arxiv.org/abs/1811.04095} {arXiv:1811.04095 [hep-ph]} \BibitemShut {NoStop}%
\bibitem [{\citenamefont {Dvorkin}\ \emph {et~al.}(2019)\citenamefont {Dvorkin}, \citenamefont {Lin},\ and\ \citenamefont {Schutz}}]{Dvorkin:2019zdi}%
  \BibitemOpen
  \bibfield  {author} {\bibinfo {author} {\bibfnamefont {C.}~\bibnamefont {Dvorkin}}, \bibinfo {author} {\bibfnamefont {T.}~\bibnamefont {Lin}},\ and\ \bibinfo {author} {\bibfnamefont {K.}~\bibnamefont {Schutz}},\ }\href {https://doi.org/10.1103/PhysRevD.99.115009} {\bibfield  {journal} {\bibinfo  {journal} {Phys. Rev. D}\ }\textbf {\bibinfo {volume} {99}},\ \bibinfo {pages} {115009} (\bibinfo {year} {2019})},\ \bibinfo {note} {[Erratum: Phys.Rev.D 105, 119901 (2022)]},\ \Eprint {https://arxiv.org/abs/1902.08623} {arXiv:1902.08623 [hep-ph]} \BibitemShut {NoStop}%
\bibitem [{\citenamefont {D'Agnolo}\ \emph {et~al.}(2020)\citenamefont {D'Agnolo}, \citenamefont {Pappadopulo}, \citenamefont {Ruderman},\ and\ \citenamefont {Wang}}]{DAgnolo:2019zkf}%
  \BibitemOpen
  \bibfield  {author} {\bibinfo {author} {\bibfnamefont {R.~T.}\ \bibnamefont {D'Agnolo}}, \bibinfo {author} {\bibfnamefont {D.}~\bibnamefont {Pappadopulo}}, \bibinfo {author} {\bibfnamefont {J.~T.}\ \bibnamefont {Ruderman}},\ and\ \bibinfo {author} {\bibfnamefont {P.-J.}\ \bibnamefont {Wang}},\ }\href {https://doi.org/10.1103/PhysRevLett.124.151801} {\bibfield  {journal} {\bibinfo  {journal} {Phys. Rev. Lett.}\ }\textbf {\bibinfo {volume} {124}},\ \bibinfo {pages} {151801} (\bibinfo {year} {2020})},\ \Eprint {https://arxiv.org/abs/1906.09269} {arXiv:1906.09269 [hep-ph]} \BibitemShut {NoStop}%
\bibitem [{\citenamefont {Hambye}\ \emph {et~al.}(2019)\citenamefont {Hambye}, \citenamefont {Tytgat}, \citenamefont {Vandecasteele},\ and\ \citenamefont {Vanderheyden}}]{Hambye:2019dwd}%
  \BibitemOpen
  \bibfield  {author} {\bibinfo {author} {\bibfnamefont {T.}~\bibnamefont {Hambye}}, \bibinfo {author} {\bibfnamefont {M.~H.~G.}\ \bibnamefont {Tytgat}}, \bibinfo {author} {\bibfnamefont {J.}~\bibnamefont {Vandecasteele}},\ and\ \bibinfo {author} {\bibfnamefont {L.}~\bibnamefont {Vanderheyden}},\ }\href {https://doi.org/10.1103/PhysRevD.100.095018} {\bibfield  {journal} {\bibinfo  {journal} {Phys. Rev. D}\ }\textbf {\bibinfo {volume} {100}},\ \bibinfo {pages} {095018} (\bibinfo {year} {2019})},\ \Eprint {https://arxiv.org/abs/1908.09864} {arXiv:1908.09864 [hep-ph]} \BibitemShut {NoStop}%
\bibitem [{\citenamefont {Heeba}\ and\ \citenamefont {Kahlhoefer}(2020)}]{Heeba:2019jho}%
  \BibitemOpen
  \bibfield  {author} {\bibinfo {author} {\bibfnamefont {S.}~\bibnamefont {Heeba}}\ and\ \bibinfo {author} {\bibfnamefont {F.}~\bibnamefont {Kahlhoefer}},\ }\href {https://doi.org/10.1103/PhysRevD.101.035043} {\bibfield  {journal} {\bibinfo  {journal} {Phys. Rev. D}\ }\textbf {\bibinfo {volume} {101}},\ \bibinfo {pages} {035043} (\bibinfo {year} {2020})},\ \Eprint {https://arxiv.org/abs/1908.09834} {arXiv:1908.09834 [hep-ph]} \BibitemShut {NoStop}%
\bibitem [{\citenamefont {Evans}\ \emph {et~al.}(2020)\citenamefont {Evans}, \citenamefont {Gaidau},\ and\ \citenamefont {Shelton}}]{Evans:2019vxr}%
  \BibitemOpen
  \bibfield  {author} {\bibinfo {author} {\bibfnamefont {J.~A.}\ \bibnamefont {Evans}}, \bibinfo {author} {\bibfnamefont {C.}~\bibnamefont {Gaidau}},\ and\ \bibinfo {author} {\bibfnamefont {J.}~\bibnamefont {Shelton}},\ }\href {https://doi.org/10.1007/JHEP01(2020)032} {\bibfield  {journal} {\bibinfo  {journal} {JHEP}\ }\textbf {\bibinfo {volume} {01}},\ \bibinfo {pages} {032}},\ \Eprint {https://arxiv.org/abs/1909.04671} {arXiv:1909.04671 [hep-ph]} \BibitemShut {NoStop}%
\bibitem [{\citenamefont {Koren}\ and\ \citenamefont {McGehee}(2020)}]{Koren:2019iuv}%
  \BibitemOpen
  \bibfield  {author} {\bibinfo {author} {\bibfnamefont {S.}~\bibnamefont {Koren}}\ and\ \bibinfo {author} {\bibfnamefont {R.}~\bibnamefont {McGehee}},\ }\href {https://doi.org/10.1103/PhysRevD.101.055024} {\bibfield  {journal} {\bibinfo  {journal} {Phys. Rev. D}\ }\textbf {\bibinfo {volume} {101}},\ \bibinfo {pages} {055024} (\bibinfo {year} {2020})},\ \Eprint {https://arxiv.org/abs/1908.03559} {arXiv:1908.03559 [hep-ph]} \BibitemShut {NoStop}%
\bibitem [{\citenamefont {Chu}\ \emph {et~al.}(2019{\natexlab{b}})\citenamefont {Chu}, \citenamefont {Kuo}, \citenamefont {Pradler},\ and\ \citenamefont {Semmelrock}}]{Chu:2019rok}%
  \BibitemOpen
  \bibfield  {author} {\bibinfo {author} {\bibfnamefont {X.}~\bibnamefont {Chu}}, \bibinfo {author} {\bibfnamefont {J.-L.}\ \bibnamefont {Kuo}}, \bibinfo {author} {\bibfnamefont {J.}~\bibnamefont {Pradler}},\ and\ \bibinfo {author} {\bibfnamefont {L.}~\bibnamefont {Semmelrock}},\ }\href {https://doi.org/10.1103/PhysRevD.100.083002} {\bibfield  {journal} {\bibinfo  {journal} {Phys. Rev. D}\ }\textbf {\bibinfo {volume} {100}},\ \bibinfo {pages} {083002} (\bibinfo {year} {2019}{\natexlab{b}})},\ \Eprint {https://arxiv.org/abs/1908.00553} {arXiv:1908.00553 [hep-ph]} \BibitemShut {NoStop}%
\bibitem [{\citenamefont {Chang}\ \emph {et~al.}(2021)\citenamefont {Chang}, \citenamefont {Essig},\ and\ \citenamefont {Reinert}}]{Chang:2019xva}%
  \BibitemOpen
  \bibfield  {author} {\bibinfo {author} {\bibfnamefont {J.~H.}\ \bibnamefont {Chang}}, \bibinfo {author} {\bibfnamefont {R.}~\bibnamefont {Essig}},\ and\ \bibinfo {author} {\bibfnamefont {A.}~\bibnamefont {Reinert}},\ }\href {https://doi.org/10.1007/JHEP03(2021)141} {\bibfield  {journal} {\bibinfo  {journal} {JHEP}\ }\textbf {\bibinfo {volume} {03}},\ \bibinfo {pages} {141}},\ \Eprint {https://arxiv.org/abs/1911.03389} {arXiv:1911.03389 [hep-ph]} \BibitemShut {NoStop}%
\bibitem [{\citenamefont {Darm\'e}\ \emph {et~al.}(2018)\citenamefont {Darm\'e}, \citenamefont {Rao},\ and\ \citenamefont {Roszkowski}}]{Darme:2017glc}%
  \BibitemOpen
  \bibfield  {author} {\bibinfo {author} {\bibfnamefont {L.}~\bibnamefont {Darm\'e}}, \bibinfo {author} {\bibfnamefont {S.}~\bibnamefont {Rao}},\ and\ \bibinfo {author} {\bibfnamefont {L.}~\bibnamefont {Roszkowski}},\ }\href {https://doi.org/10.1007/JHEP03(2018)084} {\bibfield  {journal} {\bibinfo  {journal} {JHEP}\ }\textbf {\bibinfo {volume} {03}},\ \bibinfo {pages} {084}},\ \Eprint {https://arxiv.org/abs/1710.08430} {arXiv:1710.08430 [hep-ph]} \BibitemShut {NoStop}%
\bibitem [{\citenamefont {March-Russell}\ \emph {et~al.}(2020)\citenamefont {March-Russell}, \citenamefont {Tillim},\ and\ \citenamefont {West}}]{March-Russell:2020nun}%
  \BibitemOpen
  \bibfield  {author} {\bibinfo {author} {\bibfnamefont {J.}~\bibnamefont {March-Russell}}, \bibinfo {author} {\bibfnamefont {H.}~\bibnamefont {Tillim}},\ and\ \bibinfo {author} {\bibfnamefont {S.~M.}\ \bibnamefont {West}},\ }\href {https://doi.org/10.1103/PhysRevD.102.083018} {\bibfield  {journal} {\bibinfo  {journal} {Phys. Rev. D}\ }\textbf {\bibinfo {volume} {102}},\ \bibinfo {pages} {083018} (\bibinfo {year} {2020})},\ \Eprint {https://arxiv.org/abs/2007.14688} {arXiv:2007.14688 [astro-ph.CO]} \BibitemShut {NoStop}%
\bibitem [{\citenamefont {Dodelson}\ and\ \citenamefont {Widrow}(1994)}]{Dodelson:1993je}%
  \BibitemOpen
  \bibfield  {author} {\bibinfo {author} {\bibfnamefont {S.}~\bibnamefont {Dodelson}}\ and\ \bibinfo {author} {\bibfnamefont {L.~M.}\ \bibnamefont {Widrow}},\ }\href {https://doi.org/10.1103/PhysRevLett.72.17} {\bibfield  {journal} {\bibinfo  {journal} {Phys. Rev. Lett.}\ }\textbf {\bibinfo {volume} {72}},\ \bibinfo {pages} {17} (\bibinfo {year} {1994})},\ \Eprint {https://arxiv.org/abs/hep-ph/9303287} {arXiv:hep-ph/9303287} \BibitemShut {NoStop}%
\bibitem [{\citenamefont {Shi}\ and\ \citenamefont {Fuller}(1999)}]{Shi:1998km}%
  \BibitemOpen
  \bibfield  {author} {\bibinfo {author} {\bibfnamefont {X.-D.}\ \bibnamefont {Shi}}\ and\ \bibinfo {author} {\bibfnamefont {G.~M.}\ \bibnamefont {Fuller}},\ }\href {https://doi.org/10.1103/PhysRevLett.82.2832} {\bibfield  {journal} {\bibinfo  {journal} {Phys. Rev. Lett.}\ }\textbf {\bibinfo {volume} {82}},\ \bibinfo {pages} {2832} (\bibinfo {year} {1999})},\ \Eprint {https://arxiv.org/abs/astro-ph/9810076} {arXiv:astro-ph/9810076} \BibitemShut {NoStop}%
\bibitem [{\citenamefont {Abazajian}\ \emph {et~al.}(2001)\citenamefont {Abazajian}, \citenamefont {Fuller},\ and\ \citenamefont {Patel}}]{Abazajian:2001nj}%
  \BibitemOpen
  \bibfield  {author} {\bibinfo {author} {\bibfnamefont {K.}~\bibnamefont {Abazajian}}, \bibinfo {author} {\bibfnamefont {G.~M.}\ \bibnamefont {Fuller}},\ and\ \bibinfo {author} {\bibfnamefont {M.}~\bibnamefont {Patel}},\ }\href {https://doi.org/10.1103/PhysRevD.64.023501} {\bibfield  {journal} {\bibinfo  {journal} {Phys. Rev. D}\ }\textbf {\bibinfo {volume} {64}},\ \bibinfo {pages} {023501} (\bibinfo {year} {2001})},\ \Eprint {https://arxiv.org/abs/astro-ph/0101524} {arXiv:astro-ph/0101524} \BibitemShut {NoStop}%
\bibitem [{\citenamefont {Abazajian}\ \emph {et~al.}(2012)\citenamefont {Abazajian} \emph {et~al.}}]{Abazajian:2012ys}%
  \BibitemOpen
  \bibfield  {author} {\bibinfo {author} {\bibfnamefont {K.~N.}\ \bibnamefont {Abazajian}} \emph {et~al.},\ }\href@noop {} {\  (\bibinfo {year} {2012})},\ \Eprint {https://arxiv.org/abs/1204.5379} {arXiv:1204.5379 [hep-ph]} \BibitemShut {NoStop}%
\bibitem [{\citenamefont {Abazajian}(2017)}]{Abazajian:2017tcc}%
  \BibitemOpen
  \bibfield  {author} {\bibinfo {author} {\bibfnamefont {K.~N.}\ \bibnamefont {Abazajian}},\ }\href {https://doi.org/10.1016/j.physrep.2017.10.003} {\bibfield  {journal} {\bibinfo  {journal} {Phys. Rept.}\ }\textbf {\bibinfo {volume} {711-712}},\ \bibinfo {pages} {1} (\bibinfo {year} {2017})},\ \Eprint {https://arxiv.org/abs/1705.01837} {arXiv:1705.01837 [hep-ph]} \BibitemShut {NoStop}%
\bibitem [{\citenamefont {Boyarsky}\ \emph {et~al.}(2019)\citenamefont {Boyarsky}, \citenamefont {Drewes}, \citenamefont {Lasserre}, \citenamefont {Mertens},\ and\ \citenamefont {Ruchayskiy}}]{Boyarsky:2018tvu}%
  \BibitemOpen
  \bibfield  {author} {\bibinfo {author} {\bibfnamefont {A.}~\bibnamefont {Boyarsky}}, \bibinfo {author} {\bibfnamefont {M.}~\bibnamefont {Drewes}}, \bibinfo {author} {\bibfnamefont {T.}~\bibnamefont {Lasserre}}, \bibinfo {author} {\bibfnamefont {S.}~\bibnamefont {Mertens}},\ and\ \bibinfo {author} {\bibfnamefont {O.}~\bibnamefont {Ruchayskiy}},\ }\href {https://doi.org/10.1016/j.ppnp.2018.07.004} {\bibfield  {journal} {\bibinfo  {journal} {Prog. Part. Nucl. Phys.}\ }\textbf {\bibinfo {volume} {104}},\ \bibinfo {pages} {1} (\bibinfo {year} {2019})},\ \Eprint {https://arxiv.org/abs/1807.07938} {arXiv:1807.07938 [hep-ph]} \BibitemShut {NoStop}%
%%CITATION = ARXIV:1807.07938;%%
\bibitem [{\citenamefont {Hochberg}\ \emph {et~al.}(2019{\natexlab{b}})\citenamefont {Hochberg}, \citenamefont {Kuflik},\ and\ \citenamefont {Murayama}}]{Hochberg:2018vdo}%
  \BibitemOpen
  \bibfield  {author} {\bibinfo {author} {\bibfnamefont {Y.}~\bibnamefont {Hochberg}}, \bibinfo {author} {\bibfnamefont {E.}~\bibnamefont {Kuflik}},\ and\ \bibinfo {author} {\bibfnamefont {H.}~\bibnamefont {Murayama}},\ }\href {https://doi.org/10.1103/PhysRevD.99.015005} {\bibfield  {journal} {\bibinfo  {journal} {Phys. Rev. D}\ }\textbf {\bibinfo {volume} {99}},\ \bibinfo {pages} {015005} (\bibinfo {year} {2019}{\natexlab{b}})},\ \Eprint {https://arxiv.org/abs/1805.09345} {arXiv:1805.09345 [hep-ph]} \BibitemShut {NoStop}%
\bibitem [{\citenamefont {Hochberg}\ \emph {et~al.}(2018{\natexlab{b}})\citenamefont {Hochberg}, \citenamefont {Kuflik}, \citenamefont {Mcgehee}, \citenamefont {Murayama},\ and\ \citenamefont {Schutz}}]{Hochberg:2018rjs}%
  \BibitemOpen
  \bibfield  {author} {\bibinfo {author} {\bibfnamefont {Y.}~\bibnamefont {Hochberg}}, \bibinfo {author} {\bibfnamefont {E.}~\bibnamefont {Kuflik}}, \bibinfo {author} {\bibfnamefont {R.}~\bibnamefont {Mcgehee}}, \bibinfo {author} {\bibfnamefont {H.}~\bibnamefont {Murayama}},\ and\ \bibinfo {author} {\bibfnamefont {K.}~\bibnamefont {Schutz}},\ }\href {https://doi.org/10.1103/PhysRevD.98.115031} {\bibfield  {journal} {\bibinfo  {journal} {Phys. Rev. D}\ }\textbf {\bibinfo {volume} {98}},\ \bibinfo {pages} {115031} (\bibinfo {year} {2018}{\natexlab{b}})},\ \Eprint {https://arxiv.org/abs/1806.10139} {arXiv:1806.10139 [hep-ph]} \BibitemShut {NoStop}%
\bibitem [{\citenamefont {Smirnov}\ and\ \citenamefont {Beacom}(2020)}]{Smirnov:2020zwf}%
  \BibitemOpen
  \bibfield  {author} {\bibinfo {author} {\bibfnamefont {J.}~\bibnamefont {Smirnov}}\ and\ \bibinfo {author} {\bibfnamefont {J.~F.}\ \bibnamefont {Beacom}},\ }\href {https://doi.org/10.1103/PhysRevLett.125.131301} {\bibfield  {journal} {\bibinfo  {journal} {Phys. Rev. Lett.}\ }\textbf {\bibinfo {volume} {125}},\ \bibinfo {pages} {131301} (\bibinfo {year} {2020})},\ \Eprint {https://arxiv.org/abs/2002.04038} {arXiv:2002.04038 [hep-ph]} \BibitemShut {NoStop}%
\bibitem [{\citenamefont {Dutta}\ \emph {et~al.}(2019)\citenamefont {Dutta}, \citenamefont {Ghosh},\ and\ \citenamefont {Kumar}}]{Dutta:2019fxn}%
  \BibitemOpen
  \bibfield  {author} {\bibinfo {author} {\bibfnamefont {B.}~\bibnamefont {Dutta}}, \bibinfo {author} {\bibfnamefont {S.}~\bibnamefont {Ghosh}},\ and\ \bibinfo {author} {\bibfnamefont {J.}~\bibnamefont {Kumar}},\ }\href {https://doi.org/10.1103/PhysRevD.100.075028} {\bibfield  {journal} {\bibinfo  {journal} {Phys. Rev. D}\ }\textbf {\bibinfo {volume} {100}},\ \bibinfo {pages} {075028} (\bibinfo {year} {2019})},\ \Eprint {https://arxiv.org/abs/1905.02692} {arXiv:1905.02692 [hep-ph]} \BibitemShut {NoStop}%
\bibitem [{\citenamefont {Dutta}\ \emph {et~al.}(2020)\citenamefont {Dutta}, \citenamefont {Ghosh},\ and\ \citenamefont {Kumar}}]{Dutta:2020enk}%
  \BibitemOpen
  \bibfield  {author} {\bibinfo {author} {\bibfnamefont {B.}~\bibnamefont {Dutta}}, \bibinfo {author} {\bibfnamefont {S.}~\bibnamefont {Ghosh}},\ and\ \bibinfo {author} {\bibfnamefont {J.}~\bibnamefont {Kumar}},\ }\href {https://doi.org/10.1103/PhysRevD.102.075041} {\bibfield  {journal} {\bibinfo  {journal} {Phys. Rev. D}\ }\textbf {\bibinfo {volume} {102}},\ \bibinfo {pages} {075041} (\bibinfo {year} {2020})},\ \Eprint {https://arxiv.org/abs/2007.16191} {arXiv:2007.16191 [hep-ph]} \BibitemShut {NoStop}%
\bibitem [{\citenamefont {Dutta}\ \emph {et~al.}(2022)\citenamefont {Dutta}, \citenamefont {Ghosh},\ and\ \citenamefont {Kumar}}]{Dutta:2022qvn}%
  \BibitemOpen
  \bibfield  {author} {\bibinfo {author} {\bibfnamefont {B.}~\bibnamefont {Dutta}}, \bibinfo {author} {\bibfnamefont {S.}~\bibnamefont {Ghosh}},\ and\ \bibinfo {author} {\bibfnamefont {J.}~\bibnamefont {Kumar}},\ }in\ \href@noop {} {\emph {\bibinfo {booktitle} {{Snowmass 2021}}}}\ (\bibinfo {year} {2022})\ \Eprint {https://arxiv.org/abs/2203.07786} {arXiv:2203.07786 [hep-ph]} \BibitemShut {NoStop}%
\bibitem [{\citenamefont {Fernandez}\ \emph {et~al.}(2022)\citenamefont {Fernandez}, \citenamefont {Kahn},\ and\ \citenamefont {Shelton}}]{Fernandez:2021iti}%
  \BibitemOpen
  \bibfield  {author} {\bibinfo {author} {\bibfnamefont {N.}~\bibnamefont {Fernandez}}, \bibinfo {author} {\bibfnamefont {Y.}~\bibnamefont {Kahn}},\ and\ \bibinfo {author} {\bibfnamefont {J.}~\bibnamefont {Shelton}},\ }\href {https://doi.org/10.1007/JHEP07(2022)044} {\bibfield  {journal} {\bibinfo  {journal} {JHEP}\ }\textbf {\bibinfo {volume} {07}},\ \bibinfo {pages} {044}},\ \Eprint {https://arxiv.org/abs/2111.13709} {arXiv:2111.13709 [hep-ph]} \BibitemShut {NoStop}%
\bibitem [{\citenamefont {Iles}\ \emph {et~al.}(2025)\citenamefont {Iles}, \citenamefont {Heeba},\ and\ \citenamefont {Schutz}}]{Iles:2024zka}%
  \BibitemOpen
  \bibfield  {author} {\bibinfo {author} {\bibfnamefont {E.}~\bibnamefont {Iles}}, \bibinfo {author} {\bibfnamefont {S.}~\bibnamefont {Heeba}},\ and\ \bibinfo {author} {\bibfnamefont {K.}~\bibnamefont {Schutz}},\ }\href {https://doi.org/10.1103/PhysRevLett.134.121002} {\bibfield  {journal} {\bibinfo  {journal} {Phys. Rev. Lett.}\ }\textbf {\bibinfo {volume} {134}},\ \bibinfo {pages} {121002} (\bibinfo {year} {2025})},\ \Eprint {https://arxiv.org/abs/2407.21096} {arXiv:2407.21096 [hep-ph]} \BibitemShut {NoStop}%
\bibitem [{\citenamefont {Lee}\ \emph {et~al.}(2013)\citenamefont {Lee}, \citenamefont {Lisanti},\ and\ \citenamefont {Safdi}}]{Lee:2013xxa}%
  \BibitemOpen
  \bibfield  {author} {\bibinfo {author} {\bibfnamefont {S.~K.}\ \bibnamefont {Lee}}, \bibinfo {author} {\bibfnamefont {M.}~\bibnamefont {Lisanti}},\ and\ \bibinfo {author} {\bibfnamefont {B.~R.}\ \bibnamefont {Safdi}},\ }\href {https://doi.org/10.1088/1475-7516/2013/11/033} {\bibfield  {journal} {\bibinfo  {journal} {JCAP}\ }\textbf {\bibinfo {volume} {11}},\ \bibinfo {pages} {033}},\ \Eprint {https://arxiv.org/abs/1307.5323} {arXiv:1307.5323 [hep-ph]} \BibitemShut {NoStop}%
\bibitem [{\citenamefont {Tilav}\ \emph {et~al.}(2020)\citenamefont {Tilav}, \citenamefont {Gaisser}, \citenamefont {Soldin},\ and\ \citenamefont {Desiati}}]{Tilav:2019xmf}%
  \BibitemOpen
  \bibfield  {author} {\bibinfo {author} {\bibfnamefont {S.}~\bibnamefont {Tilav}}, \bibinfo {author} {\bibfnamefont {T.~K.}\ \bibnamefont {Gaisser}}, \bibinfo {author} {\bibfnamefont {D.}~\bibnamefont {Soldin}},\ and\ \bibinfo {author} {\bibfnamefont {P.}~\bibnamefont {Desiati}} (\bibinfo {collaboration} {IceCube}),\ }\href {https://doi.org/10.22323/1.358.0894} {\bibfield  {journal} {\bibinfo  {journal} {PoS}\ }\textbf {\bibinfo {volume} {ICRC2019}},\ \bibinfo {pages} {894} (\bibinfo {year} {2020})},\ \Eprint {https://arxiv.org/abs/1909.01406} {arXiv:1909.01406 [astro-ph.HE]} \BibitemShut {NoStop}%
\bibitem [{\citenamefont {Agostini}\ \emph {et~al.}(2019)\citenamefont {Agostini} \emph {et~al.}}]{Borexino:2018pev}%
  \BibitemOpen
  \bibfield  {author} {\bibinfo {author} {\bibfnamefont {M.}~\bibnamefont {Agostini}} \emph {et~al.} (\bibinfo {collaboration} {Borexino}),\ }\href {https://doi.org/10.1088/1475-7516/2019/02/046} {\bibfield  {journal} {\bibinfo  {journal} {JCAP}\ }\textbf {\bibinfo {volume} {02}},\ \bibinfo {pages} {046}},\ \Eprint {https://arxiv.org/abs/1808.04207} {arXiv:1808.04207 [hep-ex]} \BibitemShut {NoStop}%
\end{thebibliography}%
\bibliographystyle{apsrev4-2}

\end{document}